\documentclass[longauth]{aaEC}
\usepackage{graphicx}
\usepackage{natbib}
\usepackage{scalerel}
\usepackage{multirow}
\usepackage{tabularx}
\usepackage[table]{xcolor}
\usepackage{tensor}
\usepackage{dsfont}
\usepackage[normalem]{ulem}

\usepackage{txfonts}
\usepackage[pdfencoding=auto,psdextra]{hyperref}
\hypersetup{
    colorlinks=true,
    linkcolor=blue,
    filecolor=magenta,      
    urlcolor=blue,
    citecolor=blue
}
\usepackage{lastpage}
\makeatletter
\renewcommand*\aa@pageof{, page \thepage{} of \pageref*{LastPage}}
\makeatother

\usepackage[utf8]{inputenc}

\usepackage[switch, modulo]{lineno}

\usepackage{euclid}
\usepackage[nolist,nohyperlinks]{acronym}

\newcommand*{\sqdeg}{deg$^{\,2}$\xspace}

\newcommand*{\JWST}{\textit{James Webb} Space Telescope\xspace}

\begin{document} 


   \title{\Euclid}
    \subtitle{Properties and performance of the NISP signal estimator}    


\newcommand{\orcid}[1]{} 
\author{Euclid Collaboration: F.~Cogato\orcid{0000-0003-4632-6113}\inst{\ref{aff1},\ref{aff2}}
\and B.~Kubik\orcid{0009-0006-5823-4880}\thanks{\email{b.kubik@ip2i.in2p3.fr}}\inst{\ref{aff3}}
\and R.~Barbier\inst{\ref{aff3}}
\and S.~Conseil\orcid{0000-0002-3657-4191}\inst{\ref{aff3}}
\and E.~Medinaceli\orcid{0000-0002-4040-7783}\inst{\ref{aff2}}
\and Y.~Copin\orcid{0000-0002-5317-7518}\inst{\ref{aff3}}
\and E.~Franceschi\orcid{0000-0002-0585-6591}\inst{\ref{aff2}}
\and L.~Valenziano\orcid{0000-0002-1170-0104}\inst{\ref{aff2},\ref{aff4}}
\and N.~Aghanim\orcid{0000-0002-6688-8992}\inst{\ref{aff5}}
\and B.~Altieri\orcid{0000-0003-3936-0284}\inst{\ref{aff6}}
\and S.~Andreon\orcid{0000-0002-2041-8784}\inst{\ref{aff7}}
\and N.~Auricchio\orcid{0000-0003-4444-8651}\inst{\ref{aff2}}
\and C.~Baccigalupi\orcid{0000-0002-8211-1630}\inst{\ref{aff8},\ref{aff9},\ref{aff10},\ref{aff11}}
\and M.~Baldi\orcid{0000-0003-4145-1943}\inst{\ref{aff12},\ref{aff2},\ref{aff13}}
\and A.~Balestra\orcid{0000-0002-6967-261X}\inst{\ref{aff14}}
\and S.~Bardelli\orcid{0000-0002-8900-0298}\inst{\ref{aff2}}
\and P.~Battaglia\orcid{0000-0002-7337-5909}\inst{\ref{aff2}}
\and A.~Biviano\orcid{0000-0002-0857-0732}\inst{\ref{aff9},\ref{aff8}}
\and E.~Branchini\orcid{0000-0002-0808-6908}\inst{\ref{aff15},\ref{aff16},\ref{aff7}}
\and M.~Brescia\orcid{0000-0001-9506-5680}\inst{\ref{aff17},\ref{aff18}}
\and J.~Brinchmann\orcid{0000-0003-4359-8797}\inst{\ref{aff19},\ref{aff20},\ref{aff21}}
\and S.~Camera\orcid{0000-0003-3399-3574}\inst{\ref{aff22},\ref{aff23},\ref{aff24}}
\and G.~Ca\~nas-Herrera\orcid{0000-0003-2796-2149}\inst{\ref{aff25},\ref{aff26}}
\and V.~Capobianco\orcid{0000-0002-3309-7692}\inst{\ref{aff24}}
\and C.~Carbone\orcid{0000-0003-0125-3563}\inst{\ref{aff27}}
\and J.~Carretero\orcid{0000-0002-3130-0204}\inst{\ref{aff28},\ref{aff29}}
\and S.~Casas\orcid{0000-0002-4751-5138}\inst{\ref{aff30},\ref{aff31}}
\and M.~Castellano\orcid{0000-0001-9875-8263}\inst{\ref{aff32}}
\and G.~Castignani\orcid{0000-0001-6831-0687}\inst{\ref{aff2}}
\and S.~Cavuoti\orcid{0000-0002-3787-4196}\inst{\ref{aff18},\ref{aff33}}
\and A.~Cimatti\inst{\ref{aff34}}
\and C.~Colodro-Conde\inst{\ref{aff35}}
\and G.~Congedo\orcid{0000-0003-2508-0046}\inst{\ref{aff25}}
\and C.~J.~Conselice\orcid{0000-0003-1949-7638}\inst{\ref{aff36}}
\and L.~Conversi\orcid{0000-0002-6710-8476}\inst{\ref{aff37},\ref{aff6}}
\and L.~Corcione\orcid{0000-0002-6497-5881}\inst{\ref{aff24}}
\and A.~Costille\inst{\ref{aff38}}
\and F.~Courbin\orcid{0000-0003-0758-6510}\inst{\ref{aff39},\ref{aff40},\ref{aff41}}
\and H.~M.~Courtois\orcid{0000-0003-0509-1776}\inst{\ref{aff42}}
\and R.~da~Silva\orcid{0000-0003-4788-677X}\inst{\ref{aff32},\ref{aff43}}
\and H.~Degaudenzi\orcid{0000-0002-5887-6799}\inst{\ref{aff44}}
\and G.~De~Lucia\orcid{0000-0002-6220-9104}\inst{\ref{aff9}}
\and H.~Dole\orcid{0000-0002-9767-3839}\inst{\ref{aff5}}
\and F.~Dubath\orcid{0000-0002-6533-2810}\inst{\ref{aff44}}
\and X.~Dupac\inst{\ref{aff6}}
\and S.~Dusini\orcid{0000-0002-1128-0664}\inst{\ref{aff45}}
\and A.~Ealet\orcid{0000-0003-3070-014X}\inst{\ref{aff3}}
\and S.~Escoffier\orcid{0000-0002-2847-7498}\inst{\ref{aff46}}
\and M.~Farina\orcid{0000-0002-3089-7846}\inst{\ref{aff47}}
\and R.~Farinelli\inst{\ref{aff2}}
\and F.~Faustini\orcid{0000-0001-6274-5145}\inst{\ref{aff32},\ref{aff43}}
\and S.~Ferriol\inst{\ref{aff3}}
\and F.~Finelli\orcid{0000-0002-6694-3269}\inst{\ref{aff2},\ref{aff4}}
\and N.~Fourmanoit\orcid{0009-0005-6816-6925}\inst{\ref{aff46}}
\and M.~Frailis\orcid{0000-0002-7400-2135}\inst{\ref{aff9}}
\and M.~Fumana\orcid{0000-0001-6787-5950}\inst{\ref{aff27}}
\and S.~Galeotta\orcid{0000-0002-3748-5115}\inst{\ref{aff9}}
\and K.~George\orcid{0000-0002-1734-8455}\inst{\ref{aff48}}
\and W.~Gillard\orcid{0000-0003-4744-9748}\inst{\ref{aff46}}
\and B.~Gillis\orcid{0000-0002-4478-1270}\inst{\ref{aff25}}
\and C.~Giocoli\orcid{0000-0002-9590-7961}\inst{\ref{aff2},\ref{aff13}}
\and J.~Gracia-Carpio\inst{\ref{aff49}}
\and A.~Grazian\orcid{0000-0002-5688-0663}\inst{\ref{aff14}}
\and F.~Grupp\inst{\ref{aff49},\ref{aff50}}
\and S.~V.~H.~Haugan\orcid{0000-0001-9648-7260}\inst{\ref{aff51}}
\and W.~Holmes\inst{\ref{aff52}}
\and F.~Hormuth\inst{\ref{aff53}}
\and A.~Hornstrup\orcid{0000-0002-3363-0936}\inst{\ref{aff54},\ref{aff55}}
\and P.~Hudelot\inst{\ref{aff56}}
\and K.~Jahnke\orcid{0000-0003-3804-2137}\inst{\ref{aff57}}
\and M.~Jhabvala\inst{\ref{aff58}}
\and E.~Keih\"anen\orcid{0000-0003-1804-7715}\inst{\ref{aff59}}
\and S.~Kermiche\orcid{0000-0002-0302-5735}\inst{\ref{aff46}}
\and A.~Kiessling\orcid{0000-0002-2590-1273}\inst{\ref{aff52}}
\and R.~Kohley\inst{\ref{aff6}}
\and M.~K\"ummel\orcid{0000-0003-2791-2117}\inst{\ref{aff50}}
\and M.~Kunz\orcid{0000-0002-3052-7394}\inst{\ref{aff60}}
\and H.~Kurki-Suonio\orcid{0000-0002-4618-3063}\inst{\ref{aff61},\ref{aff62}}
\and A.~M.~C.~Le~Brun\orcid{0000-0002-0936-4594}\inst{\ref{aff63}}
\and S.~Ligori\orcid{0000-0003-4172-4606}\inst{\ref{aff24}}
\and P.~B.~Lilje\orcid{0000-0003-4324-7794}\inst{\ref{aff51}}
\and V.~Lindholm\orcid{0000-0003-2317-5471}\inst{\ref{aff61},\ref{aff62}}
\and I.~Lloro\orcid{0000-0001-5966-1434}\inst{\ref{aff64}}
\and G.~Mainetti\orcid{0000-0003-2384-2377}\inst{\ref{aff65}}
\and D.~Maino\inst{\ref{aff66},\ref{aff27},\ref{aff67}}
\and E.~Maiorano\orcid{0000-0003-2593-4355}\inst{\ref{aff2}}
\and O.~Mansutti\orcid{0000-0001-5758-4658}\inst{\ref{aff9}}
\and S.~Marcin\inst{\ref{aff68}}
\and O.~Marggraf\orcid{0000-0001-7242-3852}\inst{\ref{aff69}}
\and M.~Martinelli\orcid{0000-0002-6943-7732}\inst{\ref{aff32},\ref{aff70}}
\and N.~Martinet\orcid{0000-0003-2786-7790}\inst{\ref{aff38}}
\and F.~Marulli\orcid{0000-0002-8850-0303}\inst{\ref{aff1},\ref{aff2},\ref{aff13}}
\and R.~J.~Massey\orcid{0000-0002-6085-3780}\inst{\ref{aff71}}
\and S.~Mei\orcid{0000-0002-2849-559X}\inst{\ref{aff72},\ref{aff73}}
\and Y.~Mellier\thanks{Deceased}\inst{\ref{aff74},\ref{aff56}}
\and M.~Meneghetti\orcid{0000-0003-1225-7084}\inst{\ref{aff2},\ref{aff13}}
\and E.~Merlin\orcid{0000-0001-6870-8900}\inst{\ref{aff32}}
\and G.~Meylan\inst{\ref{aff75}}
\and A.~Mora\orcid{0000-0002-1922-8529}\inst{\ref{aff76}}
\and M.~Moresco\orcid{0000-0002-7616-7136}\inst{\ref{aff1},\ref{aff2}}
\and L.~Moscardini\orcid{0000-0002-3473-6716}\inst{\ref{aff1},\ref{aff2},\ref{aff13}}
\and C.~Neissner\orcid{0000-0001-8524-4968}\inst{\ref{aff77},\ref{aff29}}
\and S.-M.~Niemi\orcid{0009-0005-0247-0086}\inst{\ref{aff78}}
\and C.~Padilla\orcid{0000-0001-7951-0166}\inst{\ref{aff77}}
\and S.~Paltani\orcid{0000-0002-8108-9179}\inst{\ref{aff44}}
\and F.~Pasian\orcid{0000-0002-4869-3227}\inst{\ref{aff9}}
\and K.~Pedersen\inst{\ref{aff79}}
\and W.~J.~Percival\orcid{0000-0002-0644-5727}\inst{\ref{aff80},\ref{aff81},\ref{aff82}}
\and V.~Pettorino\orcid{0000-0002-4203-9320}\inst{\ref{aff78}}
\and S.~Pires\orcid{0000-0002-0249-2104}\inst{\ref{aff83}}
\and G.~Polenta\orcid{0000-0003-4067-9196}\inst{\ref{aff43}}
\and M.~Poncet\inst{\ref{aff84}}
\and L.~A.~Popa\inst{\ref{aff85}}
\and F.~Raison\orcid{0000-0002-7819-6918}\inst{\ref{aff49}}
\and A.~Renzi\orcid{0000-0001-9856-1970}\inst{\ref{aff86},\ref{aff45}}
\and J.~Rhodes\orcid{0000-0002-4485-8549}\inst{\ref{aff52}}
\and G.~Riccio\inst{\ref{aff18}}
\and E.~Romelli\orcid{0000-0003-3069-9222}\inst{\ref{aff9}}
\and M.~Roncarelli\orcid{0000-0001-9587-7822}\inst{\ref{aff2}}
\and C.~Rosset\orcid{0000-0003-0286-2192}\inst{\ref{aff72}}
\and E.~Rossetti\orcid{0000-0003-0238-4047}\inst{\ref{aff12}}
\and R.~Saglia\orcid{0000-0003-0378-7032}\inst{\ref{aff50},\ref{aff49}}
\and Z.~Sakr\orcid{0000-0002-4823-3757}\inst{\ref{aff87},\ref{aff88},\ref{aff89}}
\and A.~G.~S\'anchez\orcid{0000-0003-1198-831X}\inst{\ref{aff49}}
\and D.~Sapone\orcid{0000-0001-7089-4503}\inst{\ref{aff90}}
\and B.~Sartoris\orcid{0000-0003-1337-5269}\inst{\ref{aff50},\ref{aff9}}
\and M.~Schirmer\orcid{0000-0003-2568-9994}\inst{\ref{aff57}}
\and P.~Schneider\orcid{0000-0001-8561-2679}\inst{\ref{aff69}}
\and M.~Scodeggio\inst{\ref{aff27}}
\and A.~Secroun\orcid{0000-0003-0505-3710}\inst{\ref{aff46}}
\and G.~Seidel\orcid{0000-0003-2907-353X}\inst{\ref{aff57}}
\and S.~Serrano\orcid{0000-0002-0211-2861}\inst{\ref{aff91},\ref{aff92},\ref{aff93}}
\and P.~Simon\inst{\ref{aff69}}
\and C.~Sirignano\orcid{0000-0002-0995-7146}\inst{\ref{aff86},\ref{aff45}}
\and G.~Sirri\orcid{0000-0003-2626-2853}\inst{\ref{aff13}}
\and L.~Stanco\orcid{0000-0002-9706-5104}\inst{\ref{aff45}}
\and J.~Steinwagner\orcid{0000-0001-7443-1047}\inst{\ref{aff49}}
\and P.~Tallada-Cresp\'{i}\orcid{0000-0002-1336-8328}\inst{\ref{aff28},\ref{aff29}}
\and D.~Tavagnacco\orcid{0000-0001-7475-9894}\inst{\ref{aff9}}
\and A.~N.~Taylor\inst{\ref{aff25}}
\and I.~Tereno\orcid{0000-0002-4537-6218}\inst{\ref{aff94},\ref{aff95}}
\and S.~Toft\orcid{0000-0003-3631-7176}\inst{\ref{aff96},\ref{aff97}}
\and R.~Toledo-Moreo\orcid{0000-0002-2997-4859}\inst{\ref{aff98}}
\and F.~Torradeflot\orcid{0000-0003-1160-1517}\inst{\ref{aff29},\ref{aff28}}
\and I.~Tutusaus\orcid{0000-0002-3199-0399}\inst{\ref{aff93},\ref{aff91},\ref{aff88}}
\and J.~Valiviita\orcid{0000-0001-6225-3693}\inst{\ref{aff61},\ref{aff62}}
\and T.~Vassallo\orcid{0000-0001-6512-6358}\inst{\ref{aff9}}
\and A.~Veropalumbo\orcid{0000-0003-2387-1194}\inst{\ref{aff7},\ref{aff16},\ref{aff15}}
\and Y.~Wang\orcid{0000-0002-4749-2984}\inst{\ref{aff99}}
\and J.~Weller\orcid{0000-0002-8282-2010}\inst{\ref{aff50},\ref{aff49}}
\and A.~Zacchei\orcid{0000-0003-0396-1192}\inst{\ref{aff9},\ref{aff8}}
\and F.~M.~Zerbi\inst{\ref{aff7}}
\and E.~Zucca\orcid{0000-0002-5845-8132}\inst{\ref{aff2}}
\and M.~Ballardini\orcid{0000-0003-4481-3559}\inst{\ref{aff100},\ref{aff101},\ref{aff2}}
\and M.~Bolzonella\orcid{0000-0003-3278-4607}\inst{\ref{aff2}}
\and E.~Bozzo\orcid{0000-0002-8201-1525}\inst{\ref{aff44}}
\and C.~Burigana\orcid{0000-0002-3005-5796}\inst{\ref{aff102},\ref{aff4}}
\and R.~Cabanac\orcid{0000-0001-6679-2600}\inst{\ref{aff88}}
\and M.~Calabrese\orcid{0000-0002-2637-2422}\inst{\ref{aff103},\ref{aff27}}
\and A.~Cappi\inst{\ref{aff104},\ref{aff2}}
\and T.~Castro\orcid{0000-0002-6292-3228}\inst{\ref{aff9},\ref{aff10},\ref{aff8},\ref{aff105}}
\and J.~A.~Escartin~Vigo\inst{\ref{aff49}}
\and G.~Fabbian\orcid{0000-0002-3255-4695}\inst{\ref{aff5}}
\and L.~Gabarra\orcid{0000-0002-8486-8856}\inst{\ref{aff106}}
\and J.~Garc\'ia-Bellido\orcid{0000-0002-9370-8360}\inst{\ref{aff107}}
\and V.~Gautard\inst{\ref{aff108}}
\and S.~Hemmati\orcid{0000-0003-2226-5395}\inst{\ref{aff99}}
\and J.~Macias-Perez\orcid{0000-0002-5385-2763}\inst{\ref{aff109}}
\and R.~Maoli\orcid{0000-0002-6065-3025}\inst{\ref{aff110},\ref{aff32}}
\and J.~Mart\'{i}n-Fleitas\orcid{0000-0002-8594-569X}\inst{\ref{aff111}}
\and N.~Mauri\orcid{0000-0001-8196-1548}\inst{\ref{aff34},\ref{aff13}}
\and R.~B.~Metcalf\orcid{0000-0003-3167-2574}\inst{\ref{aff1},\ref{aff2}}
\and P.~Monaco\orcid{0000-0003-2083-7564}\inst{\ref{aff112},\ref{aff9},\ref{aff10},\ref{aff8}}
\and A.~Pezzotta\orcid{0000-0003-0726-2268}\inst{\ref{aff7}}
\and M.~P\"ontinen\orcid{0000-0001-5442-2530}\inst{\ref{aff61}}
\and I.~Risso\orcid{0000-0003-2525-7761}\inst{\ref{aff7},\ref{aff16}}
\and V.~Scottez\orcid{0009-0008-3864-940X}\inst{\ref{aff74},\ref{aff113}}
\and M.~Sereno\orcid{0000-0003-0302-0325}\inst{\ref{aff2},\ref{aff13}}
\and M.~Tenti\orcid{0000-0002-4254-5901}\inst{\ref{aff13}}
\and M.~Tucci\inst{\ref{aff44}}
\and M.~Viel\orcid{0000-0002-2642-5707}\inst{\ref{aff8},\ref{aff9},\ref{aff11},\ref{aff10},\ref{aff105}}
\and M.~Wiesmann\orcid{0009-0000-8199-5860}\inst{\ref{aff51}}
\and Y.~Akrami\orcid{0000-0002-2407-7956}\inst{\ref{aff107},\ref{aff114}}
\and G.~Alguero\inst{\ref{aff109}}
\and I.~T.~Andika\orcid{0000-0001-6102-9526}\inst{\ref{aff48},\ref{aff115}}
\and G.~Angora\orcid{0000-0002-0316-6562}\inst{\ref{aff18},\ref{aff100}}
\and S.~Anselmi\orcid{0000-0002-3579-9583}\inst{\ref{aff45},\ref{aff86},\ref{aff116}}
\and M.~Archidiacono\orcid{0000-0003-4952-9012}\inst{\ref{aff66},\ref{aff67}}
\and F.~Atrio-Barandela\orcid{0000-0002-2130-2513}\inst{\ref{aff117}}
\and L.~Bazzanini\orcid{0000-0003-0727-0137}\inst{\ref{aff100},\ref{aff2}}
\and D.~Bertacca\orcid{0000-0002-2490-7139}\inst{\ref{aff86},\ref{aff14},\ref{aff45}}
\and M.~Bethermin\orcid{0000-0002-3915-2015}\inst{\ref{aff118}}
\and F.~Beutler\orcid{0000-0003-0467-5438}\inst{\ref{aff25}}
\and A.~Blanchard\orcid{0000-0001-8555-9003}\inst{\ref{aff88}}
\and L.~Blot\orcid{0000-0002-9622-7167}\inst{\ref{aff119},\ref{aff63}}
\and M.~Bonici\orcid{0000-0002-8430-126X}\inst{\ref{aff80},\ref{aff27}}
\and S.~Borgani\orcid{0000-0001-6151-6439}\inst{\ref{aff112},\ref{aff8},\ref{aff9},\ref{aff10},\ref{aff105}}
\and M.~L.~Brown\orcid{0000-0002-0370-8077}\inst{\ref{aff36}}
\and S.~Bruton\orcid{0000-0002-6503-5218}\inst{\ref{aff120}}
\and A.~Calabro\orcid{0000-0003-2536-1614}\inst{\ref{aff32}}
\and B.~Camacho~Quevedo\orcid{0000-0002-8789-4232}\inst{\ref{aff8},\ref{aff11},\ref{aff9}}
\and F.~Caro\inst{\ref{aff32}}
\and C.~S.~Carvalho\inst{\ref{aff95}}
\and Y.~Charles\inst{\ref{aff38}}
\and A.~R.~Cooray\orcid{0000-0002-3892-0190}\inst{\ref{aff121}}
\and O.~Cucciati\orcid{0000-0002-9336-7551}\inst{\ref{aff2}}
\and S.~Davini\orcid{0000-0003-3269-1718}\inst{\ref{aff16}}
\and F.~De~Paolis\orcid{0000-0001-6460-7563}\inst{\ref{aff122},\ref{aff123},\ref{aff124}}
\and G.~Desprez\orcid{0000-0001-8325-1742}\inst{\ref{aff125}}
\and A.~D\'iaz-S\'anchez\orcid{0000-0003-0748-4768}\inst{\ref{aff126}}
\and S.~Di~Domizio\orcid{0000-0003-2863-5895}\inst{\ref{aff15},\ref{aff16}}
\and J.~M.~Diego\orcid{0000-0001-9065-3926}\inst{\ref{aff127}}
\and V.~Duret\orcid{0009-0009-0383-4960}\inst{\ref{aff46}}
\and M.~Y.~Elkhashab\orcid{0000-0001-9306-2603}\inst{\ref{aff9},\ref{aff10},\ref{aff112},\ref{aff8}}
\and A.~Enia\orcid{0000-0002-0200-2857}\inst{\ref{aff2}}
\and Y.~Fang\orcid{0000-0002-0334-6950}\inst{\ref{aff50}}
\and A.~G.~Ferrari\orcid{0009-0005-5266-4110}\inst{\ref{aff13}}
\and A.~Finoguenov\orcid{0000-0002-4606-5403}\inst{\ref{aff61}}
\and A.~Fontana\orcid{0000-0003-3820-2823}\inst{\ref{aff32}}
\and A.~Franco\orcid{0000-0002-4761-366X}\inst{\ref{aff123},\ref{aff122},\ref{aff124}}
\and K.~Ganga\orcid{0000-0001-8159-8208}\inst{\ref{aff72}}
\and T.~Gasparetto\orcid{0000-0002-7913-4866}\inst{\ref{aff32}}
\and E.~Gaztanaga\orcid{0000-0001-9632-0815}\inst{\ref{aff93},\ref{aff91},\ref{aff128}}
\and F.~Giacomini\orcid{0000-0002-3129-2814}\inst{\ref{aff13}}
\and F.~Gianotti\orcid{0000-0003-4666-119X}\inst{\ref{aff2}}
\and G.~Gozaliasl\orcid{0000-0002-0236-919X}\inst{\ref{aff129},\ref{aff61}}
\and A.~Gruppuso\orcid{0000-0001-9272-5292}\inst{\ref{aff2},\ref{aff13}}
\and M.~Guidi\orcid{0000-0001-9408-1101}\inst{\ref{aff12},\ref{aff2}}
\and C.~M.~Gutierrez\orcid{0000-0001-7854-783X}\inst{\ref{aff130}}
\and A.~Hall\orcid{0000-0002-3139-8651}\inst{\ref{aff25}}
\and H.~Hildebrandt\orcid{0000-0002-9814-3338}\inst{\ref{aff131}}
\and J.~Hjorth\orcid{0000-0002-4571-2306}\inst{\ref{aff79}}
\and J.~J.~E.~Kajava\orcid{0000-0002-3010-8333}\inst{\ref{aff132},\ref{aff133}}
\and Y.~Kang\orcid{0009-0000-8588-7250}\inst{\ref{aff44}}
\and V.~Kansal\orcid{0000-0002-4008-6078}\inst{\ref{aff134},\ref{aff135}}
\and D.~Karagiannis\orcid{0000-0002-4927-0816}\inst{\ref{aff100},\ref{aff136}}
\and K.~Kiiveri\inst{\ref{aff59}}
\and J.~Kim\orcid{0000-0003-2776-2761}\inst{\ref{aff106}}
\and C.~C.~Kirkpatrick\inst{\ref{aff59}}
\and S.~Kruk\orcid{0000-0001-8010-8879}\inst{\ref{aff6}}
\and M.~Lattanzi\orcid{0000-0003-1059-2532}\inst{\ref{aff101}}
\and L.~Legrand\orcid{0000-0003-0610-5252}\inst{\ref{aff137},\ref{aff138}}
\and F.~Lepori\orcid{0009-0000-5061-7138}\inst{\ref{aff139}}
\and G.~Leroy\orcid{0009-0004-2523-4425}\inst{\ref{aff140},\ref{aff71}}
\and G.~F.~Lesci\orcid{0000-0002-4607-2830}\inst{\ref{aff1},\ref{aff2}}
\and J.~Lesgourgues\orcid{0000-0001-7627-353X}\inst{\ref{aff30}}
\and T.~I.~Liaudat\orcid{0000-0002-9104-314X}\inst{\ref{aff141}}
\and M.~Magliocchetti\orcid{0000-0001-9158-4838}\inst{\ref{aff47}}
\and A.~Manj\'on-Garc\'ia\orcid{0000-0002-7413-8825}\inst{\ref{aff126}}
\and F.~Mannucci\orcid{0000-0002-4803-2381}\inst{\ref{aff142}}
\and C.~J.~A.~P.~Martins\orcid{0000-0002-4886-9261}\inst{\ref{aff143},\ref{aff19}}
\and L.~Maurin\orcid{0000-0002-8406-0857}\inst{\ref{aff5}}
\and M.~Miluzio\inst{\ref{aff6},\ref{aff144}}
\and A.~Montoro\orcid{0000-0003-4730-8590}\inst{\ref{aff93},\ref{aff91}}
\and C.~Moretti\orcid{0000-0003-3314-8936}\inst{\ref{aff9},\ref{aff8},\ref{aff10}}
\and G.~Morgante\inst{\ref{aff2}}
\and S.~Nadathur\orcid{0000-0001-9070-3102}\inst{\ref{aff128}}
\and K.~Naidoo\orcid{0000-0002-9182-1802}\inst{\ref{aff128},\ref{aff57}}
\and P.~Natoli\orcid{0000-0003-0126-9100}\inst{\ref{aff100},\ref{aff101}}
\and A.~Navarro-Alsina\orcid{0000-0002-3173-2592}\inst{\ref{aff69}}
\and S.~Nesseris\orcid{0000-0002-0567-0324}\inst{\ref{aff107}}
\and L.~Pagano\orcid{0000-0003-1820-5998}\inst{\ref{aff100},\ref{aff101}}
\and E.~Palazzi\orcid{0000-0002-8691-7666}\inst{\ref{aff2}}
\and D.~Paoletti\orcid{0000-0003-4761-6147}\inst{\ref{aff2},\ref{aff4}}
\and F.~Passalacqua\orcid{0000-0002-8606-4093}\inst{\ref{aff86},\ref{aff45}}
\and K.~Paterson\orcid{0000-0001-8340-3486}\inst{\ref{aff57}}
\and L.~Patrizii\inst{\ref{aff13}}
\and A.~Pisani\orcid{0000-0002-6146-4437}\inst{\ref{aff46}}
\and D.~Potter\orcid{0000-0002-0757-5195}\inst{\ref{aff139}}
\and G.~W.~Pratt\inst{\ref{aff83}}
\and S.~Quai\orcid{0000-0002-0449-8163}\inst{\ref{aff1},\ref{aff2}}
\and M.~Radovich\orcid{0000-0002-3585-866X}\inst{\ref{aff14}}
\and W.~Roster\orcid{0000-0002-9149-6528}\inst{\ref{aff49}}
\and S.~Sacquegna\orcid{0000-0002-8433-6630}\inst{\ref{aff145}}
\and M.~Sahl\'en\orcid{0000-0003-0973-4804}\inst{\ref{aff146}}
\and D.~B.~Sanders\orcid{0000-0002-1233-9998}\inst{\ref{aff147}}
\and E.~Sarpa\orcid{0000-0002-1256-655X}\inst{\ref{aff11},\ref{aff105},\ref{aff10}}
\and A.~Schneider\orcid{0000-0001-7055-8104}\inst{\ref{aff139}}
\and D.~Sciotti\orcid{0009-0008-4519-2620}\inst{\ref{aff32},\ref{aff70}}
\and E.~Sellentin\inst{\ref{aff148},\ref{aff26}}
\and L.~C.~Smith\orcid{0000-0002-3259-2771}\inst{\ref{aff149}}
\and K.~Tanidis\orcid{0000-0001-9843-5130}\inst{\ref{aff106}}
\and F.~Tarsitano\orcid{0000-0002-5919-0238}\inst{\ref{aff150},\ref{aff44}}
\and G.~Testera\inst{\ref{aff16}}
\and R.~Teyssier\orcid{0000-0001-7689-0933}\inst{\ref{aff151}}
\and S.~Tosi\orcid{0000-0002-7275-9193}\inst{\ref{aff15},\ref{aff16},\ref{aff7}}
\and A.~Troja\orcid{0000-0003-0239-4595}\inst{\ref{aff86},\ref{aff45}}
\and A.~Venhola\orcid{0000-0001-6071-4564}\inst{\ref{aff152}}
\and D.~Vergani\orcid{0000-0003-0898-2216}\inst{\ref{aff2}}
\and G.~Verza\orcid{0000-0002-1886-8348}\inst{\ref{aff153},\ref{aff154}}
\and P.~Vielzeuf\orcid{0000-0003-2035-9339}\inst{\ref{aff46}}
\and S.~Vinciguerra\orcid{0009-0005-4018-3184}\inst{\ref{aff38}}
\and N.~A.~Walton\orcid{0000-0003-3983-8778}\inst{\ref{aff149}}
\and A.~H.~Wright\orcid{0000-0001-7363-7932}\inst{\ref{aff131}}}
										   
\institute{Dipartimento di Fisica e Astronomia "Augusto Righi" - Alma Mater Studiorum Universit\`a di Bologna, via Piero Gobetti 93/2, 40129 Bologna, Italy\label{aff1}
\and
INAF-Osservatorio di Astrofisica e Scienza dello Spazio di Bologna, Via Piero Gobetti 93/3, 40129 Bologna, Italy\label{aff2}
\and
Universit\'e Claude Bernard Lyon 1, CNRS/IN2P3, IP2I Lyon, UMR 5822, Villeurbanne, F-69100, France\label{aff3}
\and
INFN-Bologna, Via Irnerio 46, 40126 Bologna, Italy\label{aff4}
\and
Universit\'e Paris-Saclay, CNRS, Institut d'astrophysique spatiale, 91405, Orsay, France\label{aff5}
\and
ESAC/ESA, Camino Bajo del Castillo, s/n., Urb. Villafranca del Castillo, 28692 Villanueva de la Ca\~nada, Madrid, Spain\label{aff6}
\and
INAF-Osservatorio Astronomico di Brera, Via Brera 28, 20122 Milano, Italy\label{aff7}
\and
IFPU, Institute for Fundamental Physics of the Universe, via Beirut 2, 34151 Trieste, Italy\label{aff8}
\and
INAF-Osservatorio Astronomico di Trieste, Via G. B. Tiepolo 11, 34143 Trieste, Italy\label{aff9}
\and
INFN, Sezione di Trieste, Via Valerio 2, 34127 Trieste TS, Italy\label{aff10}
\and
SISSA, International School for Advanced Studies, Via Bonomea 265, 34136 Trieste TS, Italy\label{aff11}
\and
Dipartimento di Fisica e Astronomia, Universit\`a di Bologna, Via Gobetti 93/2, 40129 Bologna, Italy\label{aff12}
\and
INFN-Sezione di Bologna, Viale Berti Pichat 6/2, 40127 Bologna, Italy\label{aff13}
\and
INAF-Osservatorio Astronomico di Padova, Via dell'Osservatorio 5, 35122 Padova, Italy\label{aff14}
\and
Dipartimento di Fisica, Universit\`a di Genova, Via Dodecaneso 33, 16146, Genova, Italy\label{aff15}
\and
INFN-Sezione di Genova, Via Dodecaneso 33, 16146, Genova, Italy\label{aff16}
\and
Department of Physics "E. Pancini", University Federico II, Via Cinthia 6, 80126, Napoli, Italy\label{aff17}
\and
INAF-Osservatorio Astronomico di Capodimonte, Via Moiariello 16, 80131 Napoli, Italy\label{aff18}
\and
Instituto de Astrof\'isica e Ci\^encias do Espa\c{c}o, Universidade do Porto, CAUP, Rua das Estrelas, PT4150-762 Porto, Portugal\label{aff19}
\and
Faculdade de Ci\^encias da Universidade do Porto, Rua do Campo de Alegre, 4150-007 Porto, Portugal\label{aff20}
\and
European Southern Observatory, Karl-Schwarzschild-Str.~2, 85748 Garching, Germany\label{aff21}
\and
Dipartimento di Fisica, Universit\`a degli Studi di Torino, Via P. Giuria 1, 10125 Torino, Italy\label{aff22}
\and
INFN-Sezione di Torino, Via P. Giuria 1, 10125 Torino, Italy\label{aff23}
\and
INAF-Osservatorio Astrofisico di Torino, Via Osservatorio 20, 10025 Pino Torinese (TO), Italy\label{aff24}
\and
Institute for Astronomy, University of Edinburgh, Royal Observatory, Blackford Hill, Edinburgh EH9 3HJ, UK\label{aff25}
\and
Leiden Observatory, Leiden University, Einsteinweg 55, 2333 CC Leiden, The Netherlands\label{aff26}
\and
INAF-IASF Milano, Via Alfonso Corti 12, 20133 Milano, Italy\label{aff27}
\and
Centro de Investigaciones Energ\'eticas, Medioambientales y Tecnol\'ogicas (CIEMAT), Avenida Complutense 40, 28040 Madrid, Spain\label{aff28}
\and
Port d'Informaci\'{o} Cient\'{i}fica, Campus UAB, C. Albareda s/n, 08193 Bellaterra (Barcelona), Spain\label{aff29}
\and
Institute for Theoretical Particle Physics and Cosmology (TTK), RWTH Aachen University, 52056 Aachen, Germany\label{aff30}
\and
Deutsches Zentrum f\"ur Luft- und Raumfahrt e. V. (DLR), Linder H\"ohe, 51147 K\"oln, Germany\label{aff31}
\and
INAF-Osservatorio Astronomico di Roma, Via Frascati 33, 00078 Monteporzio Catone, Italy\label{aff32}
\and
INFN section of Naples, Via Cinthia 6, 80126, Napoli, Italy\label{aff33}
\and
Dipartimento di Fisica e Astronomia "Augusto Righi" - Alma Mater Studiorum Universit\`a di Bologna, Viale Berti Pichat 6/2, 40127 Bologna, Italy\label{aff34}
\and
Instituto de Astrof\'{\i}sica de Canarias, E-38205 La Laguna, Tenerife, Spain\label{aff35}
\and
Jodrell Bank Centre for Astrophysics, Department of Physics and Astronomy, University of Manchester, Oxford Road, Manchester M13 9PL, UK\label{aff36}
\and
European Space Agency/ESRIN, Largo Galileo Galilei 1, 00044 Frascati, Roma, Italy\label{aff37}
\and
Aix-Marseille Universit\'e, CNRS, CNES, LAM, Marseille, France\label{aff38}
\and
Institut de Ci\`{e}ncies del Cosmos (ICCUB), Universitat de Barcelona (IEEC-UB), Mart\'{i} i Franqu\`{e}s 1, 08028 Barcelona, Spain\label{aff39}
\and
Instituci\'o Catalana de Recerca i Estudis Avan\c{c}ats (ICREA), Passeig de Llu\'{\i}s Companys 23, 08010 Barcelona, Spain\label{aff40}
\and
Institut de Ciencies de l'Espai (IEEC-CSIC), Campus UAB, Carrer de Can Magrans, s/n Cerdanyola del Vall\'es, 08193 Barcelona, Spain\label{aff41}
\and
UCB Lyon 1, CNRS/IN2P3, IUF, IP2I Lyon, 4 rue Enrico Fermi, 69622 Villeurbanne, France\label{aff42}
\and
Space Science Data Center, Italian Space Agency, via del Politecnico snc, 00133 Roma, Italy\label{aff43}
\and
Department of Astronomy, University of Geneva, ch. d'Ecogia 16, 1290 Versoix, Switzerland\label{aff44}
\and
INFN-Padova, Via Marzolo 8, 35131 Padova, Italy\label{aff45}
\and
Aix-Marseille Universit\'e, CNRS/IN2P3, CPPM, Marseille, France\label{aff46}
\and
INAF-Istituto di Astrofisica e Planetologia Spaziali, via del Fosso del Cavaliere, 100, 00100 Roma, Italy\label{aff47}
\and
University Observatory, LMU Faculty of Physics, Scheinerstr.~1, 81679 Munich, Germany\label{aff48}
\and
Max Planck Institute for Extraterrestrial Physics, Giessenbachstr. 1, 85748 Garching, Germany\label{aff49}
\and
Universit\"ats-Sternwarte M\"unchen, Fakult\"at f\"ur Physik, Ludwig-Maximilians-Universit\"at M\"unchen, Scheinerstr.~1, 81679 M\"unchen, Germany\label{aff50}
\and
Institute of Theoretical Astrophysics, University of Oslo, P.O. Box 1029 Blindern, 0315 Oslo, Norway\label{aff51}
\and
Jet Propulsion Laboratory, California Institute of Technology, 4800 Oak Grove Drive, Pasadena, CA, 91109, USA\label{aff52}
\and
Felix Hormuth Engineering, Goethestr. 17, 69181 Leimen, Germany\label{aff53}
\and
Technical University of Denmark, Elektrovej 327, 2800 Kgs. Lyngby, Denmark\label{aff54}
\and
Cosmic Dawn Center (DAWN), Denmark\label{aff55}
\and
Institut d'Astrophysique de Paris, UMR 7095, CNRS, and Sorbonne Universit\'e, 98 bis boulevard Arago, 75014 Paris, France\label{aff56}
\and
Max-Planck-Institut f\"ur Astronomie, K\"onigstuhl 17, 69117 Heidelberg, Germany\label{aff57}
\and
NASA Goddard Space Flight Center, Greenbelt, MD 20771, USA\label{aff58}
\and
Department of Physics and Helsinki Institute of Physics, Gustaf H\"allstr\"omin katu 2, University of Helsinki, 00014 Helsinki, Finland\label{aff59}
\and
Universit\'e de Gen\`eve, D\'epartement de Physique Th\'eorique and Centre for Astroparticle Physics, 24 quai Ernest-Ansermet, CH-1211 Gen\`eve 4, Switzerland\label{aff60}
\and
Department of Physics, P.O. Box 64, University of Helsinki, 00014 Helsinki, Finland\label{aff61}
\and
Helsinki Institute of Physics, Gustaf H{\"a}llstr{\"o}min katu 2, University of Helsinki, 00014 Helsinki, Finland\label{aff62}
\and
Laboratoire d'etude de l'Univers et des phenomenes eXtremes, Observatoire de Paris, Universit\'e PSL, Sorbonne Universit\'e, CNRS, 92190 Meudon, France\label{aff63}
\and
SKAO, Jodrell Bank, Lower Withington, Macclesfield SK11 9FT, UK\label{aff64}
\and
Centre de Calcul de l'IN2P3/CNRS, 21 avenue Pierre de Coubertin 69627 Villeurbanne Cedex, France\label{aff65}
\and
Dipartimento di Fisica "Aldo Pontremoli", Universit\`a degli Studi di Milano, Via Celoria 16, 20133 Milano, Italy\label{aff66}
\and
INFN-Sezione di Milano, Via Celoria 16, 20133 Milano, Italy\label{aff67}
\and
University of Applied Sciences and Arts of Northwestern Switzerland, School of Computer Science, 5210 Windisch, Switzerland\label{aff68}
\and
Universit\"at Bonn, Argelander-Institut f\"ur Astronomie, Auf dem H\"ugel 71, 53121 Bonn, Germany\label{aff69}
\and
INFN-Sezione di Roma, Piazzale Aldo Moro, 2 - c/o Dipartimento di Fisica, Edificio G. Marconi, 00185 Roma, Italy\label{aff70}
\and
Department of Physics, Institute for Computational Cosmology, Durham University, South Road, Durham, DH1 3LE, UK\label{aff71}
\and
Universit\'e Paris Cit\'e, CNRS, Astroparticule et Cosmologie, 75013 Paris, France\label{aff72}
\and
CNRS-UCB International Research Laboratory, Centre Pierre Bin\'etruy, IRL2007, CPB-IN2P3, Berkeley, USA\label{aff73}
\and
Institut d'Astrophysique de Paris, 98bis Boulevard Arago, 75014, Paris, France\label{aff74}
\and
Institute of Physics, Laboratory of Astrophysics, Ecole Polytechnique F\'ed\'erale de Lausanne (EPFL), Observatoire de Sauverny, 1290 Versoix, Switzerland\label{aff75}
\and
Telespazio UK S.L. for European Space Agency (ESA), Camino bajo del Castillo, s/n, Urbanizacion Villafranca del Castillo, Villanueva de la Ca\~nada, 28692 Madrid, Spain\label{aff76}
\and
Institut de F\'{i}sica d'Altes Energies (IFAE), The Barcelona Institute of Science and Technology, Campus UAB, 08193 Bellaterra (Barcelona), Spain\label{aff77}
\and
European Space Agency/ESTEC, Keplerlaan 1, 2201 AZ Noordwijk, The Netherlands\label{aff78}
\and
DARK, Niels Bohr Institute, University of Copenhagen, Jagtvej 155, 2200 Copenhagen, Denmark\label{aff79}
\and
Waterloo Centre for Astrophysics, University of Waterloo, Waterloo, Ontario N2L 3G1, Canada\label{aff80}
\and
Department of Physics and Astronomy, University of Waterloo, Waterloo, Ontario N2L 3G1, Canada\label{aff81}
\and
Perimeter Institute for Theoretical Physics, Waterloo, Ontario N2L 2Y5, Canada\label{aff82}
\and
Universit\'e Paris-Saclay, Universit\'e Paris Cit\'e, CEA, CNRS, AIM, 91191, Gif-sur-Yvette, France\label{aff83}
\and
Centre National d'Etudes Spatiales -- Centre spatial de Toulouse, 18 avenue Edouard Belin, 31401 Toulouse Cedex 9, France\label{aff84}
\and
Institute of Space Science, Str. Atomistilor, nr. 409 M\u{a}gurele, Ilfov, 077125, Romania\label{aff85}
\and
Dipartimento di Fisica e Astronomia "G. Galilei", Universit\`a di Padova, Via Marzolo 8, 35131 Padova, Italy\label{aff86}
\and
Institut f\"ur Theoretische Physik, University of Heidelberg, Philosophenweg 16, 69120 Heidelberg, Germany\label{aff87}
\and
Institut de Recherche en Astrophysique et Plan\'etologie (IRAP), Universit\'e de Toulouse, CNRS, UPS, CNES, 14 Av. Edouard Belin, 31400 Toulouse, France\label{aff88}
\and
Universit\'e St Joseph; Faculty of Sciences, Beirut, Lebanon\label{aff89}
\and
Departamento de F\'isica, FCFM, Universidad de Chile, Blanco Encalada 2008, Santiago, Chile\label{aff90}
\and
Institut d'Estudis Espacials de Catalunya (IEEC),  Edifici RDIT, Campus UPC, 08860 Castelldefels, Barcelona, Spain\label{aff91}
\and
Satlantis, University Science Park, Sede Bld 48940, Leioa-Bilbao, Spain\label{aff92}
\and
Institute of Space Sciences (ICE, CSIC), Campus UAB, Carrer de Can Magrans, s/n, 08193 Barcelona, Spain\label{aff93}
\and
Departamento de F\'isica, Faculdade de Ci\^encias, Universidade de Lisboa, Edif\'icio C8, Campo Grande, PT1749-016 Lisboa, Portugal\label{aff94}
\and
Instituto de Astrof\'isica e Ci\^encias do Espa\c{c}o, Faculdade de Ci\^encias, Universidade de Lisboa, Tapada da Ajuda, 1349-018 Lisboa, Portugal\label{aff95}
\and
Cosmic Dawn Center (DAWN)\label{aff96}
\and
Niels Bohr Institute, University of Copenhagen, Jagtvej 128, 2200 Copenhagen, Denmark\label{aff97}
\and
Universidad Polit\'ecnica de Cartagena, Departamento de Electr\'onica y Tecnolog\'ia de Computadoras,  Plaza del Hospital 1, 30202 Cartagena, Spain\label{aff98}
\and
Caltech/IPAC, 1200 E. California Blvd., Pasadena, CA 91125, USA\label{aff99}
\and
Dipartimento di Fisica e Scienze della Terra, Universit\`a degli Studi di Ferrara, Via Giuseppe Saragat 1, 44122 Ferrara, Italy\label{aff100}
\and
Istituto Nazionale di Fisica Nucleare, Sezione di Ferrara, Via Giuseppe Saragat 1, 44122 Ferrara, Italy\label{aff101}
\and
INAF, Istituto di Radioastronomia, Via Piero Gobetti 101, 40129 Bologna, Italy\label{aff102}
\and
Astronomical Observatory of the Autonomous Region of the Aosta Valley (OAVdA), Loc. Lignan 39, I-11020, Nus (Aosta Valley), Italy\label{aff103}
\and
Universit\'e C\^{o}te d'Azur, Observatoire de la C\^{o}te d'Azur, CNRS, Laboratoire Lagrange, Bd de l'Observatoire, CS 34229, 06304 Nice cedex 4, France\label{aff104}
\and
ICSC - Centro Nazionale di Ricerca in High Performance Computing, Big Data e Quantum Computing, Via Magnanelli 2, Bologna, Italy\label{aff105}
\and
Department of Physics, Oxford University, Keble Road, Oxford OX1 3RH, UK\label{aff106}
\and
Instituto de F\'isica Te\'orica UAM-CSIC, Campus de Cantoblanco, 28049 Madrid, Spain\label{aff107}
\and
CEA Saclay, DFR/IRFU, Service d'Astrophysique, Bat. 709, 91191 Gif-sur-Yvette, France\label{aff108}
\and
Univ. Grenoble Alpes, CNRS, Grenoble INP, LPSC-IN2P3, 53, Avenue des Martyrs, 38000, Grenoble, France\label{aff109}
\and
Dipartimento di Fisica, Sapienza Universit\`a di Roma, Piazzale Aldo Moro 2, 00185 Roma, Italy\label{aff110}
\and
Aurora Technology for European Space Agency (ESA), Camino bajo del Castillo, s/n, Urbanizacion Villafranca del Castillo, Villanueva de la Ca\~nada, 28692 Madrid, Spain\label{aff111}
\and
Dipartimento di Fisica - Sezione di Astronomia, Universit\`a di Trieste, Via Tiepolo 11, 34131 Trieste, Italy\label{aff112}
\and
ICL, Junia, Universit\'e Catholique de Lille, LITL, 59000 Lille, France\label{aff113}
\and
CERCA/ISO, Department of Physics, Case Western Reserve University, 10900 Euclid Avenue, Cleveland, OH 44106, USA\label{aff114}
\and
Technical University of Munich, TUM School of Natural Sciences, Physics Department, James-Franck-Str.~1, 85748 Garching, Germany\label{aff115}
\and
Laboratoire Univers et Th\'eorie, Observatoire de Paris, Universit\'e PSL, Universit\'e Paris Cit\'e, CNRS, 92190 Meudon, France\label{aff116}
\and
Departamento de F{\'\i}sica Fundamental. Universidad de Salamanca. Plaza de la Merced s/n. 37008 Salamanca, Spain\label{aff117}
\and
Universit\'e de Strasbourg, CNRS, Observatoire astronomique de Strasbourg, UMR 7550, 67000 Strasbourg, France\label{aff118}
\and
Center for Data-Driven Discovery, Kavli IPMU (WPI), UTIAS, The University of Tokyo, Kashiwa, Chiba 277-8583, Japan\label{aff119}
\and
California Institute of Technology, 1200 E California Blvd, Pasadena, CA 91125, USA\label{aff120}
\and
Department of Physics \& Astronomy, University of California Irvine, Irvine CA 92697, USA\label{aff121}
\and
Department of Mathematics and Physics E. De Giorgi, University of Salento, Via per Arnesano, CP-I93, 73100, Lecce, Italy\label{aff122}
\and
INFN, Sezione di Lecce, Via per Arnesano, CP-193, 73100, Lecce, Italy\label{aff123}
\and
INAF-Sezione di Lecce, c/o Dipartimento Matematica e Fisica, Via per Arnesano, 73100, Lecce, Italy\label{aff124}
\and
Kapteyn Astronomical Institute, University of Groningen, PO Box 800, 9700 AV Groningen, The Netherlands\label{aff125}
\and
Departamento F\'isica Aplicada, Universidad Polit\'ecnica de Cartagena, Campus Muralla del Mar, 30202 Cartagena, Murcia, Spain\label{aff126}
\and
Instituto de F\'isica de Cantabria, Edificio Juan Jord\'a, Avenida de los Castros, 39005 Santander, Spain\label{aff127}
\and
Institute of Cosmology and Gravitation, University of Portsmouth, Portsmouth PO1 3FX, UK\label{aff128}
\and
Department of Computer Science, Aalto University, PO Box 15400, Espoo, FI-00 076, Finland\label{aff129}
\and
 Instituto de Astrof\'{\i}sica de Canarias, E-38205 La Laguna; Universidad de La Laguna, Dpto. Astrof\'\i sica, E-38206 La Laguna, Tenerife, Spain\label{aff130}
\and
Ruhr University Bochum, Faculty of Physics and Astronomy, Astronomical Institute (AIRUB), German Centre for Cosmological Lensing (GCCL), 44780 Bochum, Germany\label{aff131}
\and
Department of Physics and Astronomy, Vesilinnantie 5, University of Turku, 20014 Turku, Finland\label{aff132}
\and
Serco for European Space Agency (ESA), Camino bajo del Castillo, s/n, Urbanizacion Villafranca del Castillo, Villanueva de la Ca\~nada, 28692 Madrid, Spain\label{aff133}
\and
ARC Centre of Excellence for Dark Matter Particle Physics, Melbourne, Australia\label{aff134}
\and
Centre for Astrophysics \& Supercomputing, Swinburne University of Technology,  Hawthorn, Victoria 3122, Australia\label{aff135}
\and
Department of Physics and Astronomy, University of the Western Cape, Bellville, Cape Town, 7535, South Africa\label{aff136}
\and
DAMTP, Centre for Mathematical Sciences, Wilberforce Road, Cambridge CB3 0WA, UK\label{aff137}
\and
Kavli Institute for Cosmology Cambridge, Madingley Road, Cambridge, CB3 0HA, UK\label{aff138}
\and
Department of Astrophysics, University of Zurich, Winterthurerstrasse 190, 8057 Zurich, Switzerland\label{aff139}
\and
Department of Physics, Centre for Extragalactic Astronomy, Durham University, South Road, Durham, DH1 3LE, UK\label{aff140}
\and
IRFU, CEA, Universit\'e Paris-Saclay 91191 Gif-sur-Yvette Cedex, France\label{aff141}
\and
INAF-Osservatorio Astrofisico di Arcetri, Largo E. Fermi 5, 50125, Firenze, Italy\label{aff142}
\and
Centro de Astrof\'{\i}sica da Universidade do Porto, Rua das Estrelas, 4150-762 Porto, Portugal\label{aff143}
\and
HE Space for European Space Agency (ESA), Camino bajo del Castillo, s/n, Urbanizacion Villafranca del Castillo, Villanueva de la Ca\~nada, 28692 Madrid, Spain\label{aff144}
\and
INAF - Osservatorio Astronomico d'Abruzzo, Via Maggini, 64100, Teramo, Italy\label{aff145}
\and
Theoretical astrophysics, Department of Physics and Astronomy, Uppsala University, Box 516, 751 37 Uppsala, Sweden\label{aff146}
\and
Institute for Astronomy, University of Hawaii, 2680 Woodlawn Drive, Honolulu, HI 96822, USA\label{aff147}
\and
Mathematical Institute, University of Leiden, Einsteinweg 55, 2333 CA Leiden, The Netherlands\label{aff148}
\and
Institute of Astronomy, University of Cambridge, Madingley Road, Cambridge CB3 0HA, UK\label{aff149}
\and
Institute for Particle Physics and Astrophysics, Dept. of Physics, ETH Zurich, Wolfgang-Pauli-Strasse 27, 8093 Zurich, Switzerland\label{aff150}
\and
Department of Astrophysical Sciences, Peyton Hall, Princeton University, Princeton, NJ 08544, USA\label{aff151}
\and
Space physics and astronomy research unit, University of Oulu, Pentti Kaiteran katu 1, FI-90014 Oulu, Finland\label{aff152}
\and
International Centre for Theoretical Physics (ICTP), Strada Costiera 11, 34151 Trieste, Italy\label{aff153}
\and
Center for Computational Astrophysics, Flatiron Institute, 162 5th Avenue, 10010, New York, NY, USA\label{aff154}}


  \abstract{
  The \Euclid spacecraft, located at the second Lagrangian point of the Sun-Earth system, hosts the Near-Infrared Spectrometer and Photometer (NISP) instrument. NISP is equipped with a mosaic of 16 HgCdTe-based detectors (Teledyne's H2RG) to acquire near-infrared photometric and spectroscopic data. To meet the spacecraft's constraints on computational resources and telemetry bandwidth, the near-infrared signal is processed onboard via a dedicated hardware-software architecture designed to fulfil the stringent \textit{Euclid}'s data-quality requirements. A custom application software, running on the two NISP data processing units, implements the NISP signal estimator: an ad-hoc algorithm which delivers accurate flux measurements and simultaneously estimates the quality of signal estimation through the quality factor (QF) parameter.  This paper further investigates the properties of the NISP signal estimator by evaluating its performance during the early flight operations of the NISP instrument. First, we revisit the assumptions on which the inference of the near-infrared signal is based and investigate the origin of the main systematics of the signal estimator through Monte Carlo simulations. Then, we test the flight performance of the NISP signal estimator. Results indicate a systematic bias lower than $0.01\,{\rm e}^{-}\,{\rm s}^{-1}$ for $99\%$ of the NISP pixel array, well within the noise budget of the estimated signal. We also derive an analytical expression for the variance of the NISP signal estimator, demonstrating its validity, particularly when the covariance matrix is not pre-computed. Finally, we provide a robust statistical framework to interpret the QF parameter, analyse its dependence on the signal estimator bias, and show its sensitivity to cosmic ray hits on NISP detectors. Our findings corroborate previous results on the NISP signal estimator and suggest a leading-order correction based on the agreement between flight data and simulations.
  }

%
%
   \titlerunning{Properties and 
   performance of the NISP signal estimator}
   \authorrunning{Euclid Collaboration: F. Cogato et al. } 
             
   \maketitle


\section{Introduction\label{sec:intro}}
\Euclid, the second medium-class mission in the European Space Agency's Cosmic Vision programme, aims to map the geometry of the universe by observing billions of galaxies across approximately one third of the sky \citep{EuclidSkyOverview}. 
The Near-Infrared Spectrometer and Photometer \citep[NISP,][]{EuclidSkyNISP} is one of the two scientific instruments
onboard \Euclid. 
With a field of view of $0.57$\,\sqdeg in the near-infrared (NIR), NISP operates in both multi-passband photometric \citep{Schirmer-EP18} and slitless spectroscopic \citep{Gillard2025} readout mode. 
It provides imaging with a \ang{;;0.3} per pixel sampling and spectroscopic redshift measurements with precision $\sigma_z < 0.001(1+z)$ for H$\alpha$ emission-line galaxies, assuming a full width at half maximum of \ang{;;0.3} and a $3.5\sigma$ flux limit of $2\times10^{-6}$\,erg\,s$^{-1}$\,cm$^{-2}$ \citep{Scaramella-EP1}.

The instrument is equipped with a focal plane assembly (FPA) composed of a mosaic of 16 HgCdTe-based detectors (Teledyne Imaging Sensors H2RG\textsuperscript{\texttrademark}, 2048$\times$2048 pixels), whose technology enables non-destructive readout of frames during signal integration. Therefore, each NISP exposure is composed of a sequence of single-frame images (or sampling up-the-ramp frames) acquired with a frame rate of 1.45408\,s \citep[for full details see][]{Kubik2025}.

With about $67\times10^6$ pixels, the NISP FPA is the largest single NIR detector array ever deployed in space. 
For comparison, the WFC3/IR instrument \citep{WFC3_IR} onboard the \HST (HST) holds approximately $10^6$ pixels, while the NIRCam instrument \citep{NIRCam} onboard the \JWST (JWST) has about $42 \times 10^6$ pixels in its focal plane. 

While HST and JWST are designed for high-resolution and deep imaging over narrow fields\footnote{\href{https://www.stsci.edu/hst/}{WFC3/IR} observes around $0.0016$\,\sqdeg through a single H1R detector with 1024$\times$1024 pixels, achieving a \ang{;;0.13} per pixel sampling. \href{https://jwst-docs.stsci.edu}{NIRCam} observes approximately $0.0027$\,\sqdeg through eight H2RG detectors with 2048$\times$2048 pixels each, which results in a \ang{;;0.031} per pixel sampling.},
\Euclid is designed to perform a systematic survey of the extragalactic sky, observing 20 different fields daily, corresponding to $\sim 10$\,\sqdeg/day. 
In the Euclid Wide Survey, each field is observed only once during the mission lifetime, and is scanned following the \textit{Euclid}'s Reference Observation Sequence \citep[ROS,][]{Scaramella-EP1}. It consists of four dithered exposures to mitigate the effect of cosmic ray hits while also enabling compensation for physical gaps between detectors.

For each observed field, the NISP instrument acquires 17 multi-frame exposures, corresponding to more than 2500 single-frame images per field\footnote{This corresponds to the total sampling up-the-ramp frames collected by NISP during the execution of a single \textit{Euclid}'s ROS cycle. See \citet{EuclidSkyNISP} for more details.} and around 51\,000 single-frame images per day.
This translates into a data volume of around 20 Tbit/day, which exceeds the allocated data transfer budget of 290 Gbit/day. For comparison, the data downlink rates for \href{https://web.archive.org/web/20160706034142/http://hubblesite.org/the_telescope/hubble_essentials/quick_facts.php}{HST} and \href{https://webbtelescope.org/quick-facts}{JWST} do not exceed 0.02 Tbit/day and 0.5 Tbit/day, respectively. \Euclid's innovative strategy was to enable the NISP onboard data processing, which provides accurate flux estimation and reduces the NISP data volume to 248 Gbit/day, perfectly fitting the bandwidth requirements \citep{Bonoli2016, Medinaceli2020}.

In the case of NISP, the signal estimator \citep{Kubik2016} algorithm implemented in the two Data Processing Units (DPUs) optimises the computational resources, specifically regarding computational time and available memory. By simultaneously estimating the signal and computing the goodness-of-fit, i.e., the quality factor (QF), the method reduces processing complexity, accelerates the computation process, minimises power consumption, and optimises memory usage. Also, the onboard processing is done almost in parallel with \Euclid's ROS, with a very low overhead fully absorbed by the satellite slew time between two consecutive science observations.

The computational constraints of the hardware are accommodated at the cost of certain simplifications. First, the estimator is defined in the signal difference space and not in the integrated signal space.
In this domain, a substantial portion of the photon-noise-induced correlations is mitigated, resulting in a tri-diagonal covariance matrix \citep{Kubik2015}. 
Additionally, fitting in the signal-difference space offers the advantage of being insensitive to reset noise and provides a lower noise signal estimate than a typically used least squares fit \citep{Kubik2016}. Second, the method avoids the explicit pre-computation of the full covariance matrix as proposed by \citet{2014SPIE.9143E..3ZR} and \citet{2024PASP..136d5004B}, a process that would be computationally prohibitive on the NISP DPUs. Instead, the Poissonian component of the noise is treated as a free parameter within the covariance matrix.  
Moreover, accounting for the full covariance structure would require numerical maximisation of the likelihood function, which would significantly increase the computational effort. Hence, it is assumed that the covariance matrix is diagonal, effectively neglecting off-diagonal correlations.

As a result of these simplifications, the NISP signal estimator is subject to several potential sources of bias.
Specifically, treating the photon-noise component as a fit parameter rather than deriving it directly from the data, combined with the diagonal approximation of the covariance matrix, can result in sub-optimal weighting during the fitting process.
Furthermore, the NISP signal estimator requires knowledge of each pixel’s readout noise and conversion gain. However, due to the large pixel array and the onboard memory available for NISP, the processing algorithm utilises detector-averaged values instead of full pixel maps, potentially introducing a systematic bias in the signal estimates.

This work extends the analysis presented in \citet{Kubik2016}, with the primary objective of quantifying potential biases using both simulations and flight data from early NISP operations.
Section~\ref{sec:notations} introduces the notations and constants used throughout the paper. Section~\ref{sec:properties} analyses the theoretical properties of the NISP signal estimator as originally derived in \citet{Kubik2016}. We identify the associated probability distributions and revisit the assumptions of the underlying likelihood function to explore the main bias sources in both the signal and QF estimation. 
Section~\ref{sec:data_processing} provides an overview of the NISP onboard data processing, with a specific focus on the systematics related to the data processing approximations adopted by NISP.
Section~\ref{sec:flight_performance_sig} presents the evaluation of the performance of the NISP signal estimator using NISP flight data, while the in-flight performance of the QF parameter is presented in Sect.~\ref{sec:flight_performance_qf}. Section~\ref{sec:conclusions} summarises the results, discusses limitations and potential issues, and presents the main conclusions of the study. All formal derivations are provided in the Appendix sections.

\section{Notations \label{sec:notations}}
\renewcommand{\arraystretch}{1.3}
\begin{table*}
\caption{\label{tab:notations}Notations used throughout the paper.}
\begin{center}
\begin{tabular}{lll}
    \textbf{Variable/Parameter} & \textbf{Units} & \textbf{Description} \\
    \hline
    \hline
    $g$ & ${\rm e}^{-}\,{\rm group}^{-1}$ & expected flux per group\\
    $f$ & ${\rm e}^{-}\,{\rm s}^{-1}$ & expected flux per second\\
    $f_{\rm 0}$ & ${\rm e}^{-}\,{\rm s}^{-1}$ & flux value at which covariance matrix $D_{ij}$ is diagonal\\
    $f_{\rm f}$ & ${\rm e}^{-}\,{\rm s}^{-1}$ & flux value for which folding effect is negligible\\
    $t_{\rm{fr}}$ & s & single-frame readout time \\
    $\sigma_{\rm R}$ & ${\rm e}^{-}$ & \multirow{2}{20em}{pixel's readout noise} \\
    $\sigma_{\sfont R}^{\rm h}$ & ${\rm ADU}$ \\ 
    $\sigma_{\sfont{R,\,DPU}}$ & ${\rm e}^{-}$ & detector-averaged readout noise adopted by NISP \\
    $f_{\rm e}$ & ${\rm e}^{-}\,{\rm ADU}^{-1}$ & pixel's conversion gain\\
    $f_{\rm{e,}\,\sfont{DPU}}$ & ${\rm e}^{-}\,{\rm ADU}^{-1}$ & detector-averaged conversion gain adopted by NISP\\
    MACC($n_{\rm g}$, $n_{\rm f}$, $n_{\rm d}$) \ \ \ & & multi-accumulation readout mode \\
    $\gamma = \frac{2 \sigma_{\sfont R}^2}{n_{\rm f}}$ & & readout noise component of the variance of $\Delta G$ \\
    $\alpha = \frac{1\,-\,n_{\rm f}^2}{3\,n_{\rm f}(n_{\rm f}\,+\,n_{\rm  d})}$ & & correlation factor from coadding of frames \\
    $\xi = \frac{1+\alpha}{2}$ & & rescaled $\alpha$ \\
    $\beta = \frac{\gamma}{1+\alpha}$ & & rescaled $\gamma$\\
    \hline
    $G_i$ & ${\rm e}^{-}$ & \multirow{2}{20em}{a group of $n_{\rm f}$ frames, $i=1,2,3,\dots$} \\
    $H_i$ & ADU \\
    $\Delta G_i$ & ${\rm e}^{-}$ &  \multirow{2}{20em}{measured difference of two consecutive groups} \\
    $\Delta H_i$ & ${\rm ADU}$ & \\
    $\hat{g}[\Delta \vec{G}]_{n_{\rm g}}$ & ${\rm e}^{-}\,{\rm group}^{-1}$& likelihood-estimator of $g$ from $n_{\rm g}-1$ group differences\\
    $\hat{g}_x[\Delta \vec{G}]_{n_{\rm g}}$ & ${\rm e}^{-}\,{\rm group}^{-1}$ & $\chi^2$-estimator of $g$ from $n_{\rm g}-1$ group differences\\
    ${\rm QF}[\Delta \vec{G}]_{n_{\rm g}}$ & & quality factor from $n_{\rm g}-1$ group differences \\
    $D_{ij}$ & & covariance matrix of $\Delta \vec{G}$ \\
    $\hat{\rho}^2[\hat{g}]_{n_{\rm g}}$ & & estimator of the variance of $\hat{g}[\Delta \vec{G}]_{n_{\rm g}}$ \\
    $\hat{\rho}^2[\hat{g}_{x}]_{n_{\rm g}}$ & & estimator of the variance of $\hat{g}_{x}[\Delta \vec{G}]_{n_{\rm g}}$ \\
    $\hat{\rho}^2[{\rm QF}]_{n_{\rm g}}$ & & estimator of the variance of ${\rm QF}[\Delta \vec{G}]_{n_{\rm g}}$ \\
    \hline
    \hline
\end{tabular}\
\end{center}
\end{table*}
In this section, we briefly review the statistical framework in which the NISP signal estimator was derived, and define the notations adopted in the present study, which are summarised in Table~\ref{tab:notations}. 
We assume that the detectors measure the signal using the multi-accumulation (MACC) acquisition mode. An exposure begins with a reset frame and then the signal is sampled up-the-ramp (UTR) with $n_{\rm g}$ groups of $n_{\rm f}$ averaged frames. The groups are separated by $n_{\rm d}$ dropped frames \citep[see figure\,2 in][]{Kubik2025}.

\subsection{Measured flux}

The measured difference between consecutive groups, $\Delta G_{i} = G_{i+1} - G_i$, represents the flux integrated between the first frame of group $G_i$ and the first frame of group $G_{i+1}$. Throughout the paper, we use $g$ to denote the expected flux integrated in $\Delta G_i$, while $f$ denotes the expected flux per second. For clarity, the derivations in the Appendix are expressed in terms of $g$, as it is directly tied to the measured group-to-group difference $\Delta G_i$, which forms the basis of the likelihood function. The quantities $f$ and $g$ are related by
\begin{equation}
    f = \frac{g}{(n_{\rm f} + n_{\rm d})\,t_{\rm{fr}}}\,,
\end{equation}
while variables with a hat (e.g., $\hat{g}$ or $\hat{\rho}$) denote expectation values provided by estimators. 

\subsection{Likelihood and signal estimator}

\citet{Kubik2016} defined the likelihood function as
\begin{equation}\label{eq:likelihood_general}
    \mathcal{L}\,(g\,|\,\Delta G_i) = \prod_{i=1}^{n_{\rm g}-1} \frac{1}{\sqrt{2\pi D_{ii}(g) }}\exp{\left(-\frac{(\Delta G_i - g)^2 }{2 D_{ii}(g)}\right)}\;,
\end{equation}
where $D_{ii}$ are the diagonal terms of the covariance matrix from \citet{Kubik2015} for a MACC readout mode
\begin{equation}
\label{eq:Dkl}
D_{ij}(g) = 
\begin{cases} 
(1+\alpha)g + \gamma & \quad \textrm{if}\quad i=j\\
-\frac{1}{2}(\alpha\, g + \gamma)& \quad\textrm{if}\quad i=j\pm1\\
0 & \quad\textrm{otherwise}
\end{cases}
\,,
\end{equation}
where
\begin{equation}\label{eq:gamma}
\alpha = \frac{1\,-\,n_{\rm f}^2}{3\,n_{\rm f}(n_{\rm f}\,+\,n_{\rm d})}\,;\quad \gamma = \frac{2 \sigma_{\sfont R}^2}{n_{\rm f}}\,,
\end{equation}
and $\sigma_{\sfont R}$ is the single-frame readout noise. 
In this paper, the signal estimator $\hat{g}[\Delta \vec{G}]_{n_{\rm g}}$ obtained from maximising the log-likelihood is rewritten as 
\begin{equation}\label{eq:gL}
    \hat{g}[\Delta \vec{G}]_{n_{\rm g}} = \sqrt{ \xi^2 + M_2( \Delta \vec{G}, \beta, n_{\rm g} ) }  - \xi - \beta\,,
\end{equation}
with
\begin{equation}
\beta = \frac{\gamma}{1+\alpha}\,; \quad \xi = \frac{1+\alpha}{2}\,.
\end{equation}
In the notations as in Eq.\,(\ref{eq:gL}), the subscript $n_{\rm g}$ emphasises that the estimator is computed from $n_{\rm g}-1$ random variables $\Delta G_i$ and we define
\begin{equation}\label{eq:M2}
    M_2( \Delta \vec{G}, \beta, n_{\rm g} ) \equiv \frac{\sum_{i=1}^{n_{\rm g}-1} ( \Delta G_i + \beta )^2}{n_{\rm g}-1} \;.
\end{equation}

\subsection{Quality factor}
\citet{Kubik2016}  also defined a statistical factor, i.e., the QF parameter, which measures the goodness of fit of a signal estimated using Eq.\,(\ref{eq:gL}). Here, the QF is rewritten as
\begin{equation}
\label{eq:QF}
    {\rm QF}[\Delta \vec{G}]_{n_{\rm g}} = \frac{n_{\rm g}-1}{\xi}\left(\hat{g}_x[\Delta \vec{G}]_{n_{\rm g}}  - \frac{G_{n_{\rm g}} - G_1}{n_{\rm g}-1} \right)\,,
\end{equation}
where
\begin{equation}\label{eq:gx}
    \hat{g}_x[\Delta \vec{G}]_{n_{\rm g}} = \sqrt{ M_2( \Delta \vec{G}, \beta, n_{\rm g} ) } - \beta \,.
\end{equation}
We note that $\hat{g}_x[\Delta \vec{G}]_{n_{\rm g}}$ is closely related to $\hat{g}[\Delta \vec{G}]_{n_{\rm g}}$. 
The former is obtained from the $\chi^2$ minimisation, while the latter is derived from the maximisation of the log-likelihood function, where $\mathcal{L}\,\propto\,{\rm exp}(-\chi^2/2)\,$. 
Both use the same diagonal covariance matrix $D_{ii}$. The only difference lies in the likelihood normalisation term, which includes the determinant of the covariance matrix in the derivation of $\hat{g}[\Delta \vec{G}]_{n_{\rm g}}$. In practice, this approach leads to the inclusion of the term $\xi$ in $\hat{g}[\Delta \vec{G}]_{n_{\rm g}}$, which effectively reduces the bias of the signal estimator obtained from the likelihood with respect to the one obtained from $\chi^2$. All the details are provided in Sect.~\ref{sec:properties}, where we present the general properties and limitations of $\hat{g}[\Delta \vec{G}]_{n_{\rm g}}$ and QF. 

While the discussion in Sect. \ref{sec:properties} is kept general, most of the numerical results are shown for readout modes and pixel characteristics specific to the \Euclid mission. In particular, the photometric channel of the NISP instrument operates in the MACC(4,16,4) readout mode, corresponding to an integration time of 87.2\,s, whereas spectroscopic observations are performed using the MACC(15,16,11) configuration, yielding an integration time of 549.6\,s. We consider a typical \Euclid pixel, corresponding to a single-frame readout noise equal to $13\,{\rm e}^{-}$ \citep{Kubik2025}.

We express our theoretical derivations and data analysis results in electrons, while the implementation of the signal estimator in the NISP DPU application software (presented in Sect.~\ref{sec:data_processing}) operates on Analogue-to-Digital Units (ADUs), as expressed explicitly in equation (11) in \citet{Kubik2016}. The electron-based formulation is crucial for evaluating the scientific performance of the instrument, while the ADU-based representation reflects the digital processing from which the NISP signal is derived.

\section{Properties of the NISP signal estimator and QF}
\label{sec:properties}
We study the properties of the NISP signal estimator through a series of Monte Carlo (MC) simulations. The incident flux $f$ on each pixel accumulates linearly over time, and the variance associated with UTR frames convolves a Gaussian (readout noise) component and a Poissonian (shot noise) component, as described in \citet{Kubik2016}. 

It is worth noting that we assume unity quantum efficiency and ignore non-ideal effects, such as non-linearity, persistence, charge diffusion, and inter-pixel capacitance. Although simplified, this model effectively represents how a single pixel integrates the signal.

\begin{figure*}[!t]
\begin{center}
\includegraphics[scale=0.35]{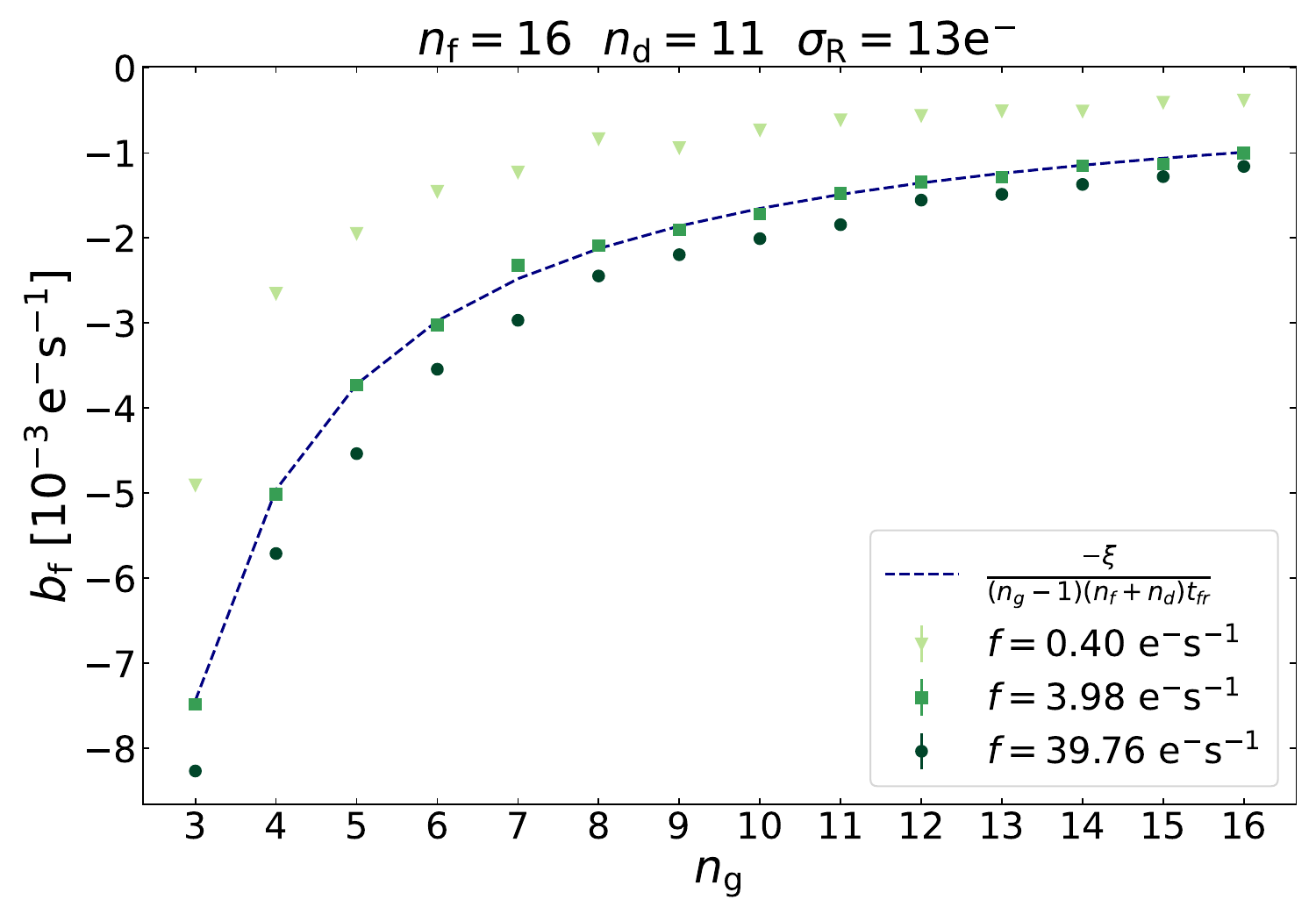}
\includegraphics[scale=0.35]{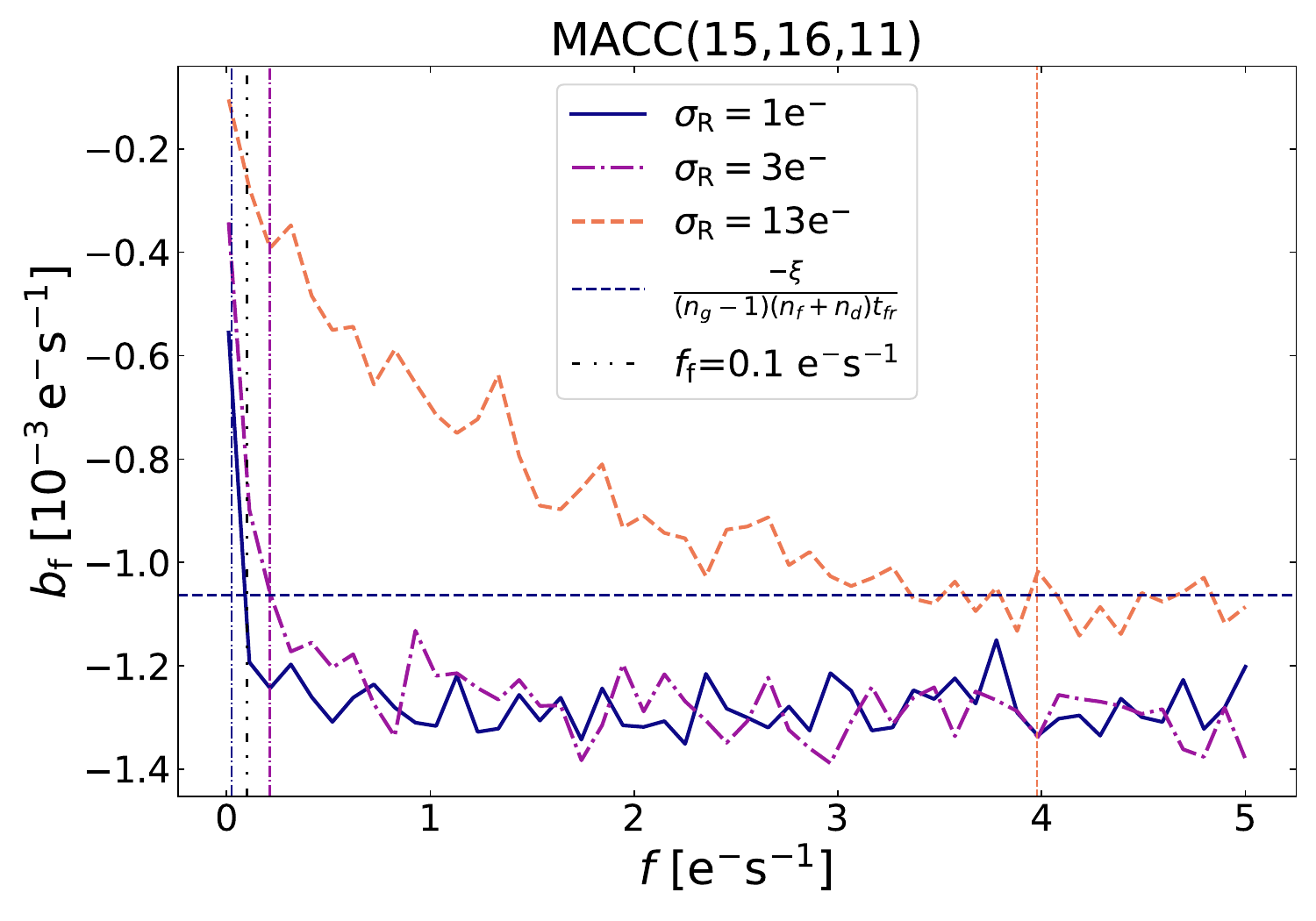}\\
\includegraphics[scale=0.35]{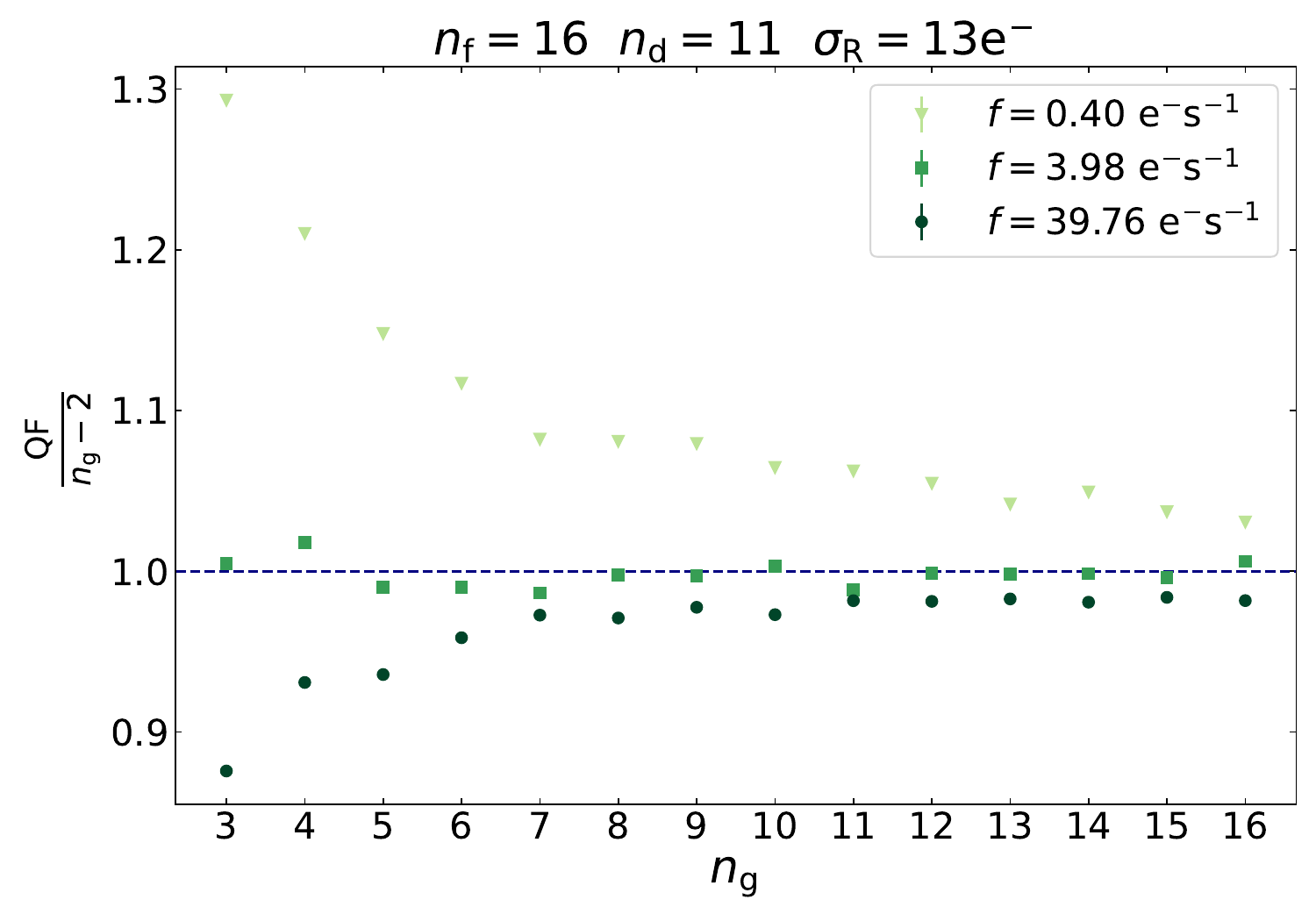}
\includegraphics[scale=0.35]{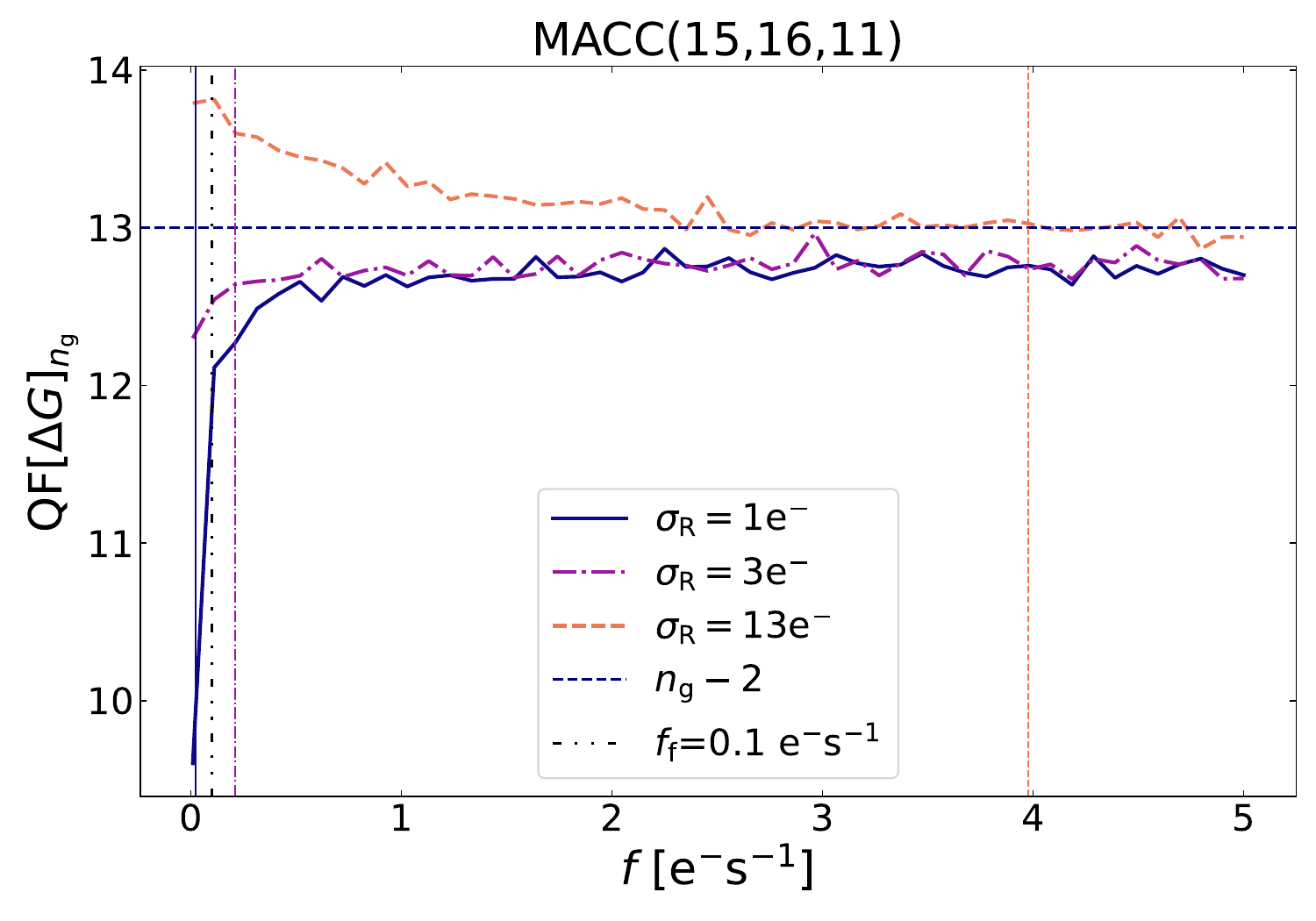}
\end{center}
\caption{\emph{Top panels:}  Bias of the signal estimator. \emph{Bottom panel:} QF values. \emph{Left panels:} Bias as function of $n_{\rm g}$ for $n_{\rm f}=16$, $n_{\rm d}=11$, and $\sigma_R=13\,{\rm e}^{-}$ in the case where $f_{\rm 0} > f_{\rm f}$ (no folding effect). For $f = f_{\rm 0}$ (green squares) for which the off-diagonal entries of the variance matrix are null, the bias is perfectly predicted by the analytical formula (dashed lines) from Eq.\,(\ref{eq:bias}) for the signal, and Eq.\,(\ref{eq:QFbias}) for the QF. For $f<f_{\rm 0}$ (light green triangles), the theoretical prediction underestimates the bias. For $f>f_{\rm 0}$ (dark green dots) the theoretical prediction overestimates the bias. \emph{Right panels:} bias of the signal estimator as a function of flux in spectroscopic readout mode. The flux values $f_{\rm 0}$ are indicated by vertical lines for each of the readout noise values. The $f_{\rm f}=0.1\,{\rm e}^{-}\,{\rm s}^{-1}$ flux value, above which the folding contribution is negligible, is indicated by a grey dash-dot-dotted vertical line. Positive contribution of the $D_{ij}$ to the covariance matrix for $f > f_{\rm 0}$ translates in the measured signal bias lower than the analytical formula, while negative contribution of the $D_{ij}$ translates in the measured signal bias higher than the analytical formula. For $f \gg f_{\rm 0}$ and $f \gg f_{\rm f}$, the bias does not depend on flux. Similar results are obtained in photometric readout mode.}
\label{fig:S_bias_fcn_ng}
\end{figure*}
Simulations indicate that, across the entire range of flux,
the NISP signal estimator exhibits a small bias $b_{\rm f}$ with respect to the expected flux $f$, that is
\begin{equation}
    b_{\rm f} = \frac{\hat{g}[\Delta \vec{G}]_{n_{\rm g}} - g}{(n_{\rm f} + n_{\rm d})\,t_{\rm{fr}}} = \frac{\hat{g}[\Delta \vec{G}]_{n_{\rm g}} }{(n_{\rm f} + n_{\rm d})\,t_{\rm{fr}}} - f\,.
\end{equation}

As shown in the top panels of Fig.~\ref{fig:S_bias_fcn_ng}, and explained in the following sections, $b_{\rm f}$ depends on the number of groups $n_{\rm g}$, the incident flux $f$, and the single frame readout noise $\sigma_{\rm R}$. It is more pronounced at very low fluxes, while it lowers with the number of groups $n_{\rm g}$, becoming almost negligible for large $n_{\rm g}$. 
Similarly, as shown in the bottom panels of Fig.~\ref{fig:S_bias_fcn_ng}, QF slightly deviates from its expected value $n_{\rm g} - 2$ (see Appendix \ref{app:QFbias} for the formal derivation of this result). 
To understand the possible origin of these deviations, in the following paragraphs we revisit the assumptions underlying the likelihood function: the Gaussian approximation for $\Delta G$, diagonal covariance, and signal dependence of $D_{ij}$. 

\subsection{Applicability of Gaussian approximation}
\label{sec:gaus}
The signal estimator was originally derived from a likelihood function modelled as the product of $n_{\rm g}-1$ independent Gaussian-distributed variables $\Delta G_i$. The
$\Delta G_i$ are constructed from reads which are affected by photon noise (Poisson statistics) and electronic readout noise (Gaussian).

The central limit theorem ensures that, for signal levels exceeding $g\approx 10\,{\rm e}^{-}\,{\rm group}^{-1}$, the Poisson distribution can be well approximated by a Gaussian, independently of the value of the pixel's readout noise. 
In the \Euclid framework this corresponds to $f\approx 0.34\,{\rm e}^{-}\,{\rm s}^{-1}$ and $f\approx0.25\,{\rm e}^{-}\,{\rm s}^{-1}$ for the photometric and spectroscopic readout mode, respectively. 
At lower flux levels, when the central limit theorem does not hold, $\Delta G_i$ are dominated by Gaussian-distributed readout noise associated with the group differences, characterised by $\gamma$.

In the case of NISP, where more than 95\% of pixels have $10\leq\sigma_{\sfont R}/{\rm e}^{-}\leq16$
\citep{Kubik2025} and $n_{\rm f} = 16$, the group differences are characterised by a typical readout noise $\sqrt{\,\gamma} \approx 4.6\,{\rm e}^{-}$. 
Consequently, even under dark conditions ($f\approx 0.01\,{\rm e}^{-}\,{\rm s}^{-1}$) the Gaussian approximation remains valid. Furthermore, it continues to hold down to readout noise levels as low as $\approx 3.0\,{\rm e}^{-}$ and $\approx 1.6\,{\rm e}^{-}$ in photometric and spectroscopic readout mode, respectively.
Thus, the NISP instrument never operates in a regime dominated by pure Poisson statistics, so using a likelihood function expressed as a product of Gaussian terms is fully justified for the \Euclid scientific case, and it cannot be at the origin of the bias $b_{\rm f}$.

\begin{figure*}[!t]
\begin{center}
\includegraphics[scale=0.35]{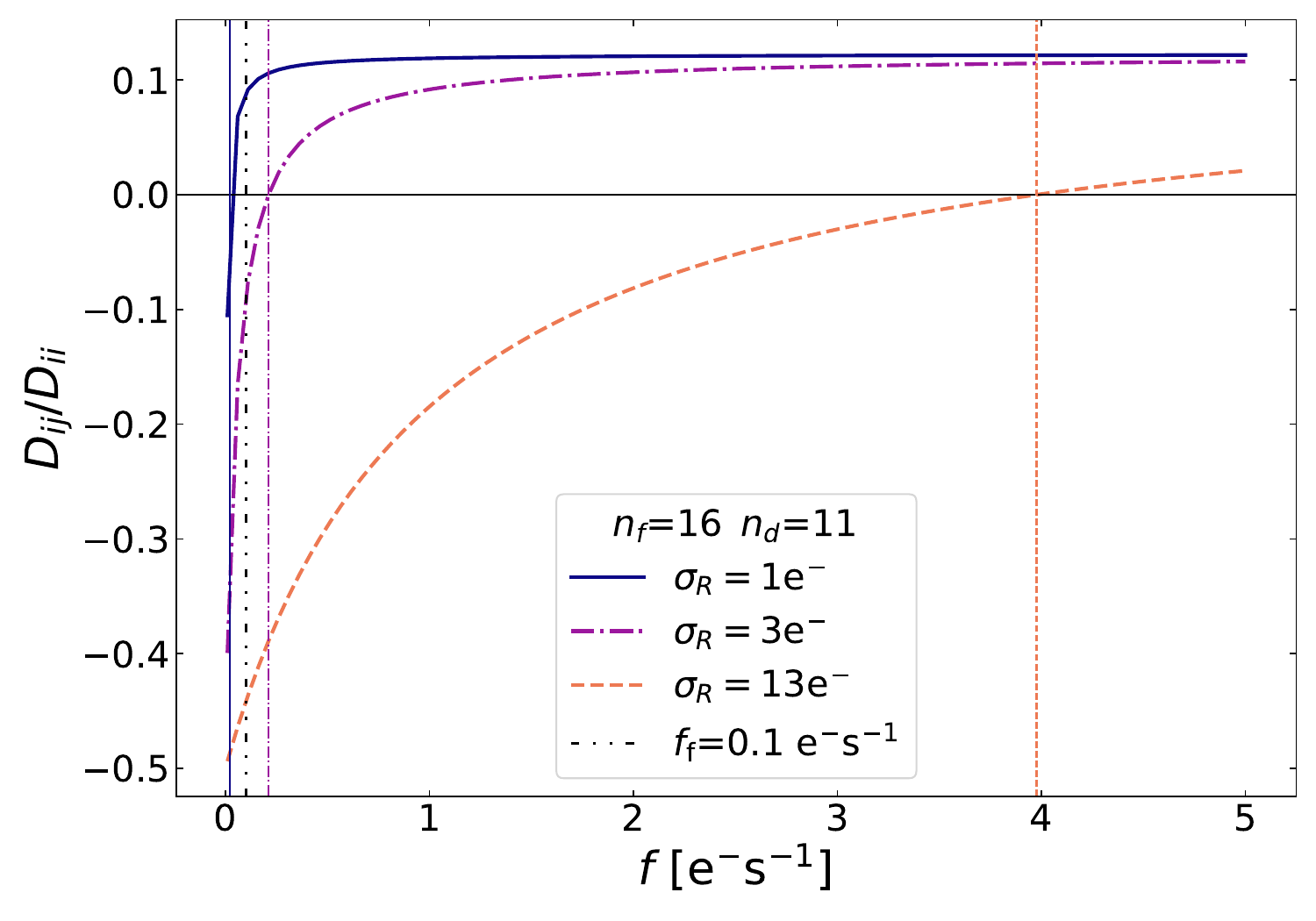}
\includegraphics[scale=0.35]{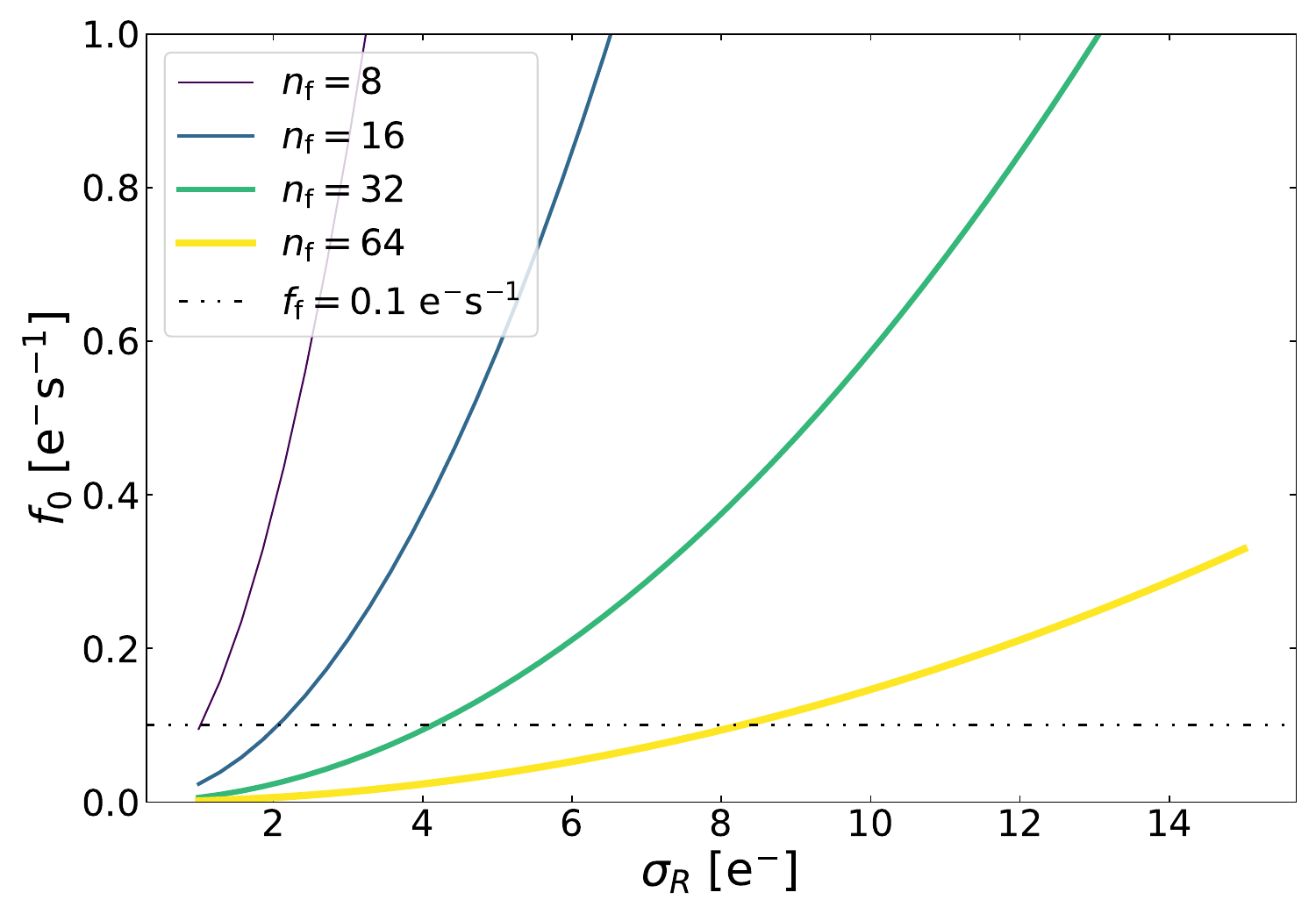}
\end{center}
\caption{\label{fig:dkl_dkk_ratio} \emph{Left panel:} Ratio of the off-diagonal terms $D_{ij}$ to the diagonal terms $D_{ii}$ in the covariance matrix of group differences as a function of flux for three different readout noise values. The flux values $f_{\rm 0}$ for which $D_{ij} = 0$ are indicated by vertical lines for each $\sigma_{\sfont R}$ value. \emph{Right panel:} Flux $f_{\rm 0}$, defined in Eq.\,(\ref{eq:f0}), plotted as function of the pixel's readout noise $\sigma_{\sfont R}$. The horizontal line indicates the limiting flux $f_{\rm f}$ defined in  Sect.~\ref{sec:gfolding} for no folding distribution at any $\sigma_{\sfont R}$ in spectroscopic readout mode.}
\end{figure*}

\subsection{Diagonal covariance\label{sec:covariance}}

In constructing the likelihood function, we assume the $\Delta G_{i}$ measurements to be statistically independent and uncorrelated, with their variance parametrised by the quantity $g$ to be estimated. Although the $\Delta G_{i}$ values are indeed independent, the neglect of correlations is justified only if the off-diagonal elements of their covariance matrix are negligible compared to the diagonal terms.

In the absence of incident flux, the ratio $ D_{ij}/D_{ii} $ reaches its maximum absolute value of 0.5 (see the left panel of Fig.~\ref{fig:dkl_dkk_ratio}), while in the high-flux limit ($g\to\infty$)
\begin{equation}
    \frac{D_{ij}}{D_{ii}} \to -\frac{\alpha}{2(1+\alpha)}\,,
\end{equation}
which varies from 0 ($n_{\rm f}=1$) to 0.25 ($n_{\rm f}\to\infty$), with $\alpha$ from Eq.\,(\ref{eq:gamma}). 

In the case of NISP ($n_{\rm f} = 16$), the ratio at high flux is approximately 0.18 for the photometric and 0.12 for the spectroscopic readout modes. Thus, the covariance terms $D_{ij}$ become significant at low flux levels, whereas at high flux they remain comparatively small -- typically a factor of four smaller than the diagonal terms.  

One can define the characteristic flux $f_{\rm 0}$, at which the off-diagonal covariance terms vanish, as follows
\begin{equation}\label{eq:f0}
    f_{\rm 0} = \frac{6\sigma_{\sfont R}^{2}}{(n_{\rm f}^{2}-1)\,t_{\rm {fr}}}\,.
\end{equation}
The dependence of $f_{0}$ on the single-frame readout noise $\sigma_{\sfont R}$ and on the number of averaged frames $n_{\rm f}$ is shown in Fig.~\ref{fig:dkl_dkk_ratio}. 

For the typical readout noise value in NISP detectors, i.e., $\sigma_{\sfont R}\geq13\,{\rm e}^{-}$, the $f_{\rm 0}$ is around $1\,{\rm e}^{-}\,{\rm s}^{-1}$. Therefore, the effects of the diagonal covariance approximation on the NISP signal estimator should be carefully addressed to avoid uncontrolled systematics. 

\subsection{Folding at low flux}
\label{sec:gfolding}

The consequence of treating the photon-noise component as a fitted parameter rather than deriving it directly from the data is a non-linear form of $\hat{g}[\Delta \vec{G}]_{n_{\rm g}}$ and $\hat{g}_x[\Delta \vec{G}]_{n_{\rm g}}$.

At very low incident flux, random fluctuations due to the readout noise may lead to negative values of $\Delta G_i$. Whenever $\Delta G_i < -\beta$, the squared term $(\Delta G_i + \beta)^2$ leads to an upward mapping of such values, resulting in an overestimation of the signal. This systematic overestimation is referred to as `folding bias'.

The occurrence of folding bias is inherently probabilistic and depends on the flux $f$, the single frame readout noise $\sigma_{\sfont R}$, and the MACC readout parameters $n_{\rm g}$, $n_{\rm f}$, and $n_{\rm d}$. The probability $P_1(\Delta G < -\beta)$ that a single group difference falls below the folding threshold $-\beta$ is given by
\begin{equation} 
    P_1(\Delta G < -\beta) = \frac{1}{2}\left[1 + \mathrm{erf} \left( \frac{- \beta - g}{\sqrt{2 D_{ii}}} \right) \right]\,.
\end{equation}
This is denoted as folding probability, and it diminishes whenever either the flux or the readout noise increases. For example, in a typical dark regime with $f\approx0.01\,{\rm e}^{-} {\rm s}^{-1}$, pixels with $\sigma_{\sfont R} \approx 3\,{\rm e}^{-}$ have a folding probability $P_1<1\%$ in both, photometric and spectroscopic readout modes, indicating minimal contribution from folding bias. 
Conversely, at $\sigma_{\sfont R}\approx 1\,{\rm e}^{-}$ -- an order of magnitude below \textit{Euclid}'s NISP readout noise level -- $P_1$ can exceed 5\% for fluxes below $0.1\,{\rm e}^{-} {\rm s}^{-1}$, reaching 25\% for $f \approx 0.01\,{\rm e}^{-} {\rm s}^{-1}$. 
Nevertheless, by averaging over $n_{\rm g} - 1$ group differences, the net contribution of any single folded value is significantly reduced, making the overall contribution of the folding bias negligible as $n_{\rm g}$ increases. 

To estimate the likelihood that a ramp is unaffected by the folding bias, we define $P_2$ as the cumulative probability that no more than 10\% of the $\Delta G_i$ values in a ramp fall below the folding threshold. 
The probability $P_2$ is modelled using the binomial cumulative distribution function
\begin{equation} 
    P_2 = \sum_{k=0}^{0.1(n_{\rm g}-1)} \binom{n_{\rm g}-1}{k}\, P_1^{k} (1 - P_1)^{n_{\rm g} - 1 - k}\,,
\end{equation}
where, when calculating $0.1(n_{\rm g}-1)$, we take the integer part. The 10\% probability threshold is chosen as a practical benchmark to facilitate approximate statistical characterisation of folding effects in the absence of a closed-form expression for their impact on the ramp slope. 

The precise influence of folded values on the signal estimate is inherently dependent on multiple factors -- including the incident flux level, pixel readout noise, and the readout configuration -- all of which affect the folding probabilities $P_1$ and $P_2$, and the bias propagation. As such, this threshold serves primarily to enable probabilistic assessments rather than to define a strict cutoff for quality control.

\begin{figure}
\begin{center}
\includegraphics[scale=0.35]{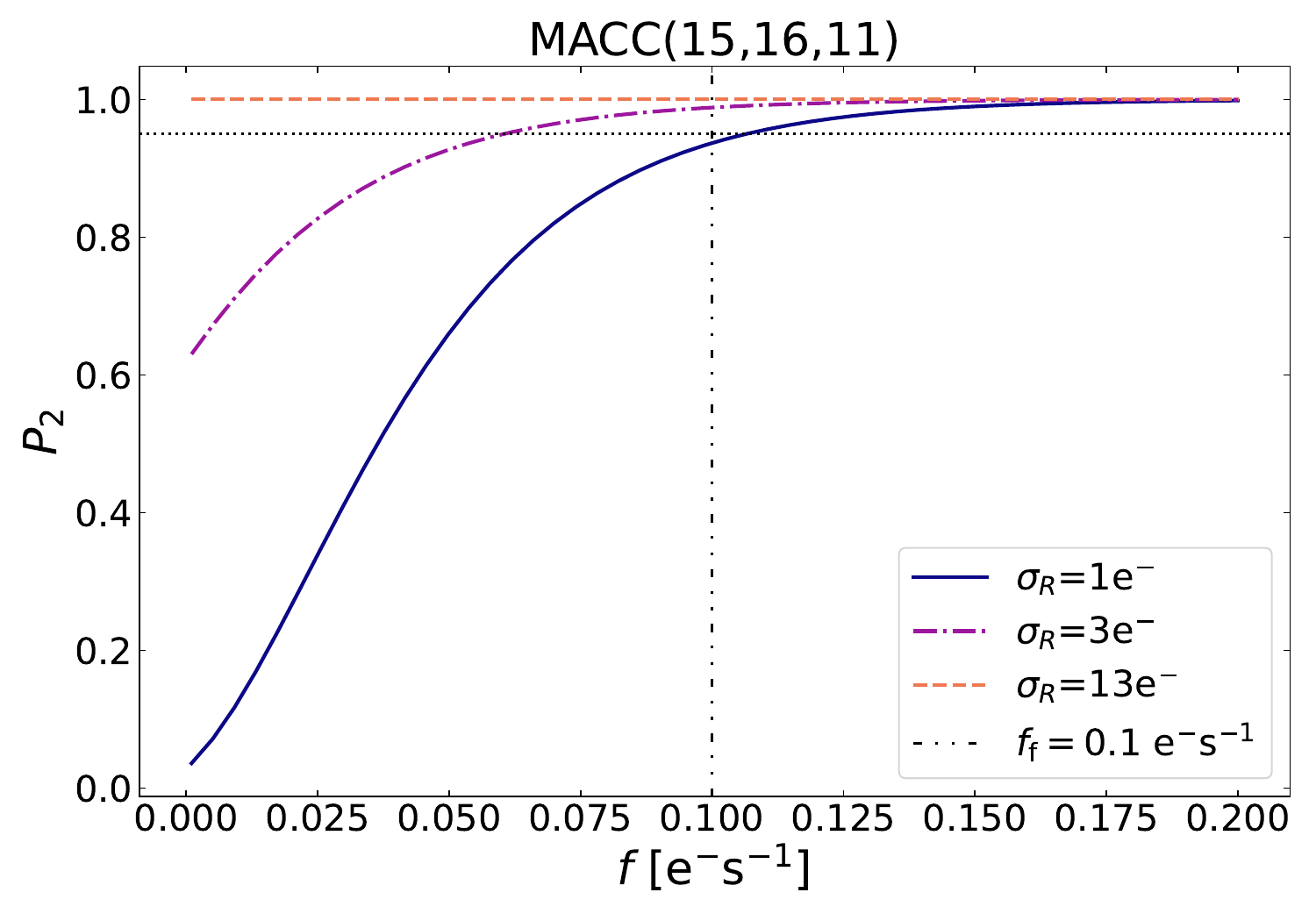}
\end{center}
\caption{\label{fig:binom_proba_spectro}Probability that no more than 10\% of $\Delta G$ fall below the folding point $-\beta$ in spectroscopic readout mode. The contribution is negligible for all readout noise values for flux higher than $f_{\rm f}=0.1\,{\rm e}^{-} {\rm s}^{-1}$. For typical \Euclid NISP pixels with readout noise $\sigma_{\sfont R}=13\,{\rm e}^{-}$, the folding contribution is negligible for the entire range of fluxes. A similar result holds for the photometric readout mode.}
\end{figure}

Figure~\ref{fig:binom_proba_spectro} shows $P_2$ over a wide range of fluxes and three, relatively low, values of the single frame readout noise for the spectroscopic readout mode. 
We define $f_{\rm f}$ as the `folding flux', that is, the flux beyond which the probability $P_{2}$ exceeds 95\% for the lowest possible readout noise value $\sigma_{\sfont R}=1\,{\rm e}^{-}$. It is a conservative threshold that shows the general limitations of the signal estimator. For example, typical pixels in NISP detectors exhibit $P_2\approx 1$ over the entire flux range, indicating the marginality of this type of error in practical application.

\subsection{Properties of the NISP signal estimator\label{sec:bias_orig}}

The signal estimator $\hat{g}[\Delta \vec{G}]_{n_{\rm g}}$ is a random variable, as it is a function of a set of $\Delta G_i$, which are themselves random variables. However, the probability distribution of $\hat{g}[\Delta \vec{G}]_{n_{\rm g}}$, as implied by the likelihood function, is not equivalent to that of the $\Delta G_i$. Specifically, the likelihood-based distribution of $\hat{g}[\Delta \vec{G}]_{n_{\rm g}}$ is only normalisable for $g>-\beta$, whereas the distribution in the $\Delta G_i$ space is normalisable from $-\infty$ to $+\infty$. This asymmetry highlights that $\hat{g}[\Delta \vec{G}]_{n_{\rm g}}$ and $\Delta G$  are not statistically equivalent variables.

The signal estimator $\hat{g}[\Delta \vec{G}]_{n_{\rm g}}$ contains a term $M_2( \Delta \vec{G}, \beta, n_{\rm g} )$, which is the sum of independent Gaussian-distributed variables (see Sect. \ref{sec:gaus}), and hence it follows a scaled non-central $\chi_{n_{\rm g}-1}^2(\lambda)$ distribution with $n_{\rm g}-1$ degrees of freedom and a non-centrality parameter
\begin{equation}
\lambda^2 = \frac{(n_{\rm g} - 1)(g + \beta)^2}{D_{ii}}\,.
\end{equation}
For $\beta$ and $g$ sufficiently large with respect to $\xi$, from Eq.\,(\ref{eq:gL}) one obtains $\hat{g}[\Delta \vec{G}]_{n_{\rm g}}\approx\sqrt{M_2( \Delta \vec{G}, \beta, n_{\rm g} )}$, and thus the signal estimator $\hat{g}[\Delta \vec{G}]_{n_{\rm g}}$ approximately follows a scaled non-central $\chi_{n_{\rm g}-1}(\lambda)$ distribution with $n_{\rm g} - 1$ degrees of freedom.

The statistical properties of the $\chi_{n_{\rm g}-1}(\lambda)$ distribution are well characterised. The mean can be written as
\begin{equation}\label{eq:noncentral_chi_mean}
\mu = \sqrt{\frac{2D_{ii}}{n_{\rm g} - 1}} \, \frac{\Gamma\left( \frac{n_{\rm g}}{2} \right)}{\Gamma\left( \frac{n_{\rm g} - 1}{2} \right)} \, {}_1F_{1}\left( \frac{-1}{2}, \frac{n_{\rm g} - 1}{2}, \frac{-\lambda^2}{2} \right),
\end{equation}
where ${}_1F_{1}(a,b,z)$ is the confluent hypergeometric function of the first kind.
The variance is given by
\begin{equation}\label{eq:noncentral_chi_variance}
\sigma^2 = \frac{D_{ii}}{n_{\rm g} - 1} \left( n_{\rm g} - 1 + \lambda^2 - \frac{\mu}{D_{ii}} \right)\,.
\end{equation}
We have verified that Eqs.\,(\ref{eq:noncentral_chi_mean}) and (\ref{eq:noncentral_chi_variance}) effectively describe the behaviour of $\hat{g}[\Delta \vec{G}]_{n_{\rm g}}$, however, a complication arises from the fact that both the mean and variance depend explicitly on $g$, which is not known a priori. Therefore, in order to establish the correction functions for the deviation of the signal estimator $\hat{g}[\Delta \vec{G}]_{n_{\rm g}}$ from the actual value of $g$, we will need some approximations.

\subsubsection{Constant bias at high flux\label{sec:gbias}}

The signal bias for $f>f_{\rm f}$ can be accurately modelled using a closed-form analytical expression and subsequently corrected. The analytical form of the bias estimator is derived in Appendix~\ref{app:Gbias} and is given by
\begin{equation}\label{eq:bias}
    \hat{b}_{\rm f} = \frac{-\xi}{(n_{\rm g}-1)(n_{\rm f} + n_{\rm d} )\,t_{\rm{fr}}}\,,
\end{equation}
where $\xi\in (\frac{1}{3},\frac{1}{2}]$ is a coefficient that depends solely on the MACC readout parameters.

The theoretical prediction of the bias is exact for $f = f_{\rm 0}$, as shown in the top left panel of Fig.~\ref{fig:S_bias_fcn_ng}, where the measured bias aligns with the analytical formula across various $n_{\rm g}$ values. In cases where the folding bias becomes significant even at $f=f_{\rm 0}$ (namely, when $f_{\rm 0}<f_{\rm f}$) -- typically corresponding to pixels with low readout noise, as shown in the right panel of Fig.~\ref{fig:dkl_dkk_ratio} -- the theoretical bias prediction no longer holds.

The bias $b_{\rm f}$ is typically small compared to both the signal and its statistical error. Nevertheless, in high signal-to-noise observations, where systematic effects become significant, this bias can be readily subtracted due to its well-defined analytical form. The fact that, even at $f_{\rm 0}$ where the off-diagonal terms cancel, a non-zero bias in the flux estimator remains, indicates that the origin of the bias is not solely due to the omission of covariances in the likelihood, but also stems from the fact that the expression for the measurement uncertainties $D_{ii}$ in the likelihood depends on the parameter to be estimated, namely the flux itself.

\subsubsection{Variance of the signal estimator}
\begin{figure*}
\begin{center}
\includegraphics[scale=0.35]{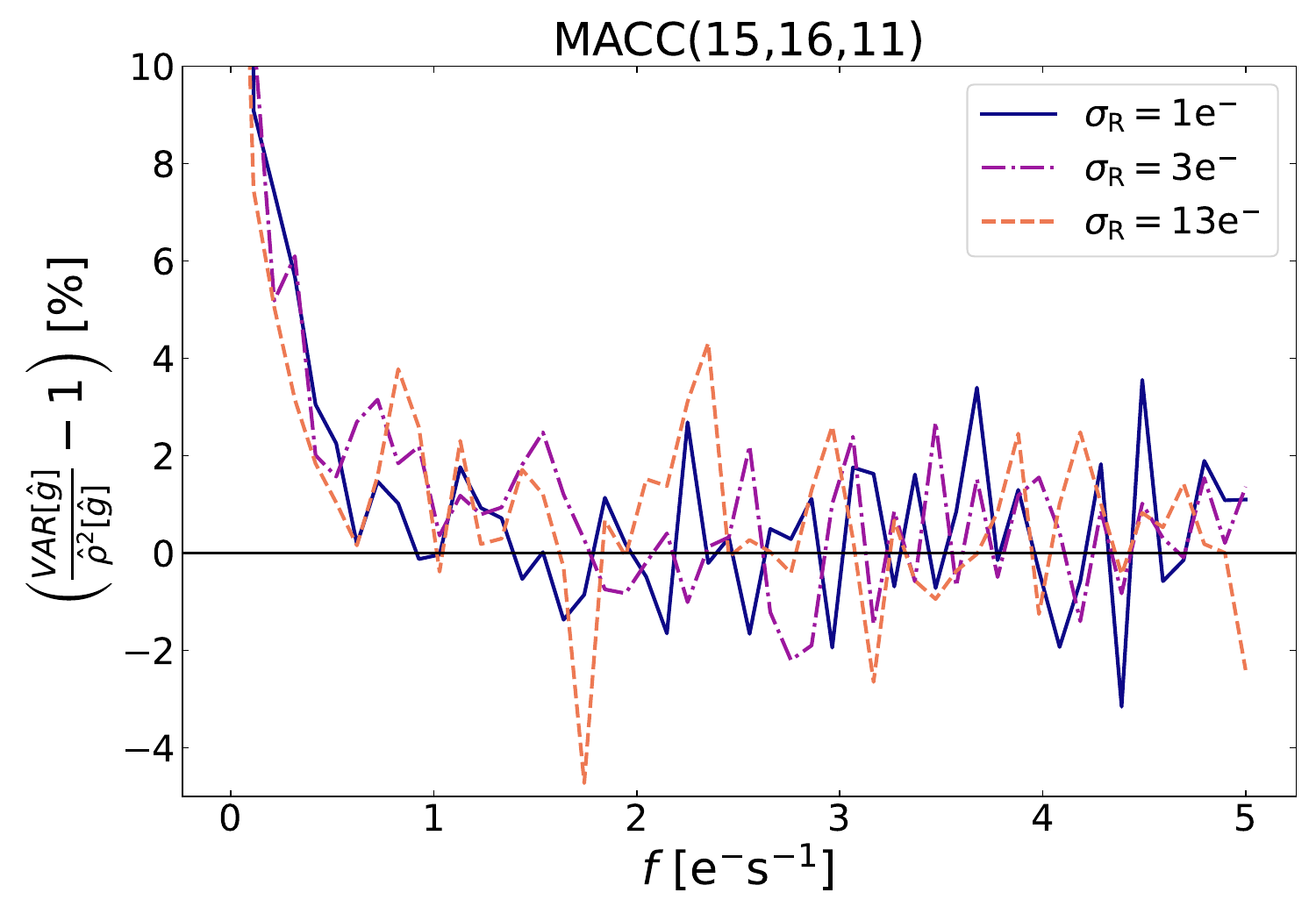}
\includegraphics[scale=0.35]{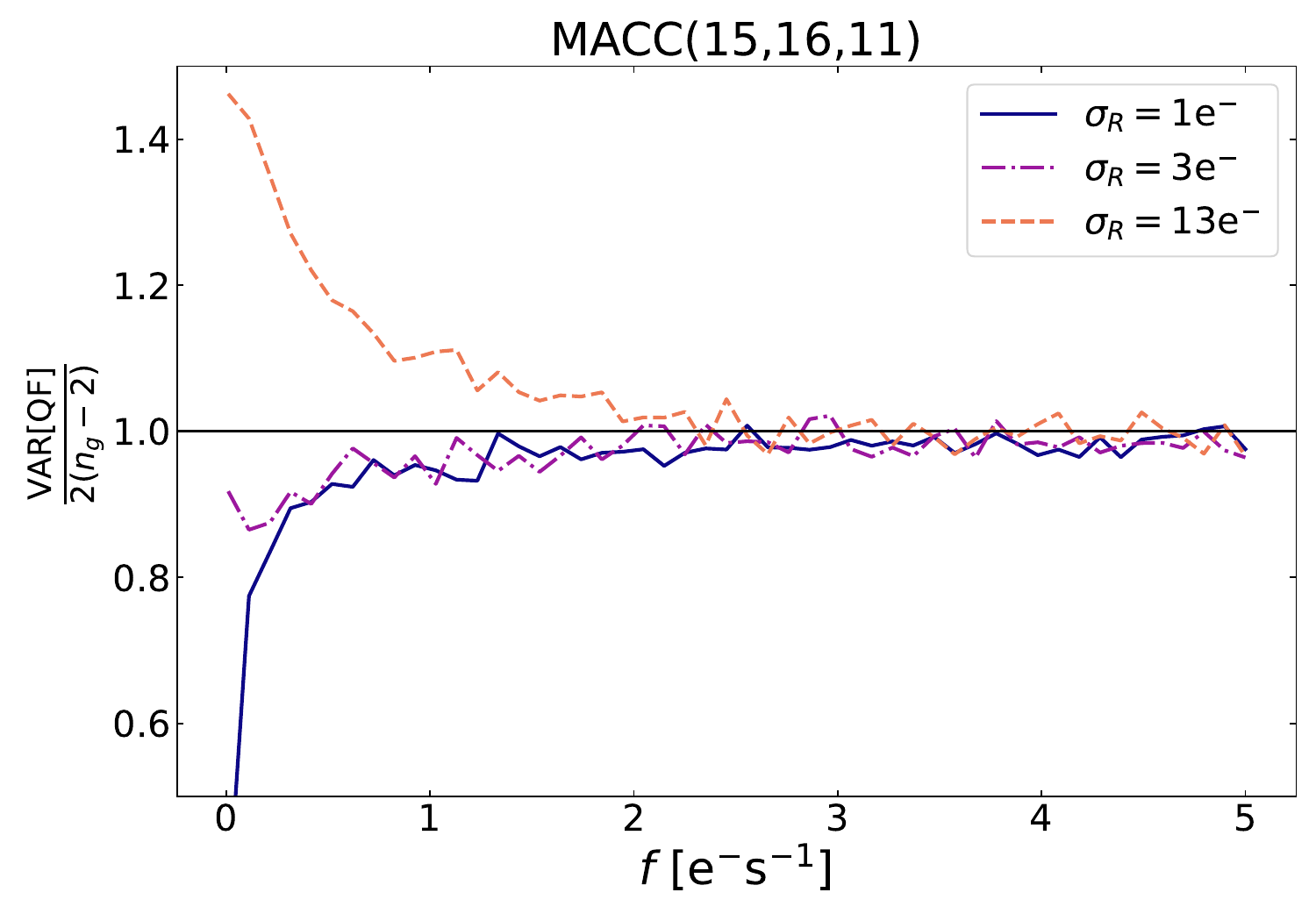}
\end{center}
\caption{\label{fig:variance}\emph{Left panel:} Variance estimator $\rho^{2}[\hat{g}]$ defined in Eq.\,(\ref{eq:var_def}) compared to the observed dispersion of the $\hat{g}[\Delta \vec{G}]_{n_{\rm g}}$ values. The theoretical estimate is in perfect agreement with the observed value for $f>0.5\,{\rm e}^{-} {\rm s}^{-1}$, independently of the readout noise and readout mode. \emph{Right panel:} QF variance normalised to $2(n_{\rm g}-2)$ as function of flux in spectroscopic readout mode.}
\end{figure*}

In equation (12) of \citet{Kubik2016}, the variance $\hat{\sigma}_{\hat{g}}^2$ of the signal estimator $\hat{g}[\Delta \vec{G}]_{n_{\rm g}}$ was defined 
as the inverse of the second derivative of the log-likelihood function evaluated at its minimum. However, this approach neglects the off-diagonal elements of the covariance matrix, leading to an inaccurate estimation of the uncertainty. Specifically, for $f < f_{\rm 0}$, where the $D_{ij}$ terms contribute negatively to the total variance, the measured variance of $\hat{g}[\Delta \vec{G}]_{n_{\rm g}}$ is smaller than the estimate $\hat{\sigma}_{\hat{g}}^2$. Conversely, for $f > f_{\rm 0}$, where the $D_{ij}$ terms contribute positively, the measured variance exceeds the estimated value. The magnitude of this discrepancy depends on both the flux and the single frame readout noise $\sigma_{\sfont R}$. 

In this work, we derive an improved variance estimator based on standard error propagation applied to the group differences $\Delta G$. The full derivation is provided in Appendix~\ref{app:Gvariance}, and the final expression is given by
\begin{equation}\label{eq:var_def}
    \hat{\rho}^{2}[\hat{g}] = \frac{[(n_{\rm g}-1)\hat{g}  + \alpha \hat{g} + \gamma]}{(n_{\rm g}-1)^{2}} \frac{(\hat{g}+\beta)^{2}}{(\hat{g}+\beta)^{2} + \xi^{2}}\,,
\end{equation}
As shown in the left panel of Fig.~\ref{fig:variance}, this new estimator accurately reproduces the measured variance of $\hat{g}[\Delta \vec{G}]_{n_{\rm g}}$ for $f > 0.5\,{\rm e}^{-} {\rm s}^{-1}$, independently of the flux, the readout noise, and the MACC parameters. 

For $f < 0.5\,{\rm e}^{-} {\rm s}^{-1}$, the $\hat{\rho}^2[\hat{g}]_{n_{\rm g}}$ estimator tends to underestimate the actual variance of the NISP signal estimator. The origin of this discrepancy is currently under investigation, but the impact on \Euclid science is negligible since the zodiacal background in NISP exposures is approximately $1\,{\rm e}^{-} {\rm s}^{-1}$.

\subsection{Properties of the NISP quality factor}
\label{sec:QF}

The QF is a statistical indicator of the reliability of the NISP signal estimator and, like the signal estimator itself, is a random variable. QF, given by Eq.\,(\ref{eq:QF}), is defined as a scaled difference between two signal estimators, namely $\hat{g}_x[\Delta \vec{G}]_{n_{\rm g}}$ and the mean group difference. Given that the mean group difference is unbiased with respect to $g$, we can infer the behaviour of QF from the detailed analysis of the signal estimator bias described in the previous sections.

\subsubsection{QF expectation value}

A first-order approximation of the QF expectation value is derived in Appendix~\ref{app:QFbias} as
\begin{equation}\label{eq:QFbias}
    \langle {\rm QF} \rangle_{n_{\rm r}} =  n_{\rm g}-2\,.
\end{equation}
This formulation is exact for $f=f_{\rm 0}$ as shown in the bottom left panel of Fig.~\ref{fig:S_bias_fcn_ng} (green squares). For $f<f_{\rm 0}$, the QF exceeds $n_{\rm g} - 2$, while for $f>f_{\rm 0}$, it falls below this value. 
The opposite trend of the QF bias with respect to $\hat{b}_{\rm f}$ comes from the fact that, as shown in Appendix~\ref{app:QFbias}, the bias $\hat{b}_{\rm{fx}}$ affecting $\hat{g}_x[\Delta \vec{G}]_{n_{\rm g}}$ has the opposite sign to the signal estimator bias $b_{\rm f}$, that is
\begin{equation}
    \label{eq:bfx}  
    \hat{b}_{\rm{fx}} = \frac{n_{\rm g}-2}{n_{\rm g}-1}\xi\,.
\end{equation}
Bottom right panel of Fig.~\ref{fig:S_bias_fcn_ng} shows the flux dependence of QF for three different readout noise values. In the specific context of NISP detectors, where the single frame readout noise is around $13\,{\rm e}^{-}$ for the majority of the pixels, and in the absence of additional systematic errors, 
the expected deviation of the average QF from its ideal value $n_{\rm g}-2$ remains below 1. For instance, under dark conditions with a flux of $0.01\,{\rm e}^{-}{\rm s}^{-1}$, the QF is expected to be approximately 2.61 in photometric mode and 13.67 in spectroscopic mode. For a zodiacal background around $1\,{\rm e}^{-}{\rm s}^{-1}$, the QF values are expected to be approximately 2.15 and 13.13 for the photometric and spectroscopic readout modes, respectively.

\subsubsection{Variance of QF}

For completeness, we also derive an analytical expression for the variance of QF, i.e., $\hat{\rho}^{2}[{\rm QF}]$. 
In Appendix~\ref{app:QFvariance}, we demonstrate that, to leading order, the variance of the quality factor is given by
\begin{equation}
    \hat{\rho}^{2}[{\rm QF}] \xrightarrow{g \to \infty} 2(n_{\rm g}-2)\,.
\end{equation}
The right panel of Fig.~\ref{fig:variance} corroborates this result using simulations of QF in spectroscopic readout mode, spanning a wide range of fluxes and three distinct readout noise values.

\section{NISP onboard data processing}
\label{sec:data_processing}

Efficient onboard processing of NIR data is essential for the \Euclid mission because NISP produce large data volumes while available computing power and telemetry bandwidth are limited. This section outlines the hardware-software architecture of the NISP onboard data processing and details the approximations it applies to balance these constraints, enabling reliable and accurate signal estimation.

A dedicated hardware-software architecture, composed of two identical DPUs and a custom application software (ASW), enables the onboard processing of the NIR signal acquired by each pixel in the NISP FPA.
Each DPU is equipped with a set of eight Detector Control Units (DCUs), each of which interfaces to a SIDECAR ASIC electronics module \citep{Holmes2022} coupled via a cryogenic flex cable with an H2RG Sensor Chip Assembly \citep[SCA,][]{Bai2018}.
The DCUs supply analogue and digital power, manage telemetry and commanding for the FPA, and carry out the first stage of static basic preprocessing, operating independently and in parallel with the detector’s live data stream.
This preprocessing step performs arithmetic averaging of frames (at 24 bits/pixel) following the MACC$(n_{\rm g},\, n_{\rm f},\,n_{\rm d})$ readout scheme. 
The resulting averaged ‘group' $H_i$ (in ADU) is then decimated\footnote{This is the computational process of lowering the signal’s sampling rate or reducing the number of data points in order to decrease the dataset’s size and resolution.} to 16 bits/pixel and passed as input for signal estimation.

To allow the onboard processing each DPU is supported by a central processing unit (Maxwell 750\textsuperscript{\texttrademark} 3$\times$SCS750\textsuperscript{\tiny\textregistered}-PPC), and a memory transit data buffer board of 6\,GB to store single groups of the MACC that allows a complete decoupling of the DCU real-time operation from the subsequent, deferred-time, processing operation along the dither sequence of exposures.
The ASW runs on a VxWorks 5.1 real-time operating system and is based on a multi-task pre-emptive scheduling algorithm, with a core task responsible for single-frame acquisition, processing, lossless compression, and transmission to the spacecraft's mass memory unit \citep{Bonoli2016, Medinaceli2020, Medinaceli2022, EuclidSkyNISP}. 

In addition to frames averaging, the pipeline of sequential steps performed by the NISP ASW comprises also corrections for group non-uniformities related to video-channel gains/offsets, and detector line random offsets. This is achieved by subtracting the reference vertical and horizontal pixels after median filtering and zonal averaging \citep{Bonoli2016}.    

Saturated pixels ($2^{16}$~ADU) are labelled for no further signal estimation. For the remaining pixels, the discrete derivative between consecutive groups ($\Delta H_i$) is evaluated. 
According to the framework described in \citet[][see Eqs. 11, 13, 19, and 20]{Kubik2016}, the signal and the QF maps are analytically derived from data in ADU, and were expressed as
\begin{equation}
\label{eq:ghat_DPU}
    \hat{h}[\vec{\Delta H}]_{n_{\rm g}} = \sqrt{\xi^{h\,2} + M_2( \Delta \vec{H}, \beta^h, n_{\rm g} )} - \xi^h - \beta^h\,;
\end{equation}
\begin{equation}
\label{eq:QF_DPU}
{\rm QF_h}[\vec{\Delta H}]_{n_{\rm g}} =  \frac{(n_{\rm g}-1)}{\xi^h} \left( \,\hat{h}_x[\vec{\Delta H}]_{n_{\rm g}} - \frac{H_{n_{\rm g}} - H_1}{n_{\rm g}-1}\right)\,,
\end{equation}
where
\begin{equation}
\label{eq:ghatX_DPU}
\hat{h}_x[\vec{\Delta H}]_{n_{\rm g}} =   \sqrt{ M_2( \Delta \vec{H}, \beta^h, n_{\rm g} )} - \beta^{\rm h}\, .
\end{equation}
\noindent
It is worth noting that $\hat{h}[\vec{\Delta H}]_{n_{\rm g}}$ and $\hat{h}_x[\vec{\Delta H}]_{n_{\rm g}}$ are expressed in units of ADU\,group$^{-1}$, and the following parameters are defined in ADU
\begin{equation}
\begin{aligned}
\sigma^{\rm h}_{\sfont R} \equiv \frac{\sigma_{\sfont R}}{f_{\rm e}}\,\,;\quad \beta^{\rm h} \equiv  \frac{\beta}{f_{\rm e}} = \frac{2f_{\rm e} \ \sigma^{\rm h\ 2}_{\sfont R}}{n_{\rm f} (1+\alpha)}\,\,;\quad
\xi^{\rm h} \equiv  \frac{\xi}{f_{\rm e}} = \frac{1+\alpha}{2f_{\rm e}}\;.
\end{aligned}
\end{equation}

The ${\rm QF_h}[\vec{\Delta H}]_{n_{\rm g}}$ image is decimated to an 8-bit resolution for the spectrometric readout mode and to a 1-bit resolution for the photometric readout mode after applying a threshold. The signal $\hat{h}[\vec{\Delta H}]_{n_{\rm g}}$ is converted to electrons in the SGS pipeline, as described in the following section.
\subsection{Data Processing Approximation (DPA)}
\label{sec:dpa}
NISP onboard processing is based on a programmable parametrisation, including the MACC readout parameters, and the readout noise and conversion gain values. 
Accurate estimation of the flux incident on a pixel requires precise knowledge of both the single-frame readout noise $\sigma_{\sfont R}$ and the conversion gain $f_{\rm e}$, as the NISP signal estimator relies on these parameters to infer $f$ from the acquired UTR frames. 
For each pixel in the NISP FPA, these two parameters were accurately measured during the ground characterisation of the NISP detectors \citep{Kubik2025}.
 
However, to compute the NISP signal and QF maps, the NISP onboard data processing utilises the detector-averaged values of these parameters, namely $\sigma_{\sfont{R,\,DPU}}\,$ and $f_{\rm{e,}\,\sfont{DPU}}$. By storing these two values in the non-volatile memory of the two NISP DPUs -- instead of the full $\sigma_{\sfont{R}}$ and $f_{\rm e}$ maps -- only 512 bits are allocated, as opposed to approximately 2 Gbits. Furthermore, because only a reduced parameter set\footnote{That is, $2\times16$ values rather than $2\times2048\times2048\times16$ values.} is copied to the DPUs' RAM, this approach significantly optimises the allocation of volatile memory during the onboard data processing.

This feature of the NISP onboard data processing is referred to as Data Processing Approximation (DPA). 
From a general perspective, if incorrect values of $\sigma_{\sfont R}$ or $f_{\rm e}$ are considered, the NISP signal estimator effectively employs an incorrect variance ($D_{ii}$ defined in Eq.\,\ref{eq:Dkl}) in the likelihood function, which may introduce a systematic bias in the signal and QF estimation.

We define the readout noise and conservation gain shifts as
\begin{equation}
\label{eq:DPA_bias_param}
    \Delta \sigma_{\sfont R} = \sigma_{\sfont R} - \sigma_{\sfont{R,\, DPU}}\; \qquad \Delta f_{\rm e} = f_{\rm e} - f_{\rm{e,}\,\sfont{DPU}}\,,
\end{equation}
while the bias affecting the NISP signal estimator is
\begin{equation}
\label{eq:DPA_bias_sig}
    b_{\rm f}^{\sigma_{\sfont R}} = \frac{\hat{g}(\sigma_{\sfont{R,\,DPU}})-g}{(n_{\rm f} + n_{\rm d})\,t_{\rm{fr}}}\,;
    \qquad 
    b_{\rm f}^{f_{\rm e}} = \frac{\hat{g}(f_{\rm{e,}\,\sfont{DPU}})-g}{(n_{\rm f} + n_{\rm d})\,t_{\rm{fr}}}\,.    
\end{equation}

\begin{figure*}[!t]
\begin{center}
\includegraphics[scale=0.35]{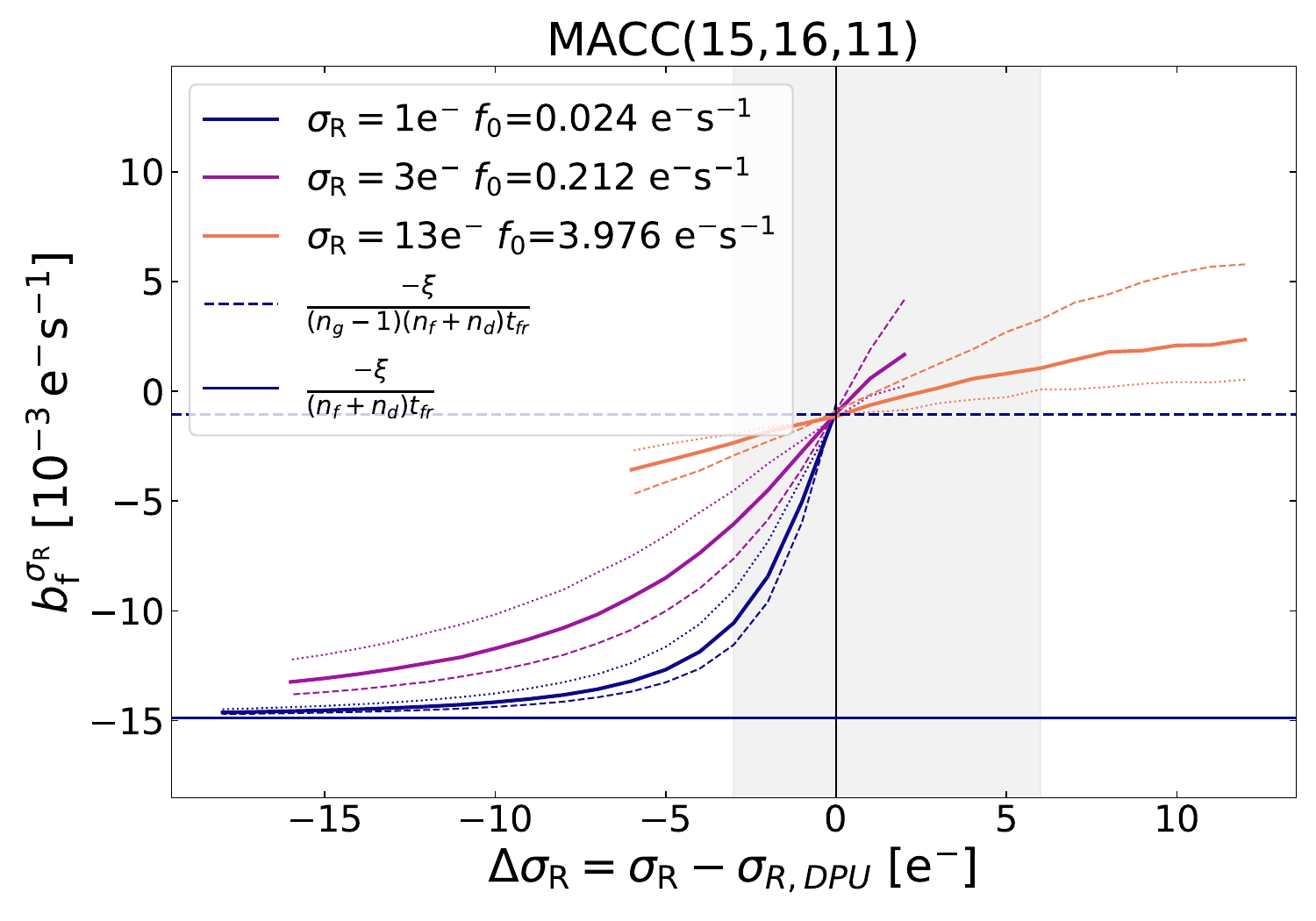}
\includegraphics[scale=0.35]{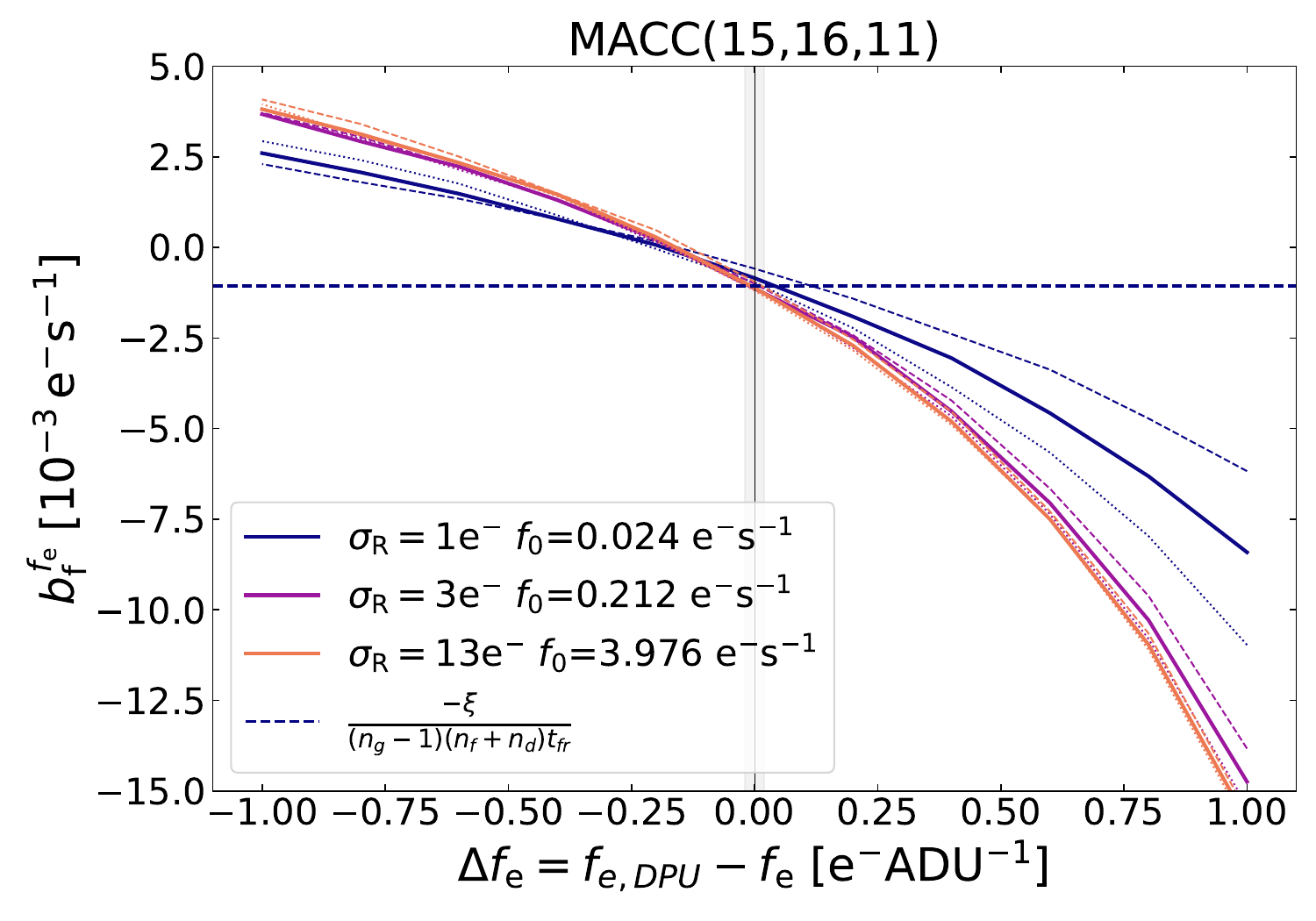}\\
\includegraphics[scale=0.35]{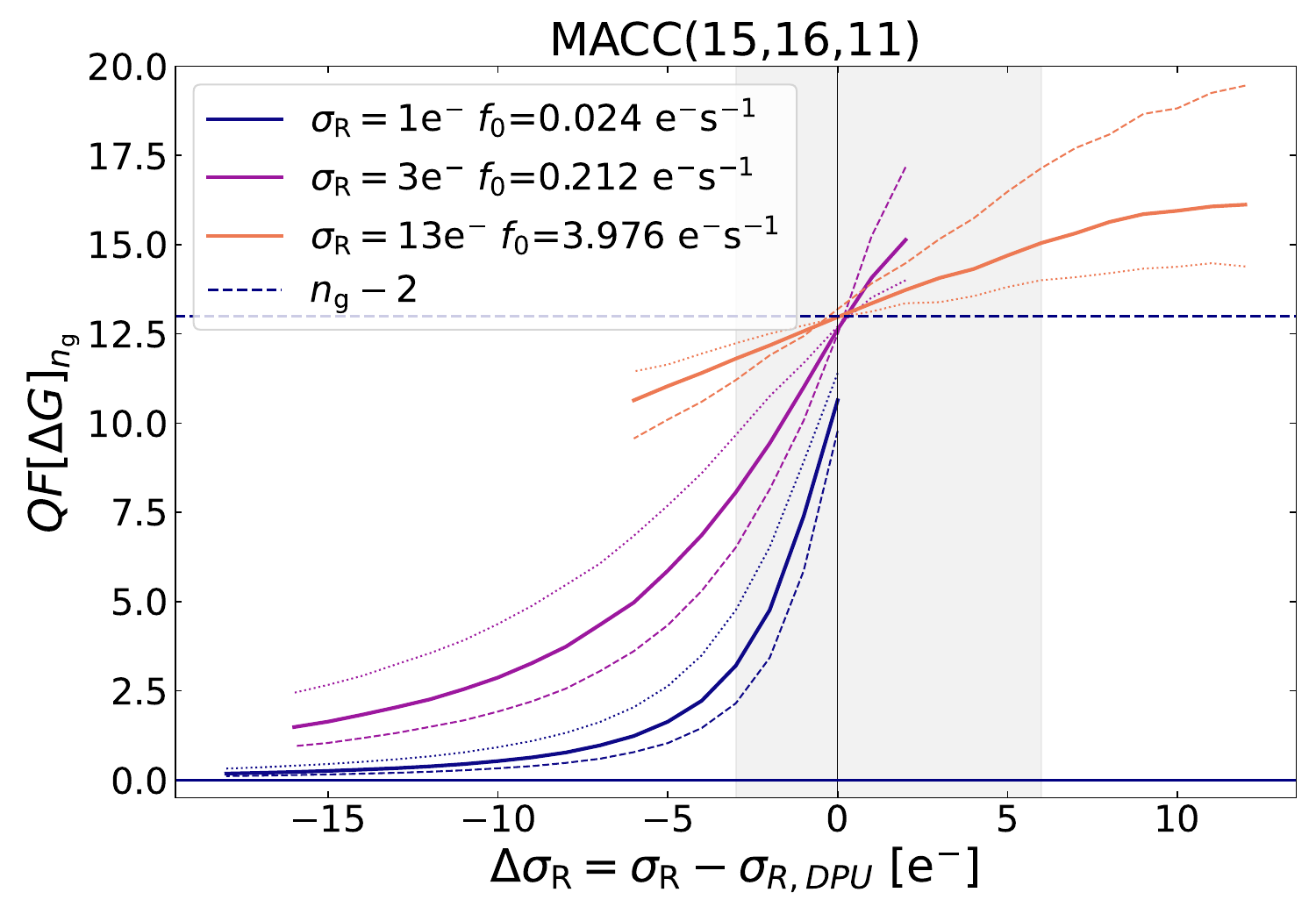}
\includegraphics[scale=0.35]{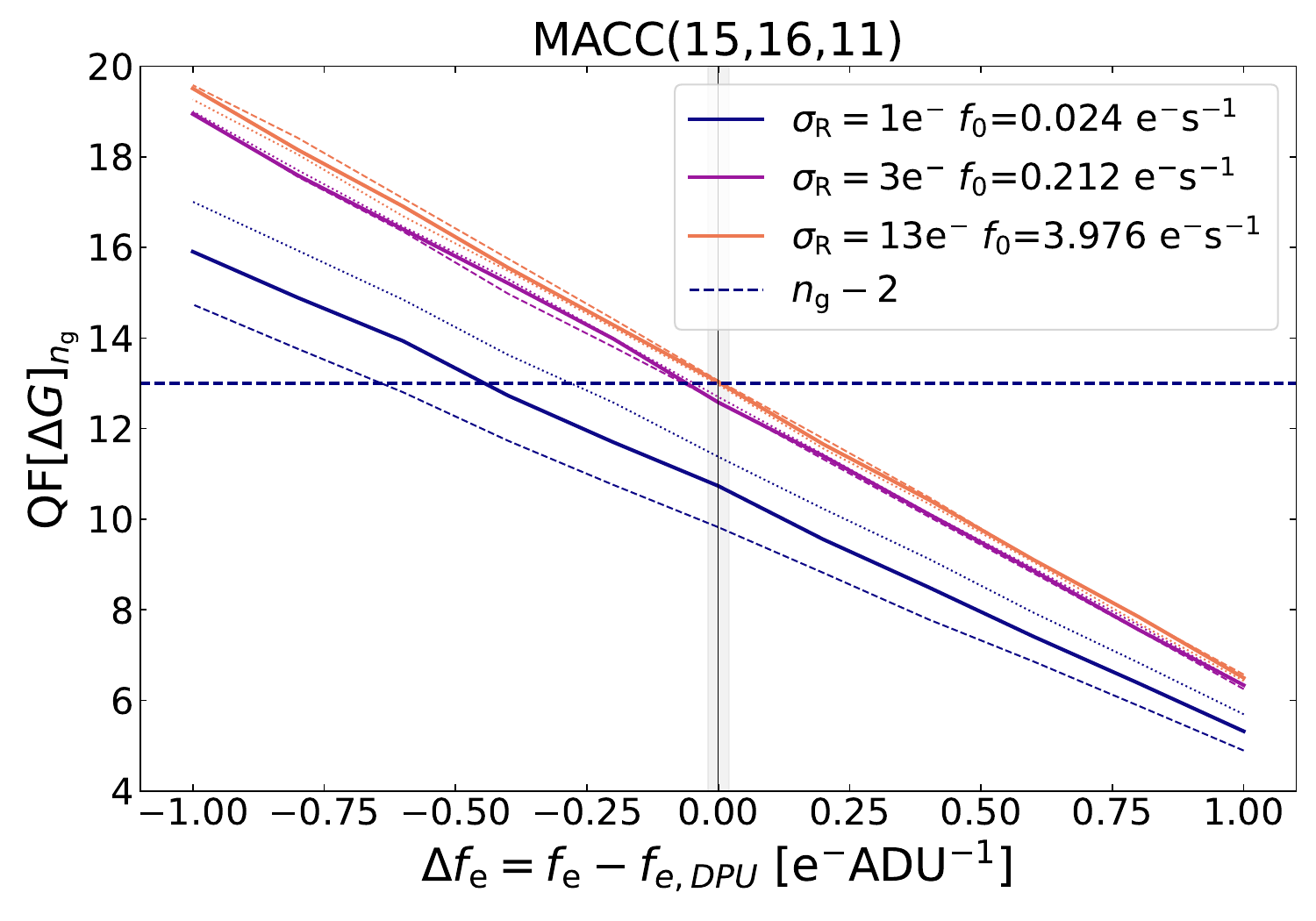}
\end{center}
\caption{DPA induced bias to the signal estimator $\hat{g}[\Delta \vec{G}]_{n_{\rm g}}$ (top panels) and to the QF (bottom panels). For each value of the pixel simulated readout noise $\sigma_{\sfont R}$, we have chosen the specific flux value $f=f_{\rm0}$ for which the off-diagonal terms in the covariance matrix vanish and the theoretical bias model conforms to the simulated values (thick solid lines). Additionally, for each $\sigma_{\sfont R}$ we perform simulations for $f=\frac{f_{\rm0}}{2}$ (dashed lines) and $f=2f_0$ (dotted lines). \emph{Left panels:} Bias due to $\Delta \sigma_{\sfont R}$. The shaded areas correspond to the typical \Euclid case of possible values of $-3 < \Delta \sigma_{\sfont R}<6$ as 99\% of pixels are in this range of $\Delta \sigma_{\sfont R}$ -- see Sect. \ref{sec:flight_performance_sig}. Typical \Euclid pixel has a readout noise $\sigma_{\sfont R}\approx 13\,{\rm e}^{-}$ and is represented by orange lines. \emph{Right panels:} Bias due to $\Delta f_{\rm e}$. The shaded area corresponds to the typical \Euclid dispersion of gain values.}
\label{fig:DPU_bias}
\end{figure*}

Figure~\ref{fig:DPU_bias} shows how the use of an incorrect readout noise, $\sigma_{\sfont{R,\,DPU}}$, biases both the signal (top left) and the QF (bottom left) estimation in spectroscopic readout mode as a function of $\Delta\sigma_{\sfont R}$. 
For each analysed value of $\sigma_{\sfont R}$, we simulate a flux $f_0$ for which the off-diagonal covariance terms vanish. Additionally, for each $\sigma_{\sfont R}$ we perform simulations for $f=f_0/2$ and $f=2f_0$. 

At $\Delta \sigma_{\sfont R}=0$ we obtain the expected theoretical bias (see Eq.~\ref{eq:bias}).  For $\Delta \sigma_{\sfont R}>0$ the flux is overestimated because analytically $\hat{g}(\sigma_{\sfont{R,\,DPU}})>\hat{g}(\sigma_{\sfont R})$ and, for very small $\sigma_{\sfont{R,\,DPU}}$, folding  further amplifies the effect. Conversely, if $\Delta\sigma_{\sfont R}<0$, a larger assumed noise induces a negative bias, which in the limit $\Delta \sigma_{\sfont R}\ll0$ cannot exceed $b_{\rm{f,\,min}}^{\sigma_{\sfont R}}=-\,\xi\, t_{\rm{fr}}^{-1}\,(n_{\rm f} + n_{\rm d})^{-1}$. 
The QF exhibits the same dependence on $\Delta \sigma_{\sfont R}$, confirming that both signal accuracy and QF performance degrade if non‐representative noise values are used. 

Similarly, if an incorrect conversion gain, $f_{\rm{e,}\,\sfont{DPU}}$, is used in the signal estimator instead of the true value $f_{\rm e}$, the resulting signal and QF will be biased, but differently compared to the $\Delta \sigma_{\sfont R}$ case.

Indeed, data are acquired and processed in ADUs, but the resulting signal, variance, and QF do not scale identically during conversion from ADUs to electrons. 

The likelihood model introduced in \citet{Kubik2016} accounts for this, and includes $f_{\rm e}$ explicitly in the estimator function, as shown in Eqs.\,(\ref{eq:ghat_DPU}) and (\ref{eq:QF_DPU}). 
In practice 
\begin{align}
    \hat{g}[\vec{\Delta G},\sigma_{\sfont R}]_{n_{\rm g}} 
        &\equiv \hat{g}[f_{\rm e} \vec{\Delta H}, f_{\rm e} \sigma_{\sfont R}^{\rm h}]_{n_{\rm g}} 
        \neq f_{\rm e}\,\hat{g}[\vec{\Delta H},\sigma_{\sfont R}^{\rm h}]_{n_{\rm g}} \,; \\
    \hat{g}_x[\vec{\Delta G},\sigma_{\sfont R}]_{n_{\rm g}} 
        &\equiv \hat{g}_x[f_{\rm e} \vec{\Delta H}, f_{\rm e} \sigma_{\sfont R}^{\rm h}]_{n_{\rm g}} 
        \neq f_{\rm e}\,\hat{g}_x[\vec{\Delta H},\sigma_{\sfont R}^{\rm h}]_{n_{\rm g}} \,; \\
    {\rm QF}[\vec{\Delta G},  \sigma_{\sfont R}]_{n_{\rm g}} 
        &\equiv {\rm QF}[f_{\rm e} \vec{\Delta H}, f_{\rm e} \sigma_{\sfont R}^{\rm h}]_{n_{\rm g}} 
        \neq f_{\rm e}\,{\rm QF}[\vec{\Delta H}, \sigma_{\sfont R}^{\rm h}]_{n_{\rm g}} \,.
\end{align}
while
\begin{align}
    \hat{g}[\vec{\Delta G}]_{n_{\rm g}} 
        &= f_{\rm e}\,\hat{h}[\vec{\Delta H}]_{n_{\rm g}} \,; \\
    \hat{g}_x[\vec{\Delta G}]_{n_{\rm g}} 
        &= f_{\rm e}\,\hat{h}_x[\vec{\Delta H}]_{n_{\rm g}} \,; \\
    {\rm QF}[\vec{\Delta G}]_{n_{\rm g}} 
        &= {\rm QF_h}[\vec{\Delta H}]_{n_{\rm g}} \,.
\end{align}

\noindent
Figure~\ref{fig:DPU_bias} shows how the use of an incorrect conversion gain introduces a non-linear bias in the signal estimation (top right panel), and in QF (bottom right panel). We note that for small $f_{\rm e}$ one obtains ${\rm QF}[\vec{\Delta G},\,  \sigma_{\sfont R}]_{n_{\rm g}} \approx f_{\rm e}\,{\rm QF}[\vec{\Delta H},\, \sigma_{\sfont R}^{\rm h}]_{n_{\rm g}}$.

\citet{Kubik2016} already demonstrated that the conversion-gain-induced DPA bias is negligible compared to the readout noise case. Moreover, the dispersion of the conversion gain within single detectors is lower than $1\%$, corresponding to a deviation lower than $0.02\,{\rm e}^{-} / {\rm ADU}$ for a typical detector-average conversion gain of $\approx 2\,{\rm e}^{-} / {\rm ADU}$.
On the other hand, the readout noise distributions display a dispersion of $\approx50\%$ around the detector-averaged values \citep{Kubik2025}.
Specifically, for a typical detector-average readout noise of $\approx 13\,{\rm e}^{-}$, the 99$\%$ of the NISP pixels are distributed within $-3\leq\Delta\sigma_{\sfont{R}}\,$/\,e$^{-}\leq6\,$, where the asymmetry towards positive values reflects the log-normal nature of the readout noise distributions.
An example of the dispersion of the readout noise distribution around the $\sigma_{\sfont{R,\,DPU}}$ value is reported in Fig. \ref{fig:RON_rawlineVSdetector}.

In light of these considerations, the analysis presented in this work quantifies the impact of DPA on the flight performance of the NISP signal estimator, focusing on the effects related to the readout noise parametrisation, under the assumption that gain-induced effects are negligible within the typical signal regimes of \textit{Euclid}’s NISP observations.

\section{Flight performance of the NISP signal estimator}
\label{sec:flight_performance_sig}
In the following sections, we evaluate the performance of the signal estimator and QF obtained from NISP data collected during early flight operations and compare it to simulations. 
We show that the signal bias is negligible across the flux range relevant for \Euclid. We also demonstrate that the bias introduced by the onboard data processing approximations aligns with predictions and affects only a tiny fraction of pixels. Finally, we assess the performance of the NISP QF, emphasising its sensitivity to space weather variations.

\subsection{Early NISP flight data\label{sec:data}}

The flight performance of the NISP signal estimator is evaluated by analysing the data acquired during the performance verification phase and part of the early survey operation phase \citep{EuclidSkyOverview}, from September\,2023 to April\,2024. 

\begin{table}[!t]
\caption{
Summary of the early NISP flight data analysed in this work. 
Dark images were acquired during the Performance Verification phase for the estimation of the photometric and spectroscopic NISP Master Dark maps \citep{Q1-TP003}.
Grism \RGE (red) and \BGE (blue) data are acquired in the NISP spectroscopic readout mode \citep{Gillard2025}. 
Filter \JE, \HE, and \YE data, corresponding to the three NISP filter passbands, are acquired in the NISP photometric readout mode \citep{Schirmer-EP18}. 
\RGE, \BGE, \JE, \HE, and \YE were acquired during the \Euclid's Early Survey Operation \citep{EuclidSkyOverview, EuclidSkyNISP}.} 
\label{tab:data}
\centering
\begin{center}      
\begin{tabular}{c|c|c}
\hline
\textbf{Observation Mode} & \textbf{Acquisition} & \textbf{Number} \\
\textbf{[\,Integration Time\,]} & \textbf{Type} & \textbf{of Exposures} \\
\hline
\hline
\multirow{3}{7em}{\shortstack{MACC(15,16,11)\\$[\,$549.6\,s$\,]$}}
& Dark & 100 \\
& Grism \RGE & 5700 \\
& Grism \BGE & 100 \\
\hline
\multirow{4}{6em}{\shortstack{MACC(4,16,4)\\$[\,$87.2\,s$\,]$}}
& Dark & 500 \\
& Filter \JE & 5700 \\
& Filter \HE & 5700 \\
& Filter \YE & 5700 \\
\hline
\end{tabular}
\end{center}
\end{table}
Table \ref{tab:data} summarises the datasets considered in this analysis, which essentially comprise two different types of data, namely dark and scientific exposures. 
Dark exposures ($f\approx 0.01\,{\rm e}^{-} {\rm s}^{-1}$) are taken with the NISP filter wheel in the closed position to measure the detectors’ thermal background. 
Scientific exposures are acquired by selecting one of the optical elements hosted in the NISP Grism and Filter wheel arrays \citep[GWA and FWA, respectively, see ][]{EuclidSkyNISP}. Depending on the element passband, these exposures collect varying levels of zodiacal light, averaging $f\approx 1\,{\rm e}^{-} {\rm s}^{-1}$. 
These data are acquired in the framework of \textit{Euclid}'s ROS. In the case of NISP, this is a sequence of spectroscopic and photometric observations used to extract both the redshift and morphology of the galaxy sample. Within a ROS observation, spectroscopic images are acquired first, using one of the red (\RGE, in one of the four dispersion directions) or blue (\BGE) grisms. This is immediately followed by photometric imaging, where the \JE, \HE, and \YE filters are used sequentially to capture broadband images \citep{EuclidSkyOverview}.

To study the performance of the NISP onboard data processing, we consider the dataset referred to as NISP Raw Frame Products\footnote{\url{https://euclid.esac.esa.int/dr/q1/dpdd/le1dpd/dpcards/le1_nisprawframe.html}} or, alternatively, NISP Level 1 (LE1) data. 
Essentially, these products are raw images (signal and QF maps) produced by the NISP onboard data processing, and transmitted to ground as telemetry packets \citep{Q1-TP001}.
The LE1 Processing Function (PF) of the Science Ground Segment (SGS) converts these telemetry packets into NISP raw data, which are subsequently ingested as inputs for the NIR and SIR PF \citep{Q1-TP003, Q1-TP006}.
Note that LE1 data are neither reduced nor calibrated, meaning that no instrumental effects have been corrected and no scientific calibration has been applied.
Although reduction and calibration are essential steps in the scientific exploitation of NISP raw data, considering raw data allows us to evaluate the performance of the signal estimator as implemented in the NISP DPUs, rather than evaluating the performance of the calibration procedures applied by the \Euclid SGS.

\begin{figure}[!t]
\begin{center}
\includegraphics[scale=0.35]{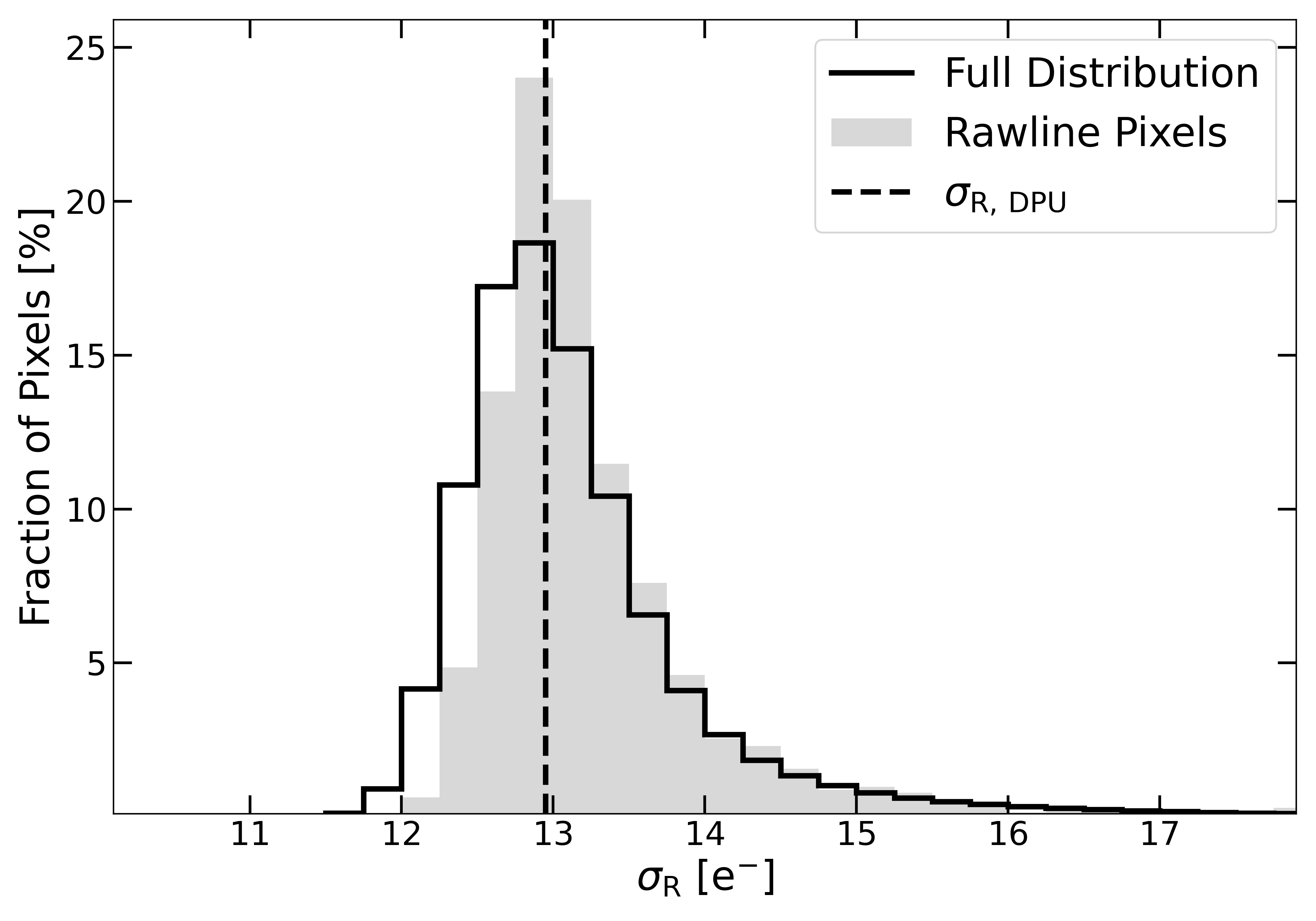}
\end{center}
\caption{Example (DET43, SCA ID 18221) of the readout noise $\sigma_{\sfont{R}}$ distribution measured during the ground characterisation campaign (TV1) of the NISP detectors \citep{Kubik2025}. The black stepped histogram depicts the $\sigma_{\sfont{R}}$ distribution for the entire pixel array ($2040\times2040$ pixels), with approximately 35$\%$ of the pixels exhibiting a readout noise around 13 e$^{-}$. The grey-filled histogram illustrates the $\sigma_{\sfont{R}}$ distribution for the rawline subsample (2040 pixels). The vertical dashed black line represents the detector-averaged value used by the NISP onboard data processing, $\sigma_{\sfont{R,\,DPU}}$. 
}
\label{fig:RON_rawlineVSdetector}
\end{figure}
The NISP Raw Frame Products include an ancillary dataset, referred to as `rawlines', that -- for each NISP exposure -- are downlinked together with the signal and QF maps.
In practice, for a subset of five rows ($5\times2048$ pixels) on each detector, the rawlines contain the full sequence of single-frame images constituting the MACC readout.

Rawlines can be located anywhere on the detector. During NISP commissioning \citep{EuclidSkyNISP, Gillard24, Cogato2024}, for each NISP detector, four rawlines were equally distributed between the top and bottom reference pixels framework \citep[full details about reference pixels in HxRG detectors are provided in][]{Loose07}, while only one rawline was placed at the detector centre within the so-called science pixels array.
Reference-pixels rawlines, 2048 pixels in length, are fundamental to verify the proper electrical behaviour of the pixel array \citep{kubik2014, Medinaceli2020, Kubik2025} and are currently used in the \Euclid's NIR data reduction pipeline to correct for detectors' channel-drifting effects \citep{Q1-TP003}.
The central rawline, denoted as science-pixels rawline, comprises 2040 light-sensitive pixels plus eight reference pixels (four at each row edge). 
Despite being only 0.05$\%$ of the entire pixel array, the science-pixels rawline is fully representative of the detector response in both signal integration and readout noise values (see Fig.\,\ref{fig:RON_rawlineVSdetector}). 

Therefore, the science-pixels rawlines distributed across the NISP FPA offer the ideal case study for our analysis: for each exposure, we can compare the NISP signal map against an independent rawline-based analysis to evaluate the flight performance of the NISP signal estimator.

\subsection{Analysis framework\label{sec:analysis_fr}}
According to Eq.\,(\ref{eq:DPA_bias_sig}), the bias introduced by the DPA, $b_{\rm{f}}^{\sigma_{\sfont{R}}}$, is defined as the deviation of the signal retrieved from the NISP signal estimator, $\hat{g}(\sigma_{\sfont{R,\,DPU}})$, with respect to $g$.
Since $g$ is not known a priori, we adapt the DPA bias estimation as follows
\begin{equation}
\label{eq:DPA_estimate}
    b_{\rm{f}}^{\sigma_{\sfont{R}}} = \frac{\hat{g}(\sigma_{\sfont{R,\,DPU}})- g_{\rm{LSF}}}{(n_{\rm{f}} + n_{\rm{d}})\,t_{\rm{fr}}}\,,
\end{equation}
where $g_{\rm{LSF}}$ is the equally-weighted linear least-squares fit (LSF) of the MACC signal integration encoded in the NISP raw lines. 
Assuming an ideal linear pixel, the LSF method provides an unbiased estimate of the signal -- independent of the adopted readout noise value, thereby serving as a reference signal estimator.

By correlating the $b_{\rm{f}}^{\sigma_{\sfont{R}}}$ with the difference between the actual per-pixel readout noise value, $\sigma_{\sfont{R}}$, and the detector-averaged value, $\sigma_{\sfont{R,\,DPU}}$, we infer the readout noise-induced effect of the DPA on the NISP signal estimator.
We consider the single-frame readout noise map, $\sigma_{\sfont{R}}$, estimated during the ground characterisation of NISP detectors \citep{Kubik2025}, and the detector-averaged values, $\sigma_{\sfont{R,\,DPU}}$, currently adopted for the NISP onboard data processing. 
The analysis is conducted according to the following procedure.
\begin{enumerate}
    \item Reconstruct the MACC($n_{\rm g}$, $n_{\rm f}$, $n_{\rm d}$) sample from the single-frame reads of the science-pixels rawlines, and convert the signal from ADU to electrons.
    
    \item Correct the electrical and thermal drifts during signal integration by applying the reference pixels correction. The method is analogous to the one used by the NISP DPUs \citep{kubik2014, Bonoli2016, Medinaceli2020}. However, since we do not have access to the entire reference pixel array, we adapt the correction method to the rawline dataset, which includes reference pixels from two rows on the top ($2\times2048$ pixels), two rows on the bottom ($2\times2048$ pixels), and four pixels on the right and left side of the science rawline ($2\times4$ pixels). In practice, we correct the 2040 pixels belonging to the science rawline by leveraging the 8200 reference pixels available within the rawlines dataset and following the same scheme defined in \citet{Kubik2025}.
    
    \item Exclude noisy and spurious pixels by applying a 5$\sigma$ clipping on $\Delta G_i$ values, in addition to the QF-thresholding technique for outlier detection (see Sect. \ref{sec:flight_performance_qf} for detailed description). 
    
    \item Fit a first-order polynomial LSF to the $G_i$ points composing the MACC signal integration and derive the LSF measurement, $g_{\rm{LSF}}$.
    
    \item For each exposure, evaluate the DPA bias $b_{\rm f}^{\sigma_{\sfont R}}$ map according to Eq.\,(\ref{eq:DPA_estimate}).
    
    \item For each pixel of the science-pixels rawlines, compute the readout noise shift $\Delta\sigma_{\sfont{R}}$ according to Eq.\,(\ref{eq:DPA_bias_param}).
    
    \item Bin the $\Delta\sigma_{\rm{R}}$ distribution with a bin-width of 1\,e$^{-}$.
    
    \item Compute the median of the $b_{\rm f}^{\sigma_{\rm R}}$ map across the entire set of exposures, $\tilde{b}_{\rm f}^{\sigma_{\sfont R}}$, to further decontaminate from random fluctuations or unidentified outliers. 
    
    \item Within each $\Delta\sigma_{\sfont{R}}$ bin, compute the mean $\langle \tilde{b}_{\rm f}^{\sigma_{\sfont R}} \rangle$ and the standard deviation of $\tilde{b}_{\rm f}^{\sigma_{\sfont R}}$ over the sixteen NISP detectors.
    
    \item Correlate $\langle \tilde{b}_{\rm f}^{\sigma_{\sfont R}} \rangle$ with the $\Delta\sigma_{\sfont{R}}$ binned distribution. Within each bin, the standard deviation represents the dispersion of the estimated bias across the NISP FPA.
\end{enumerate}

\subsection{DPA bias from early NISP flight data}
\begin{figure*}[!t]
    \centering
    \includegraphics[width=0.96\textwidth]{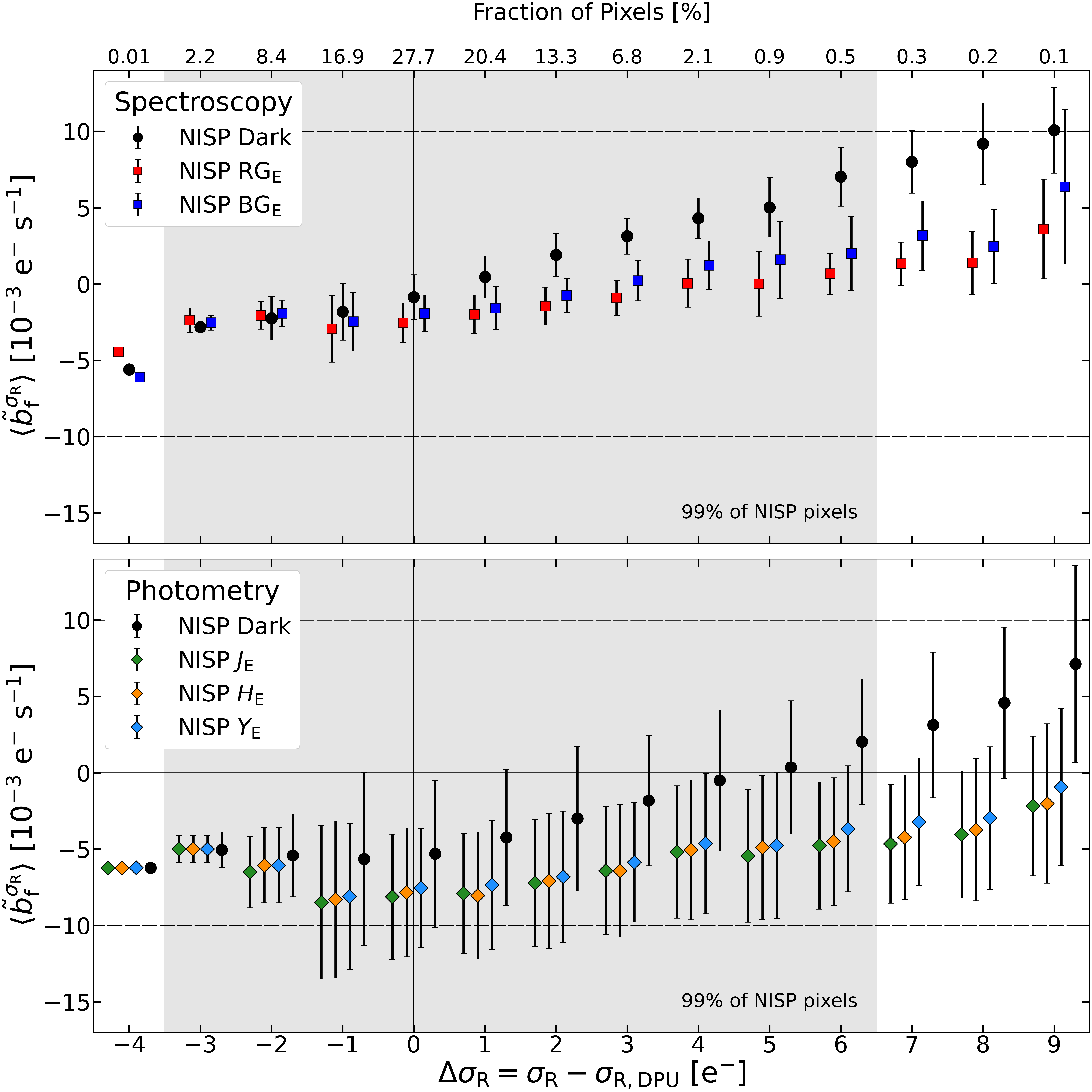} 

    \caption{The bias $\tilde{b}_{\rm f}^{\sigma_{\sfont R}}$ induced by the Data Processing Approximation (DPA) is shown as a function of the pixel readout noise shift, with a bin-width of 1\,e$^{-}$. 
    The data points and their associated error bars represent the mean and the standard deviation, respectively, computed over the 16 NISP detectors. 
    To enhance visualisation, within each $\Delta\sigma_{\sfont R}$  bin, the data points are artificially offset along the horizontal axis. 
    The top horizontal axis displays the fraction of NISP pixels contained in each $\Delta\sigma_{\sfont R}$ bin. Grey shaded area highlights the $99\%$ of NISP pixels which fall in the $\Delta\sigma_{\sfont R}$ range between $-3\,{\rm e}^{-}$ and $6\,{\rm e}^{-}$.
    Note that at the low end of $\Delta\sigma_{\sfont R}$, the standard deviation decreases because pixels with $\Delta\sigma_{\sfont R} \lesssim -2 \,{\rm e}^{-}$ are not homogeneously distributed across the detectors in the NISP focal plane.
    Black circles show the $\tilde{b}_{\rm f}^{\sigma_{\sfont R}}$ behaviour at the dark current level ($f\approx0.01\,{\rm e}^{-} {\rm s}^{-1}$), while color-coded symbols refer to the zodiacal background regime  ($f\approx1\,{\rm e}^{-} {\rm s}^{-1}$). 
    \emph{Top panel:} Results from the NISP spectroscopic readout mode, i.e., MACC(15,16,11). Red and blue squares represent the results obtained from the red (\RGE) and blue (\BGE) grism observations, respectively.
    \emph{Bottom panel:} Results from the NISP photometric readout mode, i.e., MACC(4,16,4). 
    Green, orange, and azure diamonds represent the results obtained from the \JE, \HE, and \YE observations, respectively.}
\label{fig:bias}
\end{figure*}

Figure \ref{fig:bias} illustrates the outcomes of the analysis performed on the early NISP flight data, considering the NISP rawlines dataset. 
The data points show the mean and standard deviation of the DPA bias measurements across the sixteen NISP detectors.

\subsubsection{NISP dark current}
In the case of the NISP instrument, the dark current is measured using both the photometric MACC(4,16,4) mode and the spectroscopic MACC(15,16,11) mode. These measurements are then utilised for dark subtraction in the photometric and spectroscopic data, respectively.
Additionally, dark images represent a fundamental tool for the in-flight monitoring of the hot pixels population and the characterisation of persistence effects \citep{EuclidSkyNISP, Q1-TP003}. Therefore, a precise and accurate estimate of the dark current is critical for all the NISP observations.

Results obtained from the analysis of photometric and spectroscopic dark exposures are shown in Fig. \ref{fig:bias} as black circles. 
In both cases, $\langle \tilde{b}_{\rm f}^{\sigma_{\sfont R}} \rangle$ increases as a function of $\Delta\sigma_{\sfont{R}}$.

In the $\Delta\sigma_{\sfont{R}}$ range containing the $99\%$ of NISP pixels, i.e., $-3\leq\Delta\sigma_{\sfont{R}}\,$/\,e$^{-}\leq6\,$, the DPA bias induced by the readout noise parametrisation is lower than $0.005 \,{\rm e}^{-} {\rm s}^{-1}$.
For a typical dark current level around $0.01\,{\rm e}^{-} {\rm s}^{-1}$, this represents a systematic error of $\approx50\%$ on the dark current estimation.
However, three key points must be taken into account when interpreting these results.

First, since scientific exposures typically have signal levels around $1\,{\rm e}^{-} {\rm s}^{-1}$, the dark current subtraction introduces a systematic error of about $0.5\%$ (worst case) in the calibration process.
Second, the error associated with dark measurements is one order of magnitude higher than the DPA bias. More specifically, from Table A.1 in \citet{Kubik2025}, the typical NISP photometric and spectroscopic noises are of the order of $6\,{\rm e}^{-}$ and $8\,{\rm e}^{-}$ which correspond to about $0.07\,{\rm e}^{-} {\rm s}^{-1}$ and $0.01\,{\rm e}^{-} {\rm s}^{-1}$, respectively.
Third, the subset of pixels exhibiting a systematic error higher than $0.005 \,{\rm e}^{-} {\rm s}^{-1}$ represents less than $1.5\%$ of the entire NISP pixels array, and can be easily identified from the ground characterisation of the NISP detectors \citep{Kubik2025}.
Moreover, because data and simulations agree closely (see Fig. \ref{fig:DPU_bias}), we can monitor and eventually correct these systematic effects.

By construction, pixels in the $\Delta\sigma_{\sfont{R}}=0\,{\rm e}^{-}$ bin are immune to DPA bias. Therefore, restricting our analysis to this subset allows us to isolate and measure the intrinsic bias of the NISP signal estimator, which we denote as $\langle b_{\rm f} \rangle$.
\begin{figure*}[!t]
    \centering
    \includegraphics[width=\textwidth]{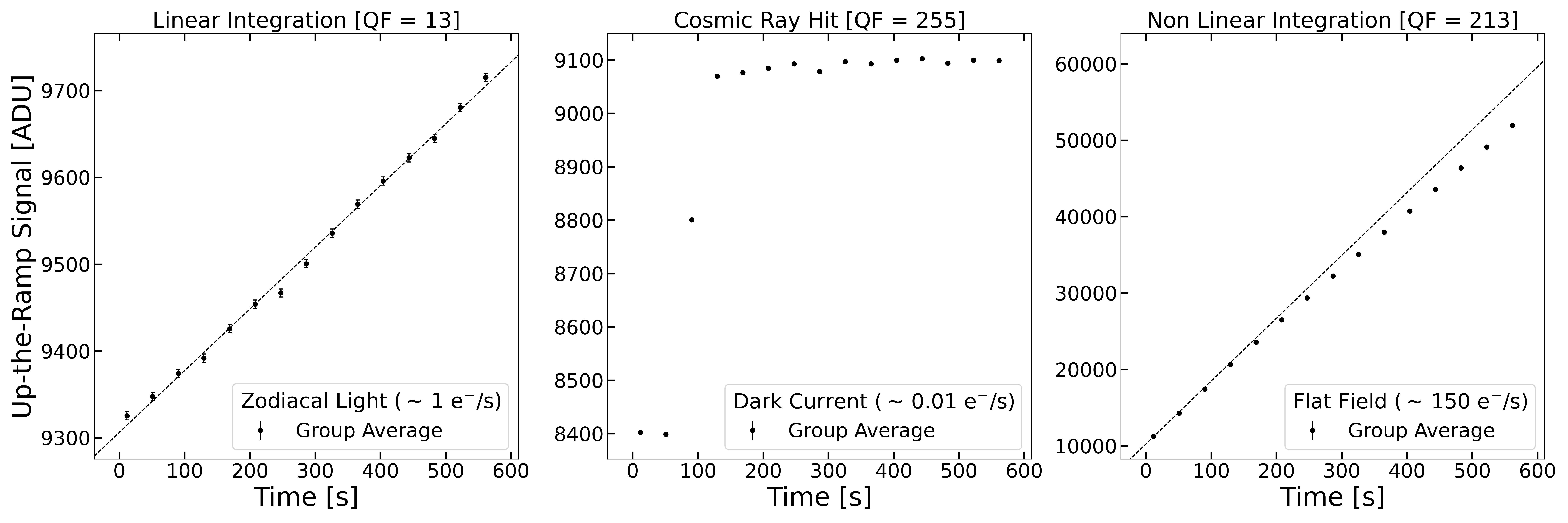}
    \caption{Examples of spectroscopic MACC(15,16,11) signal integration under different flux regimes. Each panel plots the accumulated charge (in Arbitrary Digital Units, ADU) as a function of the integration time (in seconds). 
    The data points represent the average signal for each group ($n_{\rm f}=16$), with error bars (whose size is comparable to the marker size) indicating the $\sigma_{\rm{eff}}$ as defined in \citet{Kubik2016}. The dashed lines (left and right panels) correspond to the linear integration for the corresponding flux regime.
    The figure displays the capability and effectiveness of the NISP signal estimator in accurately identifying several kinds of anomalies during the UTR signal integration, like cosmic ray hit (QF = 255, centre panel) and pixel non-linearity (QF = 213, right panel).
    \emph{Left panel:} Zodiacal light ($\approx\,1\,{\rm e}^{-} {\rm s}^{-1}$) linear integration with a QF value exactly matching the expected $\chi^2(n_{\rm g}-2)$ value, i.e, QF\,=\,13. \emph{Centre panel:} Dark current ($\approx\,0.01\,{\rm e}^{-} {\rm s}^{-1}$) integration affected by a cosmic ray event. \emph{Right panel:} Non-linear integration observed for flat field images at high flux ($\approx\,150\,{\rm e}^{-} {\rm s}^{-1}$).}
\label{fig:ex_px}
\end{figure*}

From the analysis of the early NISP flight data, the estimate of this source of bias can be summarised as
\begin{equation}
\begin{aligned}
\langle b_{\rm{f,\,sp}} \rangle &= -0.001 \pm 0.001 \,{\rm e}^{-}\,{\rm s}^{-1}\,, \quad \text{Spectroscopic Dark}\,;\\
\langle b_{\rm{f,\,ph}} \rangle &= -0.005 \pm 0.004 \,{\rm e}^{-}\,{\rm s}^{-1}\,, \quad \text{Photometric Dark}\,.
\end{aligned}
\end{equation}
These outcomes closely align with the simulation results (see Fig. \ref{fig:S_bias_fcn_ng}), and demonstrate that the photometric readout mode ($n_{\rm g}=4$) has larger bias levels compared to the spectroscopic one ($n_{\rm g}=15$).
Finally, our analysis further corroborates the findings of \citet{kubik2016b}, which showed that at very low flux levels ($f<0.1\,{\rm e}^{-}\,{\rm s}^{-1}$) the NISP signal estimator deviates by more than $10\%$ from the LSF estimate.

\subsubsection{NISP zodiacal background}
For each of those photometric and spectroscopic images acquired during the ROS observations, the NIR PF \citep{Q1-TP003} subtracts the variations in the zodiacal background. Hence, it is important to investigate the effects of the DPA bias on the NISP scientific exposures to verify the accuracy of the background estimation.

Results obtained from the analysis of photometric and spectroscopic scientific exposures are shown in Fig. \ref{fig:bias} as green (\JE), orange (\HE), azure (\YE) diamonds, and red (\RGE) and blue (\BGE) squares, respectively.
The majority of NISP pixels exhibit a negligible bias value, corresponding to less than $1\%$ error 
for a zodiacal background level around $1\,{\rm e}^{-} {\rm s}^{-1}$.

As for the dark current case, the DPA behaves as expected with 
$\langle \tilde{b}_{\rm f}^{\sigma_{\sfont R}} \rangle$ increases as a function of $\Delta\sigma_{\sfont{R}}$.
In this case, the increasing trend is less pronounced since at $\approx 1 \,{\rm e}^{-} {\rm s}^{-1}$ the noise affecting the signal integration is mostly dominated by the shot noise, and therefore the DPA bias induced by the readout noise shift is attenuated.

By restricting our analysis to pixels in the $\Delta\sigma_{\sfont{R}}=0\,{\rm e}^{-}$ bin, we quantify the intrinsic bias of the NISP signal estimator $\langle b_{\rm f} \rangle$ and confirm the simulation predictions, reported in Fig. \ref{fig:DPU_bias}. Essentially, the bias increases (in absolute values) relative to the dark current case. Also, at $f\approx1\,{\rm e}^{-}\,{\rm s}^{-1}$ the photometric readout mode exhibits a larger absolute bias than the spectroscopic mode.
The bias $\langle b_{\rm f} \rangle$ measured on early NISP data can be summarised as
\begin{equation}
\begin{aligned}
\langle b_{\rm{f,\,sp}} \rangle &= -0.002 \pm 0.001 \,{\rm e}^{-}\,{\rm s}^{-1}\,, \quad \langle \RGE,\,\BGE \rangle\,;\\
\langle b_{\rm{f,\,ph}} \rangle &= -0.008 \pm 0.004 \,{\rm e}^{-}\,{\rm s}^{-1}\,, \quad \langle \JE,\,\HE,\,\YE \rangle\,.
\end{aligned}
\end{equation}
Again, we confirm the findings obtained by \citet{kubik2016b}, i.e., at $\approx1\,{\rm e}^{-}\,{\rm s}^{-1}$ the NISP signal estimator does not present any significant bias.

\section{Flight performance of the NISP quality factor}
\label{sec:flight_performance_qf}
Under ideal conditions -- where the signal integration is not affected by anomalies, such as cosmic ray hits, random electronic noise, or non-linearity effects -- the NISP QF is expected to follow a $\chi^2_{\nu}$ distribution with $\nu = n_g - 2$ degrees of freedom, as demonstrated in Sect. \ref{sec:QF}.
The \Euclid's SGS \citep{Q1-TP001,Q1-TP003,Q1-TP006} employs the NISP QF map as a diagnostic tool to assess the NISP data quality.
In practice, deviations from linear response are identified by applying a threshold to the QF distribution, and pixels whose QF values deviate significantly from the $\chi^2_{\nu}$ distribution are flagged and excluded from the scientific analysis.

In the NISP spectroscopic readout mode, the QF map is downlinked with an 8-bit resolution. During the reduction and calibration of spectroscopic data performed by the SGS \citep{Q1-TP003, Q1-TP006}, the threshold value is set to ${\rm QF}_{\rm{sp}}=50$, corresponding to a $4.7\sigma$ clipping on a $\chi^2_{13}$ distribution.
In the NISP photometric readout mode, the QF map is a binary flag (1-bit) constructed onboard \citep{Bonoli2016, Medinaceli2020, EuclidSkyNISP} by applying a custom threshold value. Currently, the NISP onboard data processing adopts ${\rm QF}_{\rm{ph}}=10$ as a threshold, which corresponds to a conservative $2.7\sigma$ clipping on a $\chi^2_2$ distribution.

Figure \ref{fig:ex_px} shows three examples from NISP flight data illustrating the behaviour of the NISP QF in the spectroscopic readout mode. Linear ramps typically yield a QF value around the $\chi^2_{13}$ distribution mean value, $\langle\chi^2_{13}\rangle\,=\,13\,$, while pixels hit by energetic cosmic rays or exhibiting non-linear integration effects have higher QF.

\begin{figure*}[!t]
    \begin{minipage}{\textwidth}
            \centering
            \includegraphics[width=\textwidth]{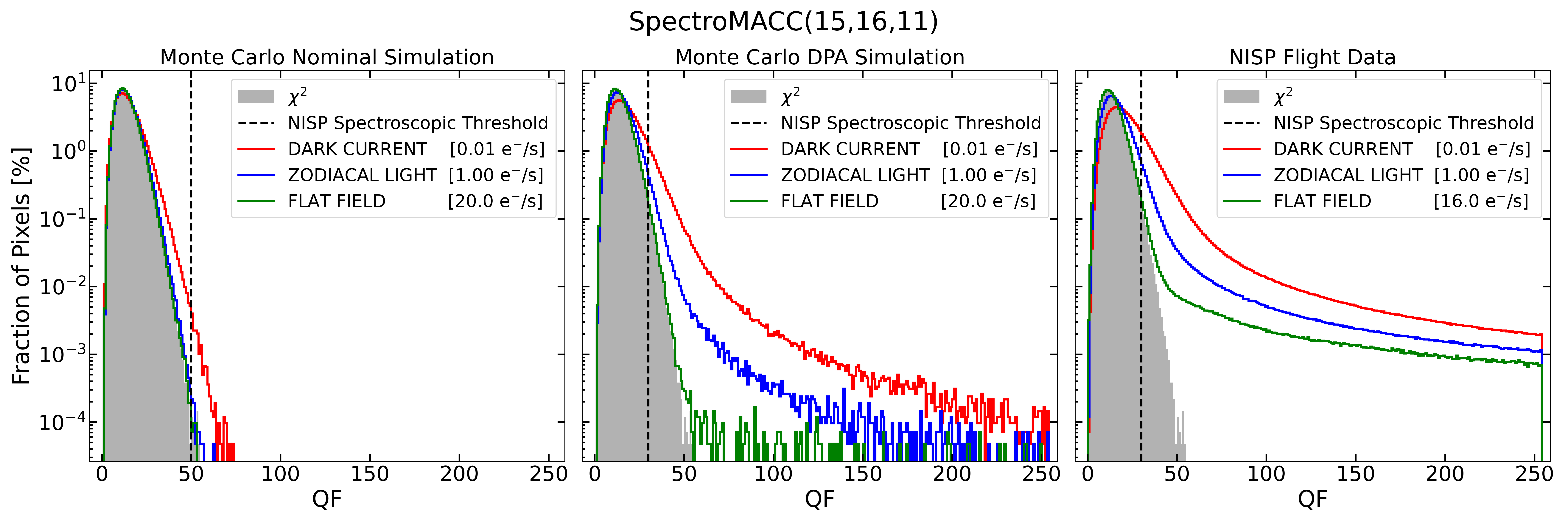} 
    \end{minipage}
    \begin{minipage}{\textwidth}
            \centering
            \includegraphics[width=\textwidth]{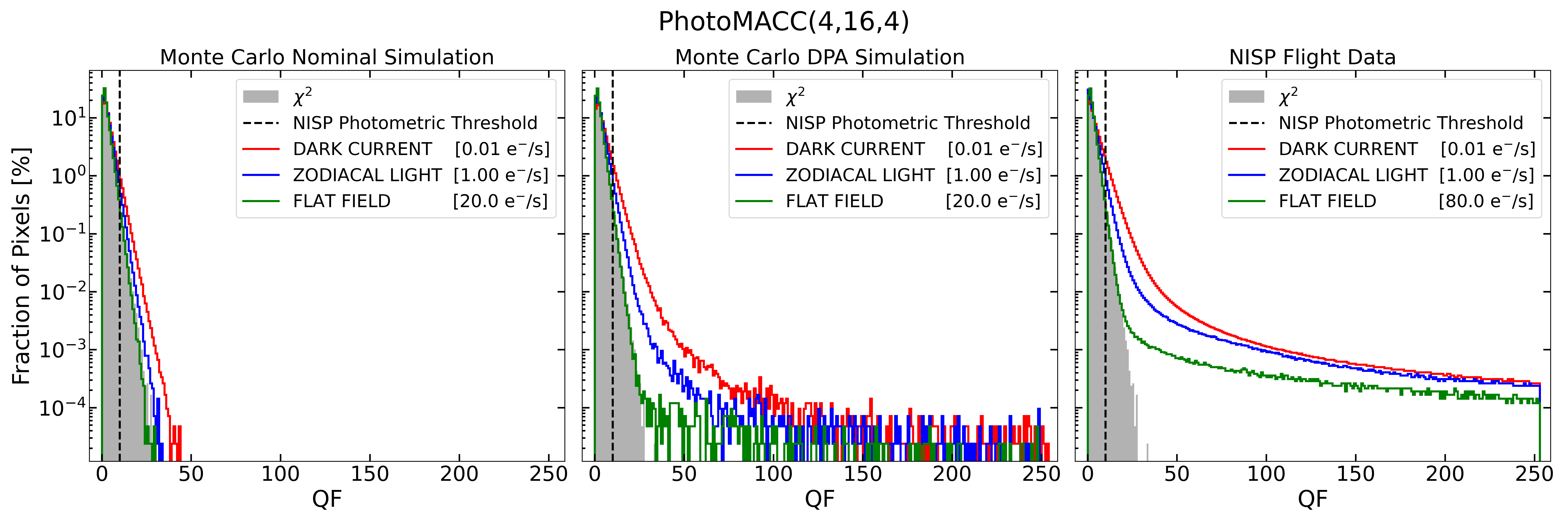} 
    \end{minipage}
\caption{Example (DET\,11, SCA 18453) of the performance of the NISP QF from simulated and early flight data. Each distribution reports the QF distribution related to a single photometric or spectroscopic exposure. The grey shaded histograms represent the $\chi^2(n_g-2)$ distribution, corresponding to the expected behaviour of the NISP QF in the absence of anomalies during the signal integration. The red, blue, and green histograms show the QF distributions for three distinct flux regimes: $0.01\,{\rm e}^{-} {\rm s}^{-1}$ (dark current), $1\,{\rm e}^{-} {\rm s}^{-1}$ (zodiacal light), and $>10\,{\rm e}^{-} {\rm s}^{-1}$ (flat field), respectively. Flat field simulations were performed at $20\,{\rm e}^{-} {\rm s}^{-1}$ for both NISP channels, while spectroscopic and photometric flight data were acquired at $16\,{\rm e}^{-} {\rm s}^{-1}$ and $80\,{\rm e}^{-} {\rm s}^{-1}$, respectively. Black dashed vertical lines show the threshold values applied to the NISP QF distributions in the spectroscopic (QF = 50) and photometric (QF = 10) channels. 
\emph{Top panels:} Spectroscopic MACC(15,16,11) mode.
\emph{Bottom panels:} Photometric MACC(4,16,4) mode.
\emph{Left panels:} Synthetic QF distributions obtained by processing the simulated data with the per-pixel readout noise map -- labelled as Nominal. At high flux ($20\,{\rm e}^{-} {\rm s}^{-1}$, green histogram) the QF distribution perfectly matches the expected $\chi^2_{\nu}$ distribution. As the flux decreases, the QF distributions progressively deviate from the $\chi^2_{\nu}$, as expected from the analysis reported in Sect.\,\ref{sec:QF}. 
\emph{Centre panels:} Synthetic QF distributions obtained by processing the simulated data with the detector-average readout noise value -- labelled as DPA.
At high flux ($20\,{\rm e}^{-} {\rm s}^{-1}$, green histogram) the QF distribution closely follows the expected $\chi^2_{\nu}$ distribution, aside from a few outliers. As the flux decreases, the QF distributions progressively deviate from the $\chi^2_{\nu}$ due to the superimposition of the intrinsic bias of the NISP signal estimator and the DPA effect. 
\emph{Right panels:}
Early flight data matching MC simulations (DPA case), if space weather effects are accounted for. The NISP QF follows the $\chi^2_{\nu}$ distribution at high flux (green histograms) while increasingly deviates from it as the flux decreases (red and blue diagrams). The outlier tail in the NISP flight data mainly results from the impact of solar energetic particles on the NISP detectors.
}
\label{fig:QF_MCvsDATA}
\end{figure*}
\subsection{The impact of DPA on the NISP QF distribution}
The response of QF to the DPA is quantified using the idealised MC model described in Sect. \ref{sec:properties}.
Considering the readout noise maps derived during the ground characterisation of the NISP detectors \citep{Kubik2025}, we simulate a single NISP exposure in both photometric and spectroscopic readout mode, under three different flux conditions, i.e., dark current ($\,0.01\,{\rm e}^{-}\,{\rm s}^{-1}$), zodiacal light ($\,1\,{\rm e}^{-}\, {\rm s}^{-1}$), and flat-field exposure ($>\,10\,{\rm e}^{-}\,{\rm s}^{-1}$). 

For each of these flux regimes, this process produces one photometric and one spectroscopic readout for each of the 66\,585\,600 science pixels composing the NISP FPA, yielding a total of $\approx9.4\times10^{10}$ synthetic UTR reads.
These synthetic exposures were processed according to the NISP signal estimator formalism described in Sect. \ref{sec:notations}, and the resulting QF distributions are shown in Fig. \ref{fig:QF_MCvsDATA}. 
Specifically, the different flux regimes are shown as red (dark current), blue (zodiacal light), and green (flat field) histograms, while the grey shaded histograms report the expected $\chi^2_{\nu}$ distributions.

To study the effects induced by DPA, we applied two distinct workflows. 
The first one, denoted as Nominal simulation (top and bottom left panels of Fig. \ref{fig:QF_MCvsDATA}), processes the UTR frames using the same readout noise map considered in the MC simulations.
The second, denoted as DPA simulation (centre panels of Fig. \ref{fig:QF_MCvsDATA}), instead processes the UTR frames using the detector-average value currently adopted in the NISP onboard data processing.
\begin{table*}[!t]
\caption{Fraction of NISP QF outliers (\%), defined as the number of pixels exceeding the NISP spectroscopic (${\rm QF}_{\rm{sp}}=50$) and photometric (${\rm QF}_{\rm{ph}}=10$) thresholds, in three different flux regimes, namely, dark current ($0.01\,{\rm e}^{-} {\rm s}^{-1}$), zodiacal light ($1\,{\rm e}^{-} {\rm s}^{-1}$), and flat field ($20\,{\rm e}^{-} {\rm s}^{-1}$). In the Nominal case, UTR frames are processed by using the actual pixels' readout noise value, while the detector-average value is used in the DPA case. The numbers quoted here refer to the average and standard deviation of the outlier fraction computed across the NISP FPA. The left and centre panels of Fig. \ref{fig:QF_MCvsDATA} show an example of the synthetic QF distributions considered in this analysis.} 
\label{tab:QF_DPA}
\centering
\begin{center}      
\begin{tabular}{c|c|c|c} 
\hline
\textbf{Observation Mode} & \textbf{Flux Regime} & \multicolumn{2}{c}{\textbf{Outlier Fraction [\%]}  } \\
\textbf{[QF Threshold]} & \textbf{$[{\rm e}^{-} {\rm s}^{-1}]$} & \textbf{Nominal Simulation} & \textbf{DPA Simulation} \\
\hline
\hline
\multirow{3}{7em}{\shortstack{Spectroscopic\\MACC(15,16,11)\\${\rm QF}_{\rm{sp}}=50$}}
& 0.01 & 0.012 $\pm$ 0.002\,\% & 0.61 $\pm$ 0.54\,\% \\
& 1.00 & < 0.001\,\% & 0.11 $\pm$ 0.05\,\% \\
& 20.0 & < 0.001\,\% & 0.02 $\pm$ 0.02\,\% \\
\hline
\multirow{3}{6em}{\shortstack{Photometric\\MACC(4,16,4)\\${\rm QF}_{\rm{ph}}=10$}}
& 0.01 & 2.02 $\pm$ 0.02\,\% & 3.45 $\pm$ 2.05\,\% \\
& 1.00 & 0.86 $\pm$ 0.06\,\% & 1.31 $\pm$ 0.58\,\% \\
& 20.0 & 0.31 $\pm$ 0.01\,\% & 0.35 $\pm$ 0.03\,\% \\
\hline
\end{tabular}
\end{center}
\end{table*}

The impact of DPA on the NISP QF distribution is particularly evident when comparing the DPA simulations to the Nominal simulations.
In the nominal case, no readout noise approximations are introduced during the data processing, and the synthetic QF distributions exhibit the expected $\chi^2_{\nu}$ behaviour, with a flux-dependent slight deviation already described in Sect. \ref{sec:properties} and shown in Fig. \ref{fig:S_bias_fcn_ng} (bottom-left panel).

Conversely, in the DPA case, the synthetic QF distributions exhibit a pronounced outlier population, which originates from using the detector-averaged readout noise $\sigma_{\sfont{R,\,DPU}}$ instead of each pixel's actual value $\sigma_{\sfont R}$.
Specifically, for pixels whose actual readout noise significantly exceeds the detector average, the effective variance of the UTR frames -- $D_{ii}$ in Eq. (\ref{eq:Dkl}) -- is systematically underestimated. During processing, the algorithm therefore misidentifies these pixels as non-linear and assigns them an artificially high QF compared to the expected $\chi^2_{\nu}$ value.

This effect is most pronounced in the low‐flux regime, where readout noise dominates the variance budget. As the incident flux increases, shot noise becomes the primary contributor to the variance, and the bias introduced by $\sigma_{\sfont R}$-related DPA diminishes accordingly. 
For example, at a high flux ($20\,{\rm e}^{-} {\rm s}^{-1}$, green histograms), the QF distributions closely follow the theoretical $\chi^2_{\nu}$ distribution curves with a negligible fraction of outliers ($<0.1\%$). 
Under zodiacal light conditions ($1\,{\rm e}^{-}\,{\rm s}^{-1}$, blue histograms) and in the dark current regime ($0.01\,{\rm e}^{-}\,{\rm s}^{-1}$, red histograms), the proportion of outliers grows as the flux decreases, reaching its maximum under dark current conditions.

Table \ref{tab:QF_DPA} compares the excess of outliers in the DPA and Nominal simulations. By applying the NISP photometric ($2.7\sigma$ clipping, ${\rm QF}_{\rm{ph}}=10$) and spectroscopic ($4.7\sigma$ clipping, ${\rm QF}_{\rm{sp}}=50$) thresholds, the impact of DPA is quantified in terms of fraction of outliers compared to the nominal behaviour of the NISP QF.
In the spectroscopic readout mode, across the entire flux range, the outliers account for less than $1\%$ of the pixel array. 
In contrast, in the photometric readout mode, the DPA bias is more significant and, at the lowest flux levels, the outlier fraction exceeds $3\%$ of the NISP pixels.

\subsection{Space weather influence on NISP QF performance}
Additionally, Fig. \ref{fig:QF_MCvsDATA} highlights a significant difference between the NISP QF distributions obtained from the DPA simulations and those obtained from early NISP flight data. 
In particular, we consider the typical QF distributions retrieved from early NISP flight data and associated with dark exposures ($\approx 0.01 \,{\rm e}^{-}\,{\rm s}^{-1}$), scientific exposures (zodiacal light, $\approx1 \,{\rm e}^{-}\,{\rm s}^{-1}$), and flat-field exposures obtained using the NISP calibration unit \citep[$> 10 \,{\rm e}^{-}\,{\rm s}^{-1}$, ][]{EuclidSkyNISPCU}

The observed excess in the tail of the QF distributions is due to any external event causing a non-linear integration ramp, e.g. cosmic rays hitting the NISP detectors.
Indeed, by orbiting the second Lagrangian point of the Sun-Earth system, the \Euclid satellite is exposed to a radiation environment characterised by temporal variations correlated with solar cycles.
Consequently, the fraction of outliers evolves with time according to the solar energetic particles flux and, in extreme cases, such as solar coronal mass ejection events, the percentage of pixels flagged by the NISP QF can exceed 50\%\,.

Considering that \Euclid operates during the peak of the 25$^{\text{th}}$ solar cycle, the influence of space weather on the mission's scientific output is not negligible (Euclid Collaboration: Schirmer et al., in prep.). In this context, the NISP QF proved to be able to identify cosmic rays hitting the NISP detectors \citep{Kohley2016, Kohley2018, Cogato2024}, thus ensuring the high quality of \Euclid's NISP data.
In particular, the processing functions responsible for the reduction and calibration of \Euclid NIR data \citep{Q1-TP003, Q1-TP006} integrate QF information with additional techniques -- such as the \texttt{L.A. Cosmics} algorithm \citep{vanDokkum01} -- to mitigate contamination from cosmic-ray hits.

\section{Conclusions\label{sec:conclusions}}

This paper revisits the statistical foundations of the signal estimation process in the \Euclid mission’s NISP instrument, with a focus on understanding and quantifying several bias sources in the signal and QF estimators derived from the likelihood-based method.
The likelihood formulation in Eq. (\ref{eq:likelihood_general}) is examined, highlighting the assumptions made and identifying key sources of bias. We confirmed that the Gaussian approximation to the likelihood holds in the \Euclid context. However, simplifying assumptions -- such as neglecting off-diagonal elements of the covariance matrix or incorporating the fitted parameter $g$ directly in the error term -- can introduce systematic biases.

One such bias arises from the non-linear form of the signal estimator, leading to the so-called folding bias at low flux levels. While this effect is analytically challenging due to its probabilistic nature, it has been evaluated through simulations and shown to be negligible in the \Euclid case -- even down to extreme scenarios involving readout noise as low as $1\,{\rm e}^{-}$ and flux as low as $0.1\,{\rm e}^{-}\,{\rm s}^{-1}$.
The dominant contribution to bias stems from the correlation between $\Delta G$ and its variance within the likelihood, producing a constant (flux-independent) bias. Though small compared to instrumental and photon noise, this effect can become noticeable in high signal-to-noise observations. An analytical correction term is derived for this bias. Additionally, smaller bias components -- beyond the scope of the analytical model -- originate from noise covariances between data points and are recognised as potential contributors to the overall residual bias.

Our work also investigates the statistical properties of both the NISP signal estimator and the QF, including their probability distributions, expected values and variances. The signal estimator approximately follows a non-central $\chi$ distribution, though its moments depend on the true value of $g$ and are thus impractical for direct use. 
In contrast, QF is shown analytically to follow a $\chi^2$ distribution with known expectation and variance. This makes it a reliable estimator for the goodness of the fit, especially because any residual bias can be modelled through simulations.
Moreover, error propagation is shown to yield a better estimate of signal variance than the second-derivative of the log-likelihood proposed in \citet{Kubik2016}, and an accurate, implementable formula is provided and already integrated into the \Euclid SGS pipeline \citep{Q1-TP003}.
Importantly, the study demonstrates that the signal estimator derived from the full log-likelihood function is systematically less biased than the one obtained via traditional $\chi^2$-minimisation, although both have comparable errors. This makes the likelihood approach the preferred method, especially in the presence of unknown or flux-dependent variances.

The effects of the various approximations made in the NISP onboard processing are also quantified. In particular, we demonstrate that any biases from averaging the pixels' readout noise and conversion gain remain negligible and well controlled, ensuring accurate signal estimates even if these parameters drift by a few percent over the mission lifetime.

Specifically, the results presented in Fig. \ref{fig:bias} indicate for the majority of pixels ($\approx 99\%$) the systematic bias is lower than $0.01 \,{\rm e}^{-}\,{\rm s}^{-1}$ over the entire range of flux levels of interest for \Euclid, namely $f\gtrsim0.01\,{\rm e}^{-}\,{\rm s}^{-1}$. Such a systematic error is well within the statistical error budget associated with the NISP exposures, and therefore does not yield a biased signal estimation. 

Nevertheless, it is important to address a few key considerations that highlight the main assumptions of the proposed method.
As highlighted in \citet{Kubik2025}, the choice of the reference signal is arbitrary and can vary depending on the scope of the analysis. 
In this study, the bias estimation was derived considering the reference signal as a first-order polynomial LSF of the signal MACC readout. This is independent of the actual pixel readout noise and provides an unbiased estimation of the actual incident flux.
Also, our analysis is based on the rawlines sample, which allows us to access the signal MACC frames and thus to independently estimate the incident flux via the LSF method. This sample is homogenously distributed over the NISP detectors and constitutes $\approx0.05\%$ of the entire pixel array, but -- despite its small size -- it is fully representative of the detector response. 
Therefore, we demonstrated that the 32\,640 science pixels of the rawlines dataset represent a robust tool for probing and testing the flight performance of the 66\,585\,600 pixels in the NISP focal plane. 

Our findings from early flight data assume the linearity of the signal integration for the entire sample of pixels. Outliers are statistically mitigated through a large data sample, namely thousands of exposures for scientific \RGE, \BGE, \JE, \HE, and \YE data, and hundreds of exposures for dark currents in both NISP readout modes. 
Spurious detections, such as cosmic ray hits or random electronic jumps, are identified with the $\sigma$-clipping method on the signal MACC readout along with the QF-thresholding technique.

The analysis presented here is conducted considering the NISP LE1 data, i.e., without any instrumental or scientific calibration. Hence, the interpretation of our results against MC simulations may be affected by additional sources of noise that are not included in our simplified signal integration model, such as inter-pixel capacitance, $1/f$ noise, or persistence effects.
Interestingly, the agreement between early flight data and simulations suggests the effectiveness of our theoretical modelling of the NISP signal estimator and provides a robust framework for deriving leading corrections of the systematic biases analysed in this work.

Finally, we investigate the performance of the QF, quantifying the effects of the approximations adopted in the NISP onboard processing and analysing their impact on the QF response in terms of outliers and spurious signals detection. Additionally, we present the typical QF distributions obtained from early NISP flight and emphasise its deviation from MC simulations -- an effect mainly driven by the particle radiation environment influencing any space-borne observatory operating at the second Lagrangian point of the Sun-Earth system.

\begin{acknowledgements}
  \AckEC
\end{acknowledgements}

\bibliographystyle{aa}
\bibliography{Euclid, Q1, paper}

\begin{appendix}
\onecolumn 
\section{Constant bias of the signal estimator\label{app:Gbias}}

We derive the formula of the constant bias of the \Euclid NISP signal estimator $\hat{g}[\Delta \vec{G}]_{n_{\rm g}}$ defined in Eq.\,(\ref{eq:gL}) at high flux regime. The estimator of the bias $\hat{b}_{g}$ of $\hat{g}[\Delta \vec{G}]_{n_{\rm g}}$ is defined as the average deviation of $\hat{g}[\Delta \vec{G}]_{n_{\rm g}}$ with respect to a true value $g$
\begin{equation}
\hat{b}_{\rm g} = \left\langle \hat{g}[\Delta \vec{G}]_{n_{\rm g}}  - g \right\rangle_{n_{\rm r}}\,,
\end{equation}
where $\langle . \rangle_{n_{\rm r}}$ is the average over $n_{\rm r}$ realisations. We can write
\begin{align}
  \left\langle \hat{g}[\Delta \vec{G}]_{n_{\rm g}} \right\rangle_{n_{\rm r}}  & = \left\langle \sqrt{\xi^2 + M_2( \Delta \vec{G}, \beta, n_{\rm g} ) }  - \xi - \beta\,\right\rangle_{n_{\rm r}} \nonumber \\
              & =\left\langle \sqrt{  \xi^2 +  M^2(\Delta\vec{G}, n_{\rm g}) + s^2[\Delta \vec{G} ]_{n_{\rm g}} + 2\beta M(\Delta\vec{G}, n_{\rm g}) + \beta^2 }\,   \right\rangle_{n_{\rm r}} - \xi - \beta   \\
                & = \left\langle \sqrt{ \xi^2 + M^2(\Delta\vec{G}, n_{\rm g}) + M(\Delta\vec{G}, n_{\rm g})\left(\frac{n_{\rm g}-2}{n_{\rm g}-1}\left(1+\alpha \right) + 2\beta\right) +  \frac{n_{\rm g}-2}{n_{\rm g}-1}\gamma + \beta^2 }\, \right\rangle_{n_{\rm r}}  - \xi - \beta\,,    \nonumber
\end{align}
where we have defined
\begin{equation}
    M( \Delta \vec{G}, n_{\rm g} ) \equiv \frac{\sum_{i=1}^{n_{\rm g}-1} \Delta G_i }{n_{\rm g}-1} \;.
\end{equation}
We have also used the relation
\begin{equation}
    M_2( \Delta \vec{G}, 0, n_{\rm g} ) = M^2( \Delta \vec{G}, n_{\rm g} ) + s^2[\Delta \vec{G} ]_{n_{\rm g}}\,,
\end{equation}
where $s^2[\Delta \vec{G} ]_{n_{\rm g}}$ is the unbiased sample variance of $n_{\rm g}-1$ group differences. It is related to the population variance $D_{ii}(g)$ by
\begin{equation}
    s^2[\Delta G ]_{n_{\rm g}} = \frac{n_{\rm g}-2}{n_{\rm g}-1}D_{ii}(g) = \frac{n_{\rm g}-2}{n_{\rm g}-1}\left[ (1+\alpha)g + \gamma \right]\,,
\end{equation}
where $D_{ii}$ is the diagonal term of the covariance matrix of group differences defined in Eq.\,(\ref{eq:Dkl}).

In order to compute the average of the square root, we use the Taylor expansion around the average value $\left\langle M( \Delta \vec{G}, n_{\rm g} ) \right\rangle_{n_{\rm r}}=g$. Only the leading term gives a non-zero contribution
\begin{align}\label{eq:limg}
 \left\langle \hat{g}[\Delta \vec{G}]_{n_{\rm g}} \right\rangle_{n_{\rm r}} \xrightarrow{g \to \infty} g + \frac{1}{2}\left( \frac{n_{\rm g}-2}{n_{\rm g}-1}(1+\alpha) + 2\beta\right) - \xi - \beta = g - \frac{\xi}{n_{\rm g}-1}\,;
\end{align}
higher derivative terms are null in high flux approximation. 
Therefore, the per-group signal bias $b_{\rm g}$ at high flux is equal to
\begin{equation}
\hat{b}_{\rm g} = - \frac{\xi}{n_{\rm g}-1}\,,
\end{equation}
or in terms of the per-second flux one obtains
\begin{equation}
\hat{b}b_{\rm f} = \frac{\hat{b}_{\rm g}}{(n_{\rm f} + n_{\rm d})\,t_{\rm{fr}}} = -\frac{\xi}{(n_{\rm g}-1)(n_{\rm f} + n_{\rm d})\,t_{\rm{fr}}} \,.
\end{equation}

\section{Estimator of the variance of the signal\label{app:Gvariance}}

The variance of the of signal estimators $\hat{\rho}^{2}[\hat{g}]$ is computed using first-order approximation
\begin{equation}\label{eq:error_def}
\hat{\rho}^{2}[\hat{g}] = \sum_{i=1}^{n_{\rm g}-1}\sum_{j=1}^{n_{\rm g}-1}\left(\left.\frac{\partial \hat{g}}{\partial \Delta G_i}\right|_{\Delta G_{i} = \hat{g}}\right) \left(\left.\frac{\partial \hat{g}}{\partial \Delta G_j}\right|_{\Delta G_{j} = \hat{g}}\right) D_{ij}(g=\hat{g})\,,
\end{equation}
where $D_{ij}$ is the covariance matrix of group differences defined in Eq.\,(\ref{eq:Dkl}).

As one has 
\begin{equation}
\left.\frac{\partial \hat{g}}{\partial \Delta G_i}\right|_{\Delta G_{i} = \hat{g}} =  \left.\frac{\partial \hat{g}}{\partial \Delta G_j}\right|_{\Delta G_{j} = \hat{g}}\,,
\end{equation}
the Eq.\,(\ref{eq:error_def}) simplifies to
\begin{equation}
\hat{\rho}^{2}[\hat{g}] = \left[(n_{\rm g}-1)\hat{g}  + \alpha \hat{g} + \gamma)\right]\left(\left.\frac{\partial \hat{g}}{\partial \Delta G_i}\right|_{\Delta G_{i} = \hat{g}}\right)^{2}\,,
\end{equation}
and the variance estimator is given by
\begin{equation}
    \hat{\rho}^{2}[\hat{g}] = \frac{[(n_{\rm g}-1)\hat{g}  + \alpha \hat{g} + \gamma]}{(n_{\rm g}-1)^{2}} \frac{(\hat{g}+\beta)^{2}}{(\hat{g}+\beta)^{2} + \xi^{2}} \approx \frac{[(n_{\rm g}-1)\hat{g}  + \alpha \hat{g} + \gamma}{(n_{\rm g}-1)^{2}}\,,
\end{equation}
where the last approximation holds for $\xi \ll \beta, \hat{g}$. Since $\xi \in (\frac{1}{3}, \frac{1}{2}]$ independently on the readout noise and flux values, the last approximation holds most of the \Euclid NISP applications. 

\section{QF expected value \label{app:QFbias}}

The QF, defined in Eq.\,(\ref{eq:QF}), is equal to the bias between $\hat{g}_x[\Delta \vec{G}]_{n_{\rm g}}$ and $\hat{g}_1 = (G_{n_{\rm g}} - G_1)(n_{\rm g}-1)^{-1}$ up to a multiplicative factor $(n_{\rm g}-1)\xi^{-1}$.
Using the facts that 
\begin{enumerate}
    \item $\hat{g}_x[\Delta \vec{G}]_{n_{\rm g}}$, defined in Eq.\,(\ref{eq:gx}), is equal to $\hat{g}[\Delta \vec{G}]_{n_{\rm g}}$, defined in Eq.\,(\ref{eq:gL}), with $\xi=0$;
    \item $\hat{g}_1$ is an unbiased estimator of $g$,
\end{enumerate}
we can use repeat the derivation given in Appendix ~\ref{app:Gbias} to show that 
\begin{align}
 \left\langle \hat{g}_x[\Delta \vec{G}]_{n_{\rm g}} \right\rangle_{n_{\rm r}} \xrightarrow{g \to \infty} g +  \frac{1}{2}\left( \frac{n_{\rm g}-2}{n_{\rm g}-1}(1+\alpha) + 2\beta \right) - \beta = g + \frac{n_{\rm g}-2}{n_{\rm g}-1}\xi\,,
\end{align}
and thus the expected value of the QF is equal to
\begin{equation}
    \langle {\rm QF} \rangle_{n_{\rm r}} = \frac{n_{\rm g}-1}{\xi}\left\langle \hat{g}_x[\Delta \vec{G}]_{n_{\rm g}}  - \hat{g}_1\right\rangle_{n_{\rm r}} \xrightarrow{g \to \infty} \frac{n_{\rm g}-1}{\xi}\frac{n_{\rm g}-2}{n_{\rm g}-1}\xi = n_{\rm g}-2\,.
\end{equation}

\section{Estimator of the QF variance\label{app:QFvariance}}

To derive the variance estimator of the QF we need to use the second-order approximation
\begin{equation}
    \left.\hat{\rho}^{2}[{\rm QF}] \approx \left(\frac{n_{\rm g}-1}{\xi}\right)^2 \left[ \left(\vec{\nabla} \hat{g}_x - \vec{\nabla} \hat{g}_1\right)^T\tens{D}\left(\vec{\nabla} \hat{g}_x - \vec{\nabla} \hat{g}_1\right) + \frac{1}{2}{\rm Tr}\left[\left(\tens{H}_x - \tens{H}_1\right)\tens{D}\left(\tens{H}_x - \tens{H}_1\right)\tens{D}\right] \right]\right|_{\Delta G = g}\,,
\end{equation}
where 
\begin{equation}
    \nabla \hat{g}_{xi}=\frac{\partial \hat{g}_{x}}{\partial \Delta G_i}\,;\quad \nabla \hat{g}_{1i}=\frac{\partial \hat{g}_{1}}{\partial \Delta G_i}\,,
\end{equation} 
are the elements of the gradient vectors of $\hat{g}_{x}$ and $\hat{g}_{1}$, $\tens{D}$ is the covariance matrix defined in Eq.\,(\ref{eq:Dkl}) and  the matrices $\tens{H}_{x}$ and $\tens{H}_{1}$ are the Hessian matrices of $\hat{g}_{x}[\Delta \vec{G}]_{n_{\rm g}}$ and $\hat{g}_{1}[\Delta \vec{G}]_{n_{\rm g}}$ respectively. Their elements are given by
\begin{equation}
    H_{x,ij} = \frac{\partial^2 \hat{g}_{x,1}}{\partial \Delta G_i\partial \Delta G_j}\,;\quad 
        H_{1,ij} = \frac{\partial^2 \hat{g}_{1,1}}{\partial \Delta G_i\partial \Delta G_j}\,.
\end{equation}
Noting that $\nabla \hat{g}_{x}|_{\Delta G = g} = \nabla \hat{g}_{1}|_{\Delta G = g}$, and that $H_{1,ij}\equiv 0$, the variance of QF simplifies to
\begin{equation}\label{eq:varQFsimplified}
    \hat{\rho}^{2}[{\rm QF}] = \left.\frac{1}{2}\left(\frac{n_{\rm g}-1}{\xi}\right)^2 {\rm Tr}[\tens{H}_x\tens{D}\tens{H}_x\tens{D}]\right|_{\Delta G = g}\,.
\end{equation}
Using the explicit form of the elements of $\tens{H}_{x}|_{\Delta G = g}$
\begin{equation}
H_{ij}(g_x) = 
\begin{cases} 
\frac{1}{(n_{\rm g}-1)(g+\beta)}\left( 1 - \frac{1}{n_{\rm g}-1} \right) & \quad \textrm{if}\quad i=j \\
 \frac{-1}{(n_{\rm g}-1)^2(g+\beta)} & \quad\textrm{if}\quad i\neq j
\end{cases}
\,,
\end{equation}
one has 
\begin{equation}
{\rm Tr}[\tens{H}_x\tens{D}\tens{H}_x\tens{D}]|_{\Delta G = g} = \frac{{\rm Tr}[\tens{J}\tens{D}\tens{J}\tens{D}]}{(n_{\rm g}-1)^4(g+\beta)^2} - \frac{2{\rm Tr}[\tens{D}\tens{J}\tens{D}]}{(n_{\rm g}-1)^3(g+\beta)^2} + \frac{{\rm Tr}[\tens{D}\tens{D}]}{(n_{\rm g}-1)^2(g+\beta)^2}
\end{equation}
where $\tens{J}$ is the matrix of ones. Expanding explicitly the above equation with elements of $\tens{H}$ given in Eq.\,(\ref{eq:Dkl}) and inserting it into Eq.\,(\ref{eq:varQFsimplified}) on can show that
\begin{equation}
    \hat{\rho}^{2}[{\rm QF}] \xrightarrow{g \to \infty} 2(n_{\rm g}-2)
\end{equation}
as expected for a variable following $\chi^{2}$ distribution with $n_{\rm g}-2$ degrees of freedom.

\end{appendix}
\end{document}